%% file: Technical3PNPaper.tex
\documentclass[a4paper,twoside]{article} 

\usepackage{ifthen}
\newboolean{adp}
\setboolean{adp}{false}
\newboolean{arxiv}
\setboolean{arxiv}{true}
\newboolean{prd}
\setboolean{prd}{false}
\newboolean{publishing}
\setboolean{publishing}{false}

\newboolean{prdandpublishing}
\setboolean{prdandpublishing}{false}
\ifprd
\ifpublishing
\setboolean{prdandpublishing}{true}
\fi
\fi
\newboolean{notadp}
\setboolean{notadp}{true}
\ifadp
\setboolean{notadp}{false}
\fi
\newboolean{notarxiv}
\setboolean{notarxiv}{true}
\ifarxiv
\setboolean{notarxiv}{false}
\fi
\newboolean{notprd}
\setboolean{notprd}{true}
\ifprd
\setboolean{notprd}{false}
\fi

\newboolean{notpublishing}
\setboolean{notpublishing}{true}
\ifpublishing
\setboolean{notpublishing}{false}
\fi

\newboolean{notprdandpublishing}
\setboolean{notprdandpublishing}{true}
\ifprdandpublishing
\setboolean{notprdandpublishing}{false}
\fi

\usepackage{times}

\usepackage[utf8]{inputenc}
\usepackage[american]{babel}

\ifnotadp
\ifnotprd
\usepackage{a4wide}
\input{fleqn.clo}
\PassOptionsToPackage{fleqn}{amsmath}
\PassOptionsToPackage{fleqn}{amstex}
\fi
\fi

\usepackage{amsmath}
\usepackage{amsfonts}
\usepackage{amssymb}
\allowdisplaybreaks

\ifnotprdandpublishing
  \ifnotarxiv
    \usepackage{xcolor}
    \xdefinecolor{mylinkcolor}{rgb}{0,0,0.5}
    \usepackage[
      bookmarksnumbered, bookmarksopen, bookmarksopenlevel=2,
      breaklinks=true, colorlinks=true,
      filecolor=mylinkcolor, citecolor=mylinkcolor, linkcolor=mylinkcolor,
      urlcolor=mylinkcolor, menucolor=mylinkcolor,
    ]{hyperref}
  \fi

  \ifarxiv 
    \usepackage{xcolor}
    \xdefinecolor{mylinkcolor}{rgb}{0,0,0.5}
    \usepackage[
      bookmarksnumbered, bookmarksopen, bookmarksopenlevel=1,
      colorlinks=true,
      filecolor=mylinkcolor, citecolor=mylinkcolor, linkcolor=mylinkcolor,
      urlcolor=mylinkcolor, menucolor=mylinkcolor,
    ]{hyperref}
  \fi
\fi

\ifnotadp
\ifnotprd
\usepackage[merge,square,comma,numbers,sort&compress]{natbib}
\fi
\fi
\newcommand{\mycite}[2]{
\ifnotadp\cite{#2}\fi%
\ifadp\cite{#1}\fi}

\newcommand{\Sec}[1]{Sect.\ \ref{#1}}

\providecommand{\inst}[1]{$\,^{#1}$}

\def\switchonwide{\ifprd\begin{widetext}\fi}
\def\switchoffwide{\ifprd\end{widetext}\fi}

\input{macros.tex}

\begin{document}

\def\tit{Next-to-next-to-leading order post-Newtonian linear-in-spin binary Hamiltonians}
\def\titleShort{NNLO PN linear-in-spin binary Hamiltonians}



\def\Fauthor{Johannes Hartung}
\def\FauthorShort{J.\ Hartung}
\def\Sauthor{Jan Steinhoff}
\def\SauthorShort{J.\ Steinhoff}

\def\Tauthor{Gerhard Sch\"afer}
\def\TauthorShort{G. Sch\"afer}

\def\FauthorFootnote{Corresponding author\quad
	E-mail:~\textsf{johannes.hartung@uni-jena.de},
	Phone: +49\,3641\,947\,106,
	Fax: +49\,3641\,947\,102}

\def\FauthorInst{1}
\def\SauthorInst{1,2}
\def\TauthorInst{1}

\def\Fauthoremail{johannes.hartung@uni-jena.de}
\def\Sauthoremail{jan.steinhoff@uni-jena.de}
\def\Tauthoremail{gos@tpi.uni-jena.de}


\def\Fadr{Theoretisch--Physikalisches Institut, \\
	Friedrich--Schiller--Universit\"at, \\
	Max--Wien--Platz 1, 07743 Jena, Germany, EU}

\def\Sadr{Centro Multidisciplinar de Astrof\'isica (CENTRA), Departamento de F\'isica, \\
	Instituto Superior T\'ecnico (IST), Universidade T\'ecnica de Lisboa, \\ 
	Avenida Rovisco Pais 1, 1049-001 Lisboa, Portugal, EU}

\def\abs{
The next-to-next-to-leading order post-Newtonian spin-orbit and spin(1)-spin(2) Hamiltonians for 
binary compact objects in general relativity are derived. The Arnowitt-Deser-Misner canonical formalism and 
its generalization to spinning compact objects in general relativity are presented and a fully reduced matter-only 
Hamiltonian is obtained. Several simplifications using integrations by parts are discussed.
Approximate solutions to the constraints and evolution equations of motion are provided. Technical details of the integration
procedures are given including an analysis of the short-range behavior of the integrands around the sources.
The Hamiltonian of a test-spin moving in a stationary Kerr spacetime  is obtained by rather simple approach
and used to check parts of the mentioned results. Kinematical consistency checks by using the global 
(post-Newtonian approximate) Poincar\'{e} algebra are applied.
Along the way a self-contained overview for the computation of the 3PN ADM point-mass Hamiltonian is provided, too.
}

\def\pacsnr{04.25.Nx, 04.20.Fy, 04.25.-g, 97.80.-d, 45.50.Jf}
\def\keyw{post-Newtonian approximation; canonical formalism;
	approximation methods; equations of motion; binary stars}

\def\ack{
This work is supported by the Deutsche Forschungsgemeinschaft (DFG) through projects
GRK 1523 ``Quanten- und Gravitationsfelder,'' STE 2017/1-1,  and SFB/TR7.
We gratefully acknowledge P.~Jaranowski for sharing insight in 3PN point-mass calculations, for useful discussions
on integration procedures, for providing several testintegrals, and for useful discussions on
calculating the spin-precession frequency. 
In particular we thank T.~Damour for very useful comments and hints regarding the calculation
of $\dim$-dimensional short-range behavior of certain integrals. 
Many useful suggestions by an anonymous referee are gratefully acknowledged.
We additionally thank M. Levi for pleasant collaboration on the comparison of our result 
with the potential obtained within the effective field theory approach.
We also thank S.~Hergt and M.~Tessmer for many
useful discussions about the Poincar\'{e} algebra and orbital parameterization issues, respectively.
We further gratefully acknowledge useful discussions 
with D.~Brizuela on {\scshape xPert} and doing perturbation theory in 
arbitrary dimensions,
with M.~Q.~Huber on three-body integral related Appell $F_4$ functions,
with H.~Witek on numerical relativity and literature on some special topics, and 
with T.~J.~Rothe on implementation issues in {\scshape Mathematica} and for giving
us some hints to mathematical theorems. 
}

\ifnotadp
  \title{\tit}
  \ifnotprd
      \author{{\Large \Fauthor\thanks{\FauthorFootnote} \inst{\FauthorInst},
	\Sauthor\inst{\SauthorInst}, and \Tauthor\inst{\TauthorInst} } \medskip \\
	\inst{1} \Fadr \\
	\inst{2} \Sadr}
      \maketitle
      \begin{abstract}
	\abs
      \end{abstract}
      \noindent PACS numbers: \pacsnr      
  \fi
  \ifprd
    \date{\today}
    \author{\Fauthor\inst{\FauthorInst}}
    \email{\Fauthoremail}
    \author{\Sauthor\inst{\SauthorInst}}
    \email{\Sauthoremail}
    \author{\Tauthor\inst{\TauthorInst}}
    \email{\Tauthoremail}
    \pacs{\pacsnr}
    \affiliation{${}^1$\Fadr \\ ${}^2$\Sadr}
    \begin{abstract}
      \abs     
    \end{abstract}
    
    \maketitle    
  \fi
\fi

\ifadp
\title[\titleShort]{\tit}
\author[\FauthorShort]{\Fauthor\footnote{\FauthorFootnote}\inst{\FauthorInst}}
\author[\SauthorShort]{\Sauthor\inst{\SauthorInst}}
\author[\TauthorShort]{\Tauthor\inst{\FauthorInst}}
\address[\inst{1}]{\Fadr}
\address[\inst{2}]{\Sadr}
\begin{abstract}
\abs
\end{abstract}
\maketitle
\fi

\ifnotprdandpublishing
  \ifnotarxiv
    \hypersetup{pdftitle={\tit}, pdfauthor={\Fauthor, \Sauthor, \Tauthor}}
  \fi
\fi

\newcommand{\descappendix}{{Appendix}}
\newcommand{\descappendices}{{Appendices}}
%
%


\section{Introduction}\label{sec:intro}
Since Einstein discovered the theory of general relativity 
\cite{Einstein:1915:1,Einstein:1915:2,Einstein:1915:3,Einstein:1915:4,Einstein:1916}, many attempts to
solve the field equations were undertaken. Yet only a few analytical 
solutions to the full field equations are known at the time being, mostly for highly symmetric matter or field configurations
\cite{Stephani:Kramer:MacCallum:Hoenselaers:Herlt:2003,Griffiths:Podolsky:2009}. The most famous solutions
of the field equations are the Schwarzschild \cite{Schwarzschild:1916} and the Kerr \cite{Kerr:1963} ones.
Even so, an analytic solution for binary systems of black holes (or even neutron stars)
is missing and unlikely to be found in the future. Nevertheless, such binaries are very
interesting. In particular they constitute the most promising and strongest
sources for gravitational waves, one of the most fascinating predictions of
the theory of general relativity \cite{Cutler:Apostolatos:Bildsten:others:1993,Reisswig:Husa:Rezzolla:Dorband:Pollney:Seiler:2009}.

To observe gravitational waves one needs very
sensitive detectors because of the tiny cross section of the waves with matter. 
There exist several ground-based detector projects like e.g. Geo600, VIRGO, and LIGO 
\cite{GEO600home:2011, VIRGOhome:2011, advancedLIGOhome:2011} for this purpose. Their sensitivity increased
during the last years due to continuous upgrades and they probably will detect gravitational waves directly within the next {\em few} years.
The mentioned large-scale detectors were started to be built after gravitational waves had been observed indirectly 
by measuring the change of the radial orbital period e.g.
for the binary pulsar PSR 1913+16 by Hulse and Taylor \cite{Hulse:Taylor:1975} in 1978 (Nobel prize 1993).
Further precise observations of  orbital period decays are the {\em double} pulsar
PSR J0737--3039, \cite{Kramer:Stairs:Manchester:McLaughlin:Lyne:others:2006, Kramer:Wex:2009},
and the white dwarf binary J0651+2844 \cite{Hermes:others:2012}.
Their theoretical prediction is based on the famous quadrupole formula (see e.g. \cite[Eq. (4.12)]{Schafer:1985})
which gives an expression for the orbital averaged energy loss of the whole system due
to gravitational radiation and can thus be translated into a decay of the orbital period.

If gravitational wave signals are detected in the near future, a great effort in data analysis will be necessary to extract
them from the noisy raw data. Necessary ingredients to achieve this goal on the way towards gravitational wave astronomy are 
predictions on which kind of signals can be expected, e.g. from binary sources. The theoretical form of 
the signals -- which should incorporate all known physical effects to some specific order of magnitude 
-- is called template and relies on a solution of the field equations and therefore on the physical parameters 
of the system. 
%
%
Unfortunately, after around one hundred years of research there are no dynamical analytical solutions of the 
field equations for an $n$-body system ($n\ge2$) known. There are two possibilities to circumvent this problem. 
The first one is to rely on numerical simulations. The second possibility is to rely on
approximation methods to extract solutions for these kind of problems from 
the field equations. These include in particular the post-Minkowskian approximation, the post-Newtonian (PN) approximation, extreme mass ratios
(including testmass case), and the effective-one-body approach.

One of the most successful approximation methods is the post-Newtonian approximation, a slow motion and wide
separation approximation. It is used to treat the finite propagation velocity of the gravitational field 
approximately as an instantaneous effect and therefore ``freeze" its dynamics.
Thus the field degrees of freedom are eliminated in this approximation scheme. 
Afterwards one is left with ordinary differential equations for positions, momenta, and spins of the objects
in the system only. It is convenient to encode these equations of motion into a Lagrangian or a Hamiltonian.

To get a more quantitative understanding of the post-Newtonian approximation, consider a gravitationally bound binary system in the Newtonian limit. In this case
one can relate the Newtonian kinetic energy and the Newtonian gravitational potential through the Virial theorem, namely
\begin{align}\label{eq:PNcount}
 \frac{v^2}{c^2} \sim \frac{\gravthree M}{c^2 r} \sim \frac{r_s}{r}\,,
\end{align}
with $v$ denoting a typical orbital velocity in the system, $c$ meaning the speed of light, $\gravthree$ Newton's gravitational constant,
$M$ the total mass of the system, $r$ a typical distance between the gravitating objects, and $r_s$ being the Schwarzschild radius 
of the total system. Every order in $v^2/c^2$ is denoted as one relative
post-Newtonian order. The post-Newtonian approximation was used to obtain matter equations of motion for some special systems
already shortly after general relativity had been developed, see e.g. the 1PN binary Lagrangian in \mycite{Lorentz:Droste:1917:1,Lorentz:Droste:1917:2}{Lorentz:Droste:1917:1,*Lorentz:Droste:1917:2}
and the 1PN equations of motion in e.g. \cite{Einstein:Infeld:Hoffmann:1938}. Furthermore previous results for non-spinning objects
obtained within the formalism used in the present article, namely the canonical formalism of Arnowitt, Deser, and Misner (ADM)
\mycite{Arnowitt:Deser:Misner:1962,Arnowitt:Deser:Misner:2008}{Arnowitt:Deser:Misner:1962,*Arnowitt:Deser:Misner:2008}, are 
the 2PN \cite{Ohta:Okamura:Kimura:Hiida:1974, Damour:Schafer:1985, Damour:Schafer:1988}, 2.5PN 
\cite{Schafer:1985,Schafer:1986}, 3PN \cite{Jaranowski:Schafer:1998, Kimura:Toiya:1972,
Jaranowski:Schafer:1999, Damour:Jaranowski:Schafer:2000,Damour:Jaranowski:Schafer:2001}, 
and 3.5PN \cite{Jaranowski:Schafer:1997,Konigsdorffer:Faye:Schafer:2003} Hamiltonians. 
For various other (non-canonical) derivations of post-Newtonian results for non-rotating objects
see \cite{Blanchet:2006, Futamase:Itoh:2007, Pati:Will:2000, Damour:EspositoFarese:1995,
Goldberger:Rothstein:2006, Gilmore:Ross:2008, Kol:Smolkin:2009, Foffa:Sturani:2011, Foffa:Sturani:2012} 
and references therein.

Regarding the spin corrections to the post-Newtonian approximation, the leading
order can be found in \cite{Barker:OConnell:1975, DEath:1975,
Barker:OConnell:1979, Thorne:Hartle:1985}. Interestingly the leading order equations of
motion were already obtained earlier within the (more general) post-Minkowskian approximation
\cite{Goenner:Gralewski:Westpfahl:1967, Bennewitz:Westpfahl:1971}.
The post-Newtonian next-to-leading order in spin was only tackled more recently
{\mycite{Tagoshi:Ohashi:Owen:2001, Faye:Blanchet:Buonanno:2006, Damour:Jaranowski:Schafer:2008:1, Steinhoff:Hergt:Schafer:2008:2,
Steinhoff:Hergt:Schafer:2008:1, Perrodin:2010,
Porto:2010, Levi:2010, Porto:Rothstein:2008:1, Porto:Rothstein:2008:1:err, Levi:2008}
{Tagoshi:Ohashi:Owen:2001, Faye:Blanchet:Buonanno:2006, Damour:Jaranowski:Schafer:2008:1, Steinhoff:Hergt:Schafer:2008:2,
Steinhoff:Hergt:Schafer:2008:1, Perrodin:2010,
Porto:2010, Levi:2010, Porto:Rothstein:2008:1, *Porto:Rothstein:2008:1:err, Levi:2008}}.
The post-Newtonian next-to-next-to-leading order spin-orbit and spin(1)-spin(2) Hamiltonians are
the subject of the present paper. For very rapidly rotating objects they can be comparable in
strength to 3.5PN and 4PN corrections, respectively.
Half a post-Newtonian order above them the first (leading order) spin-dependent \emph{radiative} Hamiltonians appear.
The spin-orbit and spin(1)-spin(2) ones were already obtained recently \cite{Wang:Steinhoff:Zeng:Schafer:2011}.
At the 3.5PN level one should further include all Hamiltonians cubic
in the spins derived in \cite{Hergt:Schafer:2008:2,
Hergt:Schafer:2008}. Notice that these cubic Hamiltonians are only known for
binary black holes so far, whereas all other mentioned Hamiltonians (including the
ones derived in the present paper) are valid for general compact objects [or have
been generalized to this case, see \mycite{Poisson:1998, Porto:Rothstein:2008:2, 
Steinhoff:Schafer:2009:1, Hergt:Steinhoff:Schafer:2010:1}
{Poisson:1998, Porto:Rothstein:2008:2,
*Porto:Rothstein:2008:2:err, Steinhoff:Schafer:2009:1,
Hergt:Steinhoff:Schafer:2010:1} for the spin(1)-spin(1) level].
However, the Hamiltonians in the present paper are only obtained for binary
systems (but many results for three and more objects already exist \cite{Schafer:1987, Ohta:Kimura:Hiida:1975, Lousto:Nakano:2008, Chu:2009, Hartung:Steinhoff:2010}).
Besides spin effects, tidal contributions to the post-Newtonian approximation become very important for general
compact objects like neutron stars \cite{Damour:Nagar:2009, Vines:Flanagan:2010, Bini:Damour:Faye:2012}, also see e.g.\ \cite{Steinhoff:Puetzfeld:2012} for the extreme mass ratio case.

The present article provides a detailed exposition of how the next-to-next-to-leading order (NNLO) spin-orbit
and spin(1)-spin(2) post-Newtonian Hamiltonians can be derived from an extension
of the canonical formalism of Arnowitt, Deser, and Misner
\mycite{Arnowitt:Deser:Misner:1962,Arnowitt:Deser:Misner:2008}{Arnowitt:Deser:Misner:1962,*Arnowitt:Deser:Misner:2008}.
The mentioned extension of the ADM formalism refers to a generalization from (non-rotating) point-masses (PM) to rotating objects 
\cite{Steinhoff:Schafer:2009:2}, see also \cite{Steinhoff:2011, Steinhoff:Wang:2009, Steinhoff:Schafer:Hergt:2008}.
Results for these Hamiltonians were already presented in this journal \cite{Hartung:Steinhoff:2011:1,Hartung:Steinhoff:2011:2}. 
Their technicalities are also discussed in the (german) PhD thesis of JH \cite{Hartung:2012}.
A corresponding Lagrangian potential for the NNLO spin(1)-spin(2) interaction was derived simultaneously by M. Levi in
\cite{Levi:2011} via an effective field theory (EFT) approach. A comparisons between EFT and ADM results at NNLO spin(1)-spin(2) was not yet undertaken, as it is not straightforward (see \cite{Hergt:2011, Hergt:Steinhoff:Schafer:2011} for a discussion at NLO).
However, in \cite{Marsat:Bohe:Faye:Blanchet:2012,Bohe:Marsat:Faye:Blanchet:2012} the equations of motion in harmonic gauge were calculated at NNLO spin-orbit level and agreement with our Hamiltonian was found.

For unbound systems one can in general not relate the velocities of the objects to the strength of
the gravitational coupling, i.e.\ the basic relation \eqref{eq:PNcount} of the post-Newtonian approximation is not applicable.
For these kinds of systems (e.g. scattering of black holes) a useful approximation is the so called post-Minkowskian approximation, which is an
expansion in powers of the gravitational coupling constant $G$ only and thus also appropriate for very high velocities and weak fields.
The first post-Minkowskian approximation for non-rotating objects was used to derive the Hamiltonian in \cite{Ledvinka:Schafer:Bicak:2008} within the ADM formalism. In principle
the expressions given in \cite{Schafer:1986} can be used to derive the ADM Hamiltonian in the second post-Minkowskian
approximation if there is no incoming radiation. But the integrals have not been given in closed form yet and since we are only 
interested in gravitationally bound systems, we retreat to the post-Newtonian approximation.


The Hamiltonians derived in this article are not the end of the journey. In order to extract useful 
information (i.e. the parameters of a binary) from the gravitational waves, one needs the solution of the post-Newtonian equations of motion for 
the binary.\footnote{The following literature and in particular \cite{Tessmer:2011} gives a complete overview over the research area of parameterization.
See e.g.\ \cite{Damour:Deruelle:1985} for a point-mass 1PN parameterization, 
\mycite{Schafer:Wex:1993,Schafer:Wex:1993:err}{Schafer:Wex:1993,*Schafer:Wex:1993:err} for a point-mass 2PN parameterization, 
\cite{Memmesheimer:Gopakumar:Schafer:2004} for a quasi-Keplerian 3PN point-mass parameterization,
\cite{Wex:1995,Konigsdorffer:Gopakumar:2005} for point-mass parameterizations under leading order spin-orbit coupling, 
\cite{Konigsdorffer:Gopakumar:2006} for using a 3PN point-mass parameterization including radiative dynamics for the phasing of gravitational waves, 
\cite{Tessmer:2009} for a post-equal-mass parameterization of a binary at 3PN point-mass level under leading order spin-orbit coupling, and finally
\cite{Tessmer:Hartung:Schafer:2010} for a parameterization up to 2.5PN with orbital angular momentum aligned spins.
\cite{Tessmer:Hartung:Schafer:2012} incorporates the linear-in-spin Hamiltonians given in the present article into the orbital elements
for orbital momentum aligned spins.} 
For known orbital parameterizations one can further calculate the far-zone radiation 
field (see e.g \cite{Bonnor:1959,Thorne:1980,Blanchet:Damour:1986}
for general formalism of treating radiation in general relativity,
\mycite{Blanchet:Schafer:1989, Blanchet:Schafer:1989:err}{Blanchet:Schafer:1989, *Blanchet:Schafer:1989:err}
for higher order radiation losses in point-mass binaries,
\mycite{Kidder:1995,Blanchet:Buonanno:Faye:2006,
Blanchet:Buonanno:Faye:2006:err,Blanchet:Buonanno:Faye:2006:err:2,Buonanno:Faye:Hinderer:2012}%
{Kidder:1995,Blanchet:Buonanno:Faye:2006,
*Blanchet:Buonanno:Faye:2006:err,*Blanchet:Buonanno:Faye:2006:err:2,Buonanno:Faye:Hinderer:2012}
for spin effects on the radiation,
\cite{Blanchet:Buonanno:Faye:2011} for spin-dependent tail effects,
and \cite{Porto:Ross:Rothstein:2012,Porto:Ross:Rothstein:2010} for multipole moments including spin up to 2.5PN).
In case of eccentric orbits the radiation consists of
several modes which may be extracted by a mode decomposition (see
e.g.\ \cite{Turner:Will:1978, Galtsov:Matiukhin:Petukhov:1980,Pierro:Pinto:Spallicci:Laserra:Recano:2001}
for decomposition of the radiation field in tensor spherical harmonics, and computing and solving 
ordinary differential equations for mean motion and eccentricity in a binary without spin; see also
\cite{Junker:Schafer:1992, MorenoGarrido:Buitrago:Mediavilla:1994,MorenoGarrido:Mediavilla:Buitrago:1995,
Gopakumar:Iyer:2002, Tessmer:Gopakumar:2006, Tessmer:Schafer:2010, Tessmer:Schafer:2011}
for higher order mode decomposition of multipole moments also for point-mass binaries only).
Further for known parameterizations one is also able to calculate the loss of energy and angular momentum during the
inspiral process, see e.g. \cite{Gopakumar:Iyer:1997,Gopakumar:Iyer:2002,Damour:Gopakumar:Iyer:2004} for time evolution effects on the template banks and
\cite{Chandrasekhar:Esposito:1970, Wang:Steinhoff:Zeng:Schafer:2011} for the near-zone luminosity.
This is necessary to construct the mentioned template banks to extract the physical parameters from a noisy signal
via a matched filtering procedure. There the ``fitting factor'' is a very sensitive indicator for the performance of 
the template bank vis-{\`a}-vis the real signal. An introduction to matched filtering can be found in
\cite{Finn:1992,Apostolatos:1995}.
There exists a plethora of articles referring to circular inspiral without spin, e.g. \mycite{Damour:Iyer:Sathyaprakash:2000,Damour:Iyer:Sathyaprakash:2001,Damour:Iyer:Sathyaprakash:2001:err,
Ajith:Babak:Chen:others:2008,Ajith:Babak:Chen:others:2008:err,Buonanno:Chen:Vallisneri:2003,Buonanno:Chen:Vallisneri:2003:err}{Damour:Iyer:Sathyaprakash:2000,Damour:Iyer:Sathyaprakash:2001,*Damour:Iyer:Sathyaprakash:2001:err,
Ajith:Babak:Chen:others:2008,*Ajith:Babak:Chen:others:2008:err,Buonanno:Chen:Vallisneri:2003,*Buonanno:Chen:Vallisneri:2003:err}.

Though the effects considered in the present article are very small,
they still probably have a serious impact on future template banks. The reason for this is
that even tiny contributions to the binary's interaction accumulate during the long inspiral phase (where the post-Newtonian approximation is still valid)
and thus may become observable for potentially planned space based detectors in the future.
During the {\em very} late inspiral phase these effects will become more important, but the 
post-Newtonian approximation will break down
due to the highly nonlinear behavior of the dynamics and high velocities ($v/c \gtrsim 1/3$). To overcome this problem it is most convenient to
extrapolate to this nonlinear regime by resumming the post-Newtonian series. Such a resummation was successfully
implemented into the effective-one-body approach, which analytically provides
complete gravitational waveforms for binary inspiral that are in good agreement with numerical relativity. 
This succeeded so far for point-masses \cite{Damour:Nagar:2009:2, Buonanno:etal:2009} and for non-precessing spins \cite{Pan:etal:2009} 
by calibrations to full numerical simulations, but more work is needed for precessing spins \cite{Damour:2001, Barausse:Buonanno:2009}. 
Here the Hamiltonians derived in the present paper should be very useful, and the spin-orbit one was indeed already incorporated into the 
effective-one-body approach \cite{Nagar:2011,Barausse:Buonanno:2011}.
See also \cite{Taracchini:Pan:Buonanno:Barausse:Boyle:Chu:Lovelace:Pfeiffer:Scheel:2012} for a very complete overview of 
the literature on the effective-one-body approach.
Alternative ways of resumming the post-Newtonian series by Pad\'{e} approximants are possible,
which is most interesting for certain gauge invariant quantities \cite{Damour:Jaranowski:Schafer:2000:2,Damour:Jaranowski:Schafer:2000:3}.
Within the overlap region of post-Newtonian approximation and numerical relativity
in which the gravitational field is not too strong and the number of orbits can
be handled by numerical simulations the results of both approaches can be compared.
The mentioned resummation methods can make these approximate results competitive to numerical relativity%
\footnote{See e.g.\ \cite{Faber:2009,Duez:2010,Rosswog:2010} for reviews, \cite{Shibata:Uryu:2000}
for the first simulation of binary neutron stars, and e.g. 
\cite{Thierfelder:Bernuzzi:Brugmann:2011, Gold:Bernuzzi:Thierfelder:Brugmann:Pretorius:2011,Bernuzzi:Nagar:Thierfelder:Brugmann:2012}
for very recent studies about them.
The first simulations of coalescences of binary black holes were performed in \cite{Pretorius:2005,Baker:Centrella:Choi:Koppitz:vanMeter:2006}.
Furthermore see e.g.\ \cite{Lousto:Zlochower:2008,Campanelli:Lousto:Zlochower:2008,Galaviz:Brugmann:Cao:2010} 
for numerical simulations of a system of more than two black holes and the recent publication \cite{Galaviz:2011} about its chaotic behavior.} 
also in the late inspiral phase
\cite{Damour:Nagar:2009:2, Buonanno:etal:2009}.
Last but not least, a powerful interface between self-force calculations and the ADM Hamiltonians derives from the 
redshift observable and the first law of binary dynamics \cite{LeTiec:Barausse:Buonanno:2012}, which was extended 
to include spin recently \cite{Blanchet:Buonanno:LeTiec:2012}.

For all computations we used {\scshape xTensor} \cite{MartinGarcia:2002}, a free package
for {\scshape Mathematica} \cite{Wolfram:2003}, especially because of its fast index
canonicalizer based on the package {\scshape xPerm} \cite{MartinGarcia:2008}. We also used
the package {\scshape xPert} \cite{Brizuela:MartinGarcia:MenaMarugan:2009}, which is part of {\scshape xTensor},
for performing the perturbative part of our calculations. Furthermore we wrote several 
{\scshape Mathematica} packages ourselves for the various steps of the computation including evaluating integrals.

The article is organized as follows. \Sec{sec:d1splitADM}
shows how to split the spacetime in a $(\dim+1)$ manner, derives the
Hamilton constraint and the momentum constraint from the Einstein-Hilbert action,
and shows the definition of the ADM Hamiltonian. In \Sec{sec:expansionconstraints} the
constraint equations are expanded using a post-Newtonian power counting scheme and
several integrations by parts to simplify subsequent integrations.
Afterwards a transition from the ADM Hamiltonian to a Routhian via a Legendre transform
is performed in \Sec{sec:routhian_waveequation}.
The integrands are composed of various fields and their sources. 
These are explained in detail in \Sec{sec:sources}. From the sources
one can obtain the solution of the constraint equations and wave equations in \Sec{sec:fieldsolutions} and is
ready to integrate. Due to its importance a detailed explanation of the ultraviolet analysis is also provided there.
In \Sec{sec:results} the resulting next-to-next-to-leading order Hamiltonians are given.
The kinematic consistency check of the Hamiltonians, namely the global post-Newtonian
approximate Poincar\'e algebra, is discussed in \Sec{sec:kinematicalconsistency}. 
Also the center-of-mass vectors are uniquely determined from an ansatz in the same section.
Another check is provided in \Sec{sec:testspin}. There
the Hamiltonians from \Sec{sec:results} are compared with the Hamiltonian
of a test-spin in a stationary exterior gravitational field.
After that in \Sec{sec:conclusions} the conclusions are presented
and in the \descappendices\ more details on some of the calculation procedures
used in the former sections are provided. 

The spacetime has $\dim$ spatial dimensions, $1$ time dimension, and metric signature $\dim-1$.
In this article a restriction to $\dim=3$ is explicitly written, if this is 
not the case $\dim$ is always generic. All calculations at the level of the field equations
are performed in arbitrary dimensions, because of the necessary
ultraviolet(UV) analysis concerning the short-range decay of fields around the sources.
Vectors are written in boldface and their components are
denoted by Latin indices from the middle of the alphabet. The scalar product 
between two vectors $\vct{a}$ and 
$\vct{b}$ is denoted by $\scpm{\vct{a}}{\vct{b}} \equiv (\vct{a} \cdot \vct{b})$.
Our units are such that $c=1$. There is no special convention for Newton's gravitational
constant $\grav$. (Notice that $\grav$ is the $\dim$-dimensional coupling strength of
the gravitational field. It has the same numerical value as $\gravthree$ in $\dim=3$
dimensions, but different units.) In the results and the expressions for the sources, $\vmom{a}$ 
denotes the canonical linear momentum of the $a$th object, $\vx{a}$ the canonical 
conjugate position of the object, $m_a$ the mass of the object, 
$\vspin{a}$ and $\spin{a}{i}{j}$ the spin vector and the spin tensor of the object, 
$\relab{ab}=|\vx{a} - \vx{b}|$ the relative distance between two objects, 
and $\vnxa{ab} = (\vx{a} - \vx{b})/\relab{ab}$ 
the direction vector pointing from object $b$ to object $a$. Also important are the
distance between source $a$ and field point, $\relab{a} = |\vct{x} - \vx{a}|$,
the unit vector pointing from source $a$ to the field point $\vnxa{a} = (\vct{x} - \vx{a})/\relab{a}$, and
the circumference of the triangle of source $a$, source $b$ and the field point $\vct{x}$,
given by $s_{ab} = \relab{a} + \relab{b} + \relab{ab}$.
In the binary case the object labels $a, b$ take only the values $1$ and $2$.
The round brackets around the indices of the canonical spin tensor
$\spin{a}{i}{j}$ indicate that its components are given in a local Lorentz
basis, which is essential for the canonical formalism, see \cite{Steinhoff:Schafer:2009:2, Steinhoff:2011}.


\section{ADM Canonical Formalism}\label{sec:d1splitADM}
In the following we introduce the ADM canonical formalism
\mycite{Arnowitt:Deser:Misner:1962,Arnowitt:Deser:Misner:2008,DeWitt:1967,Regge:Teitelboim:1974}
{Arnowitt:Deser:Misner:1962,*Arnowitt:Deser:Misner:2008,DeWitt:1967,Regge:Teitelboim:1974}.
We pay special attention to the dependence on the spatial dimensions $\dim$, as
dimensional regularization is important for the consistency of the post-Newtonian
approximation when point-like sources are utilized \cite{Damour:Jaranowski:Schafer:2001}. The ADM formalism is
extended from non-spinning to spinning point-like sources valid to linear order
in spin here, see \cite{Steinhoff:Schafer:2009:2,Steinhoff:2011} for the case
$\dim=3$.

\subsection{Setting the Canonical Formalism}
Canonical methods usually require one to single out a time coordinate, thus
splitting spacetime into space and time. The geometrically favored way for such
a splitting 
gives rise to the line element
\begin{align}
 \text{d}s^2 &= -\lapse^2 \text{d}t^2 +  \gamma_{ij} (\shiftup{i} \text{d}t + \text{d}x^i) (\shiftup{j} \text{d}t + \text{d}x^j)\label{eq:metricdecomp1}\,,
\end{align}
which corresponds to a $(\dim+1)$-dimensional metric tensor field given by
	\begin{align}
	 g_{\mu\nu} = \left(\begin{array}{cc}
			-\lapse^2 + \shift{i} \shiftup{i} & \shift{i} \\
			\shift{j} & \gamma_{ij}
	              \end{array}\right)\,, \qquad
	 g^{\mu\nu} = \left(\begin{array}{cc}
			-1/\lapse^2 & \shiftup{i}/\lapse \\
			\shiftup{j}/\lapse & \gamma^{ij} - \frac{\shiftup{i} \shiftup{j}}{\lapse^2}
	              \end{array}\right)\label{eq:metricdecomp2}\,,
	\end{align}
where $\lapse$ is the lapse function, $\shiftup{i}$ the shift vector,  $\shift{i} = \gamma_{ij} \shiftup{j}$, and $\gamma_{ij}$ the metric of the spatial slices, $\gamma^{ij}$ being
its inverse \cite{Misner:Thorne:Wheeler:1973,Poisson:2002,Gourgoulhon:2007}. Notice that rewriting the metric tensor field in terms of
lapse, shift, and the $\dim$-dimensional metric $\gamma_{ij}$ corresponds only to another representation of the metric, since together they
have the same number of degrees of freedom. On the one hand a symmetric rank two tensor field like $g_{\mu\nu}$ in $\dim+1$ dimensions has $(\dim+1)(\dim+2)/2$ independent entries and on the other hand, a symmetric rank two tensor in $\dim$ dimensions like $\gamma_{ij}$ has $\dim(\dim+1)/2$ degrees of freedom, the vector $\shiftup{i}$ has $\dim$ independent entries, and the scalar $\lapse$ represents one degree of freedom. Obviously the degrees of freedom match.

The action of the gravitational field is given by the usual Einstein--Hilbert action, namely
\begin{align}
 W_{\field} = \int \text{d}^{\dim+1}x\, \mathcal{L}_{\field} \,, \qquad
 \mathcal{L}_{\field} = \frac{1}{16\pi\grav} \sqrt{-g} \ricci{(\dim+1)}{} \,,
\end{align}
where $g = \det(g_{\mu\nu})$ and $\ricci{(\dim+1)}{}$ is the $(\dim+1)$-dimensional Ricci scalar,
which can be split in a $(\dim+1)$ manner resulting in 
\begin{align}
 \mathcal{L}_{\field} &= \frac{1}{16\pi\grav} \lapse \sqrt{\gamma} \left[\ricci{(\dim)}{} + K_{ij}K^{ij} - (\gamma_{ij} K^{ij})^2\right] + \text{(td)} \,. \label{eq:Ldecomp}
\end{align}
Here $\ricci{(\dim)}{}$ is the $\dim$-dimensional spatial Ricci scalar,
$K_{ij}$ is the extrinsic curvature, $\gamma = \det(\gamma_{ij})$, and (td) denotes a total divergence which
we ignore for now (it will be discussed in \Sec{sec:ADMHam}). We define
\begin{align}
\pi^{ij} &= 16\pi \grav \frac{\partial \mathcal{L}_\field}{\partial \gamma_{ij,0}} 
	= \sqrt{\gamma} (\gamma^{ij}\gamma^{kl} - \gamma^{ik}\gamma^{jl})K_{kl}\,,
\end{align}
where $2 \lapse K_{ij} = - \gamma_{ij,0} + 2 N_{(i;j)}$ 
was used (which deviates from the convention for $K_{ij}$ used in \cite{Poisson:2002}),
a comma denotes a partial derivative, and a
semicolon denotes the $\dim$-dimensional covariant derivative. The inversion reads
\begin{align}
K_{ij} = \frac{1}{\sqrt{\gamma}} \left( \frac{1}{\dim-1} \gamma_{ij}\gamma_{kl} - \gamma_{ik}\gamma_{jl} \right) \pi^{kl} \,,
\end{align}
and the dimension finally enters the calculation explicitly. In order to put the field
equations in the form of Hamilton's canonical equations, we perform a Legendre
transform,
\begin{equation}
\mathcal{L}_{\field} = \frac{1}{16\pi\grav} \pi^{ij} \gamma_{ij,0} - \lapse \srcfield{} + \shiftup{i} \srcfield{i} + \text{(td)} \,,
\end{equation}
where
\begin{subequations}
\begin{align}
	\srcfield{} &= - \frac{1}{16\pi\sqrt{\gamma} \grav} \left[ \gamma \ricci{(\dim)}{} 
		- \gamma_{ij} \gamma_{k l} \pi^{ik} \pi^{jl}
		+ \frac{1}{\dim-1} \left( \gamma_{ij} \pi^{ij} \right)^2 \right]\,,\label{eq:hamconstr} \\
	\srcfield{i} &= \frac{1}{8\pi \grav} \gamma_{ij} \pi^{jk}_{~~ ; k} \label{eq:momconstr}\,.
\end{align}
\end{subequations}
Variation of the action with respect to $\gamma_{ij}$ and $\pi^{ij}$ leads to
$\dim(\dim+1)$ first order evolution equations for these variables. The $\dim$-metric
$\gamma_{ij}$ and the field momentum $\tfrac{1}{16\pi\grav}\pi^{ij}$ are canonically conjugate variables in
the vacuum case.
However, the action does not contain time derivatives of $\lapse$ and $\shiftup{i}$, so a
variation with respect to these variables leads to constraint-type equations
(i.e.\ containing no time derivatives). Variation of $\lapse$ gives only one equation
called the Hamilton constraint, while variation of $\shiftup{i}$ gives $\dim$ equations
called momentum constraint. We have arrived at a canonical
formalism without any restriction on the coordinates, or gauge, but in the
presence of constraints. In the following sections we will eliminate these
constraints by simultaneously fixing the gauge.

In order to couple the ADM formalism to point-like objects possessing masses $m_a$
and $(\dim+1)$-dimensional spin tensors $S_{a\,\mu\nu} = - S_{a\,\nu\mu}$, it is best to start with
a matter action of the form
\begin{align}
 W_{\matter}
 = \sum_a \int \text{d}\tau_a \biggl[
  p_{a\,\mu} u^\mu_a 
  + \frac{1}{2} S_{a\,\mu\nu} \Omega^{\mu\nu}_a 
  - \lambda_a (g^{\mu\nu} p_{a\,\mu} p_{a\,\nu} + m^2_a)
 \biggr]\,,
\end{align}
where $\tau_a$ is a worldline parameter, $p_{a\,\mu}$ the $(\dim+1)$-momentum,
$u^{\mu}_a = \text{d} z_a^{\mu} / \text{d} \tau_a$, $z_a^{\mu}$ the position, $\Omega^{\mu\nu}_a = - \Omega^{\nu\mu}_a$
the angular velocity tensor, and $\lambda_a$ a Lagrange multiplier belonging to the mass-shell constraint
$g^{\mu\nu} p_{a\,\mu} p_{a\,\nu} + m^2_a = 0$ (all for the $a$-th object), see
\cite{Steinhoff:Schafer:2009:2,Steinhoff:2011} for details.
Notice that the angular velocity tensor is built from a Lorentz
matrix encoding the orientation of a body-fixed frame and its covariant
derivative. This means one has to handle a derivative coupling of the 
metric because of Christoffel symbols or Ricci rotation coefficients 
appearing in the covariant derivative of the Lorentz matrix.
Further constraints must be fulfilled for the mentioned Lorentz matrix and the spin, the
latter reads $S^{\mu\nu}_a p_{a\,\mu} = 0$.
It is important that this action is valid to linear order in spin and is now
considered for a generic spatial dimension $\dim$. The equations of motion following
from this matter action are the well-known Mathisson-Papapetrou equations
\mycite{Mathisson:1937,Mathisson:2010,Papapetrou:1951}{Mathisson:1937,*Mathisson:2010,Papapetrou:1951}, see also \cite{Tulczyjew:1959,Dixon:1979},
\begin{align}
 \frac{\text{D} S^{\mu\nu}_a}{\text{d}\tau_a} &= 2 p^{[\mu}_a u^{\nu]}_a \,, \quad 
  \frac{\text{D} p_{a\mu}}{\text{d}\tau_a} = -\frac{1}{2}\; \ricci{(\dim+1)}{\mu\rho\beta\alpha} u^\rho_a S^{\beta\alpha}_a \,,
\end{align}
where $\ricci{(\dim+1)}{\mu\rho\beta\alpha}$ is the $(\dim+1)$-dimensional Riemann tensor,
and the source of the gravitational field equations is given by Tulczyjew's
singular energy-momentum tensor density \cite{Tulczyjew:1959}
\begin{align}
  \sqrt{-g} T^{\mu\nu} &= \sum_a \int \text{d}\tau_a \biggl[
    u^{(\mu}_a p^{\nu)}_a \dl{(\dim+1)a} 
    - \left( S^{\alpha(\mu}_a u^{\nu)}_a \dl{(\dim+1)a} \right)_{\parallel \alpha}
  \biggr]\,,
\end{align}
where $\dl{(\dim+1)a} = \delta(x^{\mu}-z^{\mu}_a)$. For a review of spin in relativity
see e.g.\ \cite{Westpfahl:1967,Westpfahl:1969:1}. Further details on a corresponding
action principle can be found in \cite{Goenner:Westpfahl:1967, Romer:Westpfahl:1969,
Westpfahl:1969:2, Hanson:Regge:1974, Bailey:Israel:1975}.

Next the matter constraints are
eliminated from the action with suitable gauge choices, e.g.\ $\tau_a = t$ where
$t$ is the coordinate time and the matter variables are transformed to (reduced)
canonical variables denoted by a hat, e.g.\ $\hat{z}^i_a$ or $\hat{p}_{ai}$. This is
completely analogous to \cite{Steinhoff:Schafer:2009:2,Steinhoff:2011} and will
therefore not be repeated here. But the following facts are important for the
present article. The spatial dimension $\dim$ is not entering the just mentioned
calculations explicitly, whereas it appears in the gravitational part of the action.
Further, the matter action becomes linear in lapse and shift, so the gravitational
constraints following from their variation now contain matter source terms $\src{}$
and $\src{i}$,
\begin{equation}
\srcfield{} + \src{} = 0 \,, \quad
\srcfield{i} + \src{i} = 0 \label{eq:matterfielddecomp} \,.
\end{equation}
Finally, due to the coupling of the spin to derivatives of the metric, time derivatives of the
gravitational field are present in the matter part. This necessitates a matter
contribution $\pimati{ij}$ to the \emph{canonical} field momentum, which now reads
\begin{align}\label{eq:picandef}
 \picani{ij} & = \pi^{ij} + \pimati{ij} \,.
\end{align}
Explicit expressions for $\src{}$, $\src{i}$, and $\pimati{ij}$ are provided in
\Sec{sec:sources} and have the same form as for $\dim = 3$ in \cite{Steinhoff:Schafer:2009:2,Steinhoff:2011}.

The next goal is the elimination of the field constraints within the so called
ADM transverse-traceless gauge (ADMTT). The corresponding gauge conditions read
\begin{subequations}
\label{eq:diffgaugeconditionsADM}
\begin{align}
 \gamma_{ij,j} - \frac{1}{\dim} \gamma_{jj,i} &= 0 \label{eq:metricdiffADM}\,, \\
 \picani{ii} &= 0 \label{eq:momentumdiffADM} \,.
\end{align}
\end{subequations}
We proceed to work out field decompositions suitable for an elimination of the
field constraints.



\subsection{Metric Decomposition}
The differential gauge condition \eqref{eq:metricdiffADM} for the metric is solved by
\begin{align}
 \gamma_{ij} &= \Psi \delta_{ij} + \htt_{ij} \,,
\end{align}
which can be seen from a transverse-traceless (TT) decomposition of $\gamma_{ij}$.
The first part is the conformally flat part of the metric and the last part can be interpreted in the far-zone as the radiation field,
which is transverse-traceless, i.e.\ $\htt_{ii} = 0 = \htt_{ij,j}$.
Due to the requirement of maximal simple curvature density expression \cite{Faye:Jaranowski:Schafer:2004}, one can 
determine the conformal part of the metric in a form appropriate for a post-Newtonian expansion. Consider
the static (i.e. momentum independent) part of the Hamilton constraint using \eqref{eq:hamconstr} and \eqref{eq:matterfielddecomp}
setting $\pi^{ij} = 0$,
\begin{align}
 \sqrt{\gamma}\; \ricci{(\dim)}{} &= 16\pi\grav \src{}\,.
\end{align}
If one inserts the truncation $\gamma_{ij} = \Psi \delta_{ij}$ with the ansatz $\Psi = \psi^\beta$
into the static Hamilton constraint, one obtains
\switchonwide
\begin{align}
 -\frac{1}{4} \beta (\dim-1) \psi^{-2 + \frac{1}{2} \beta (\dim - 2)} \left(4 \psi \Delta \psi 
+ (-4 + \beta(\dim-2)) (\psi_{,i})^2\right) &= 16\pi\grav\src{}\,,
\end{align}
\switchoffwide
where $\Delta = \partial_i \partial_i$ and $\partial_i$ denotes a partial coordinate derivative.
Demanding that the nonlinear gradient term $(\psi_{,i})^2$ disappears (yielding a Poisson-type equation)
or the $\psi$ in front has a vanishing exponent, both leads to
\begin{align}
 \beta &= \frac{4}{\dim-2}\,,
\end{align}
and thus gives the most simple expression for the curvature density $\sqrt{\gamma}\; \ricci{(\dim)}{}$.
Using this solution, the static Hamilton constraint reduces to
\begin{align}
 -\frac{4(\dim-1)}{\dim-2}\psi \Delta \psi &= 16\pi\grav\src{}\,.
\end{align}
Now one can further set $\psi = 1 + \alpha \phi$ and demand that the static part
of the Hamilton constraint linear in $\phi$ reduces to a Poisson-type
equation without any $\dim$-dependence. This leads to
\begin{align}
 \alpha &= \frac{\dim-2}{4(\dim-1)}\,,
\end{align}
and
\begin{align}
 -\left(1+\frac{\dim-2}{4(\dim-1)}\phi\right) \Delta \phi &= 16\pi\grav\src{}\,.
\end{align}
Finally, the optimized metric decomposition reads
\begin{align}
 \gamma_{ij} &= \left(1+\frac{\dim-2}{4(\dim-1)}\phi\right)^{4/(\dim-2)} \delta_{ij} + \htt_{ij}\,, \label{eq:gammadecomp}
\end{align}
see, e.g., \cite{Damour:Jaranowski:Schafer:2001}.
One can also borrow the conformal factor from the $\dim$-dimensional isotropic Schwarzschild metric given in, e.g.,
\cite[Eq. (22)]{Xanthopoulos:Zannias:1989}. For convenience we introduce 
\begin{align}
 \bar{\phi} &= \frac{\dim-2}{4(\dim-1)} \phi\,, 
\end{align}
in order to absorb certain dimension-depending factors in subsequent calculations.

\subsection{Momentum Decomposition}
For convenience we introduce some differential-integral operators called $\tproj{}{}$-operators, given by
\begin{align}
 \tproj{a}{ij} = \delta_{ij} + (a - 1)\partial_i \partial_j\invlapl{1} = \tproj{a}{ji}\,,
\end{align}
where $\invlapl{1}$ is the inverse Laplacian with usual boundary conditions.
These have the following nice properties
\begin{subequations}
\begin{align}
 \frac{1}{n}\sum_n \tproj{a_n}{ij} &= \tproj{\frac{1}{n}\sum_n a_n}{ij}\,,\eanl
 \tproj{a}{ik} \tproj{b}{kj} &= \tproj{ab}{ij}\,,\eanl
 \tproj{a^{-1}}{ij} &= (\tproj{a}{ij})^{-1}\,,
\end{align}
\end{subequations}
and further
\begin{subequations}
\begin{align}
 \tproj{1}{ij} &= \delta_{ij}\,,\eanl 
 \delta_{ij} \tproj{a}{ij} &= \dim+a-1\,,\eanl
 \partial_i \tproj{a}{ij} &= a \partial_j\,.
\end{align}
\end{subequations}
Note that for $a=0$ one gets the transverse projector
\begin{align}
 \transv{ij} &\defdby \tproj{0}{ij}\,,
\end{align}
which has vanishing divergence and is not invertible. From the multiplication property of the $\tproj{}{}$-operators the projection property
$\transv{ik}\transv{kj} = \transv{ij}$ follows immediately. Also a longitudinal projector can be constructed, namely
\begin{align}
 \longit{ij} &\defdby \tproj{1}{ij} - \tproj{0}{ij} = \delta_{ij} - \transv{ij}\,.
\end{align}
Obviously the longitudinal projector also fulfills the projector condition $\longit{ik}\longit{kj} = \longit{ij}$.
According to \cite{Steinhoff:Wang:2009,Steinhoff:2011} one can decompose the field momentum as
\begin{align}
 \pi^{ij} &= \pitti{ij} + \pitildei{ij} + \pihati{ij}\,. \label{eq:pidecomp}
\end{align}
The different parts are given by
\switchonwide
\begin{subequations}
\begin{align}
 \pitti{ij} &= \biggl[\transv{k(i}\transv{j)\ell} - \frac{1}{\dim-1}\transv{ij}\transv{k\ell}\biggr]\pi^{k\ell} \bydefd \TTproj{k\ell}{ij} \pi^{k\ell}\,,\eanl
 \pitildei{ij} &= \biggl[\longit{k(i}\delta_{j)\ell} + \delta_{k(i} \longit{j)\ell} - \longit{k(i}\longit{j)\ell}-\frac{1}{\dim-1}\transv{ij}\longit{k\ell}\biggr]\pi^{k\ell} \bydefd \LTproj{k\ell}{ij} \pi^{k\ell}\,,\eanl
 \pihati{ij} &= \frac{1}{\dim-1}\transv{ij} \delta_{k\ell} \pi^{k\ell} \bydefd \Trproj{k\ell}{ij} \pi^{k\ell}\,.
\end{align}
\end{subequations}
\switchoffwide
Obviously $\pitti{ij}$ has no divergence and is trace free (it is transverse-traceless), $\pitildei{ij}$ is trace free and its divergence contains
the divergence of the whole field momentum, and $\pihati{ij}$ is divergence free and its trace is the trace of the whole 
field momentum. This decomposition is essentially the usual transverse-traceless-decomposition of a symmetric rank two tensor field, but rearranged in a way more suitable for the present computations. For example,
$\pihati{ij}$ is fixed by the gauge condition \eqref{eq:momentumdiffADM} as
\begin{equation}\label{eq:pitrfix}
\pihati{ij} = - \frac{1}{\dim-1} \transv{ij} \pimati{kk}\,,
\end{equation}
which follows from its definition and the trace of Eq.\ \eqref{eq:picandef}.
Furthermore one can decompose $\pitildei{ij}$ into two different vector potentials,
$\pitildei{i}$ and $V^{i}$. The decompositions read
\begin{align}
 \pitildei{ij} &= \pitildei{i}_{,j} + \pitildei{j}_{,i} - \frac{1}{\dim-1}\tproj{\dim-1}{ij}\pitildei{k}_{,k}\,,\label{eq:momentumtildepitilde}\\
	&= V^i_{,j} + V^j_{,i} - \frac{2}{\dim} \delta_{ij} V^k_{,k}\label{eq:momentumtildevpot}\,.
\end{align}
Of course $\pitildei{i}$ and $V^{i}$ are interrelated,
\begin{equation}
\pitildei{i} = \tproj{2\frac{\dim-1}{\dim}}{ij} V^j \,, \qquad
V^{i} = \tproj{\frac{\dim}{2(\dim - 1)}}{ij} \pitildei{j} \,, \label{eq:pitildevpot}
\end{equation}
and it holds
\begin{equation}
\pitildei{i} = \invlapl{1} \pitildei{ij}_{,j} \,. \label{eq:pisolve}
\end{equation}
These two vector potentials have different advantages. From $V^i$ it is possible to compute $\pitildei{ij}$ without any integration.
On the other hand $\pitildei{i}$ has a simpler structure and can more easily be obtained from the momentum constraint using \eqref{eq:pisolve} [cf.\ \eqref{eq:pitilde3sol}, \eqref{eq:vpot3sol}, and \eqref{eq:momentumtilde3sol}].
The transverse-traceless parts of metric $\gamma_{ij}$, $\htt_{ij}$ and of the canonical field momentum $\picani{ij}$, $\picantti{ij}$ 
are denoted as transverse-traceless degrees of freedom or propagating field degrees of freedom in the following. 
The latter denotation will become clear in the next section.

\subsection{ADM Hamiltonian}\label{sec:ADMHam}
We are now principally able to solve the $\dim+1$ constraint equations
\eqref{eq:matterfielddecomp} for the $\dim+1$ variables $\phi$ and $\pitildei{i}$.
Though this involves solving a system of nonlinear partial differential
equations, it can be tackled analytically within the post-Newtonian approximation up to a
certain order. Notice that $\pihati{ij}$ is fixed by \eqref{eq:pitrfix} and that
$\pitti{ij}$ can be replaced by
\begin{align}
 \pitti{ij} = \picantti{ij} - \TTproj{k\ell}{ij} \pimati{k\ell} \,. \label{eq:piTTreplace}
\end{align}
The independent variables are thus the \emph{reduced} canonical field variables
$\htt_{ij}$ and $\picantti{ij}$ with Poisson brackets
\begin{align}
 \{\htt_{ij}(\vct{x}), \picantti{k\ell}(\vct{x}^\prime)\} &= 16\pi\grav \TTproj{k\ell}{ij}\delta(\vct{x}-\vct{x}^\prime)\,,
\end{align}
as well as reduced canonical matter variables which enter via the matter parts
of the constraint equations \eqref{eq:matterfielddecomp} and are discussed in more
detail in \Sec{sec:sources}.

It was shown in \mycite{Arnowitt:Deser:Misner:1962,Arnowitt:Deser:Misner:2008,DeWitt:1967,Regge:Teitelboim:1974}
{Arnowitt:Deser:Misner:1962,*Arnowitt:Deser:Misner:2008,DeWitt:1967,Regge:Teitelboim:1974}
using different methods that the fully reduced Hamiltonian after gauge fixing
is given by the ADM energy
\begin{align}
E_{\text{ADM}} = \frac{1}{16\pi\grav} \oint \text{d}^{\dim-1} s_i [\gamma_{ij,j} - \gamma_{jj,i}]\,,
\end{align}
where $\oint \text{d}^{\dim-1} s_i$ denotes an integral over the asymptotic
boundary of a spatial hypersurface at fixed time. 
(The identical expression for the energy follows from the Landau-Lifshitz 
superpotential which is related with the well-known Landau-Lifshitz 
stress-energy-momentum pseudotensor of the gravitational field, see \cite{Landau:Lifshitz:Vol2:4}.)
More precisely, the ADM energy
$E_{\text{ADM}}$ turns into the ADM Hamiltonian $H_{\text{ADM}}$ if it is
expressed in terms of the mentioned reduced canonical variables. Inserting the
metric decomposition \eqref{eq:gammadecomp}, the surface integral can be
transformed into a volume integral,
\begin{align}
H_{\text{ADM}} = -\frac{1}{16\pi\grav}\int \text{d}^\dim x\,\Delta \phi\,, \label{eq:HADM}
\end{align}
where the asymptotic behavior of certain field components was used, see
e.g.\ \cite{Regge:Teitelboim:1974}. It was argued by Regge and Teitelboim \cite{Regge:Teitelboim:1974}
that one must modify the total divergence in \eqref{eq:Ldecomp} such that it
leads to the surface integral expression for the ADM energy $E_{\text{ADM}}$,
otherwise the variational principle is not well posed. Indeed, for the same
reason one should add the (regularized) York--Gibbons--Hawking, or ``trace K,'' surface term
\cite{York:1972, Gibbons:Hawking:1977, Wald:1984, York:1986} already to the
Einstein--Hilbert action. The ADM energy then directly arises from a surface term
contained in the complete action, see \cite{Brown:York:1993,Hawking:Horowitz:1996} and also \cite{Poisson:2002}.
Notice that the Einstein equations can be followed from a variation of the action by disregarding all surface terms, but this is in general not allowed if the variation of the action has prescribed nontrivial behavior at the boundary.


The ADM Hamiltonian still depends on the reduced canonical field variables
$\htt_{ij}$ and $\picantti{ij}$. In \Sec{sec:routhian_waveequation} also these
remaining field variables will be eliminated through a Routhian approach to arrive
at a conservative matter-only Hamiltonian.


\section{Expansion of the Constraints and Integrations by Parts}\label{sec:expansionconstraints}
As already stated in \Sec{sec:intro} we will use the post-Newtonian approximation throughout this article. Furthermore 
as mentioned above we utilize the spin-extended ADM formalism. So first of all we have to solve the constraint
equations order by order. This requires to expand them in powers of the post-Newtonian approximation parameter \eqref{eq:PNcount} with the
decompositions \eqref{eq:gammadecomp} and \eqref{eq:pidecomp} (which are adapted to the ADMTT gauge)
inserted. This is the task performed in the following subsection.
\subsection{Order Counting}\label{subsub:ordercounting}
The field and source expansions starting at their leading order are given by
\switchonwide
\begin{subequations}
\label{eq:fieldexpansions}
\begin{align}
 \phi & = \phis{2} + \phis{4} + \phis{6} + \phis{8} + \phis{10} + \dots \,,\\
 \pitildei{ij} & = \momls{3}{ij} + \momls{5}{ij} + \momls{7}{ij} + \dots \,,\\
 \src{} & = \srcs{2} + \srcs{4} + \srcs{6} + \srcs{8} + \srcs{10} + \dots \,, \\
 \src{i} & = \srcis{3}{i} + \srcis{5}{i} + \srcis{7}{i} + \dots \,,
\end{align}
where the subscript in round brackets denotes the $\cInv{1}$ order. A similar order counting
is also valid for all derived fields, like vector potentials.
At a later stage of the calculation we also need to expand $\htt_{ij}$ and $\picantti{ij}$,
\begin{align}
 \htt_{ij} & = \htt_{(4)ij} + \htt_{(6)ij} + \dots\,,\label{eq:httorders}\\
 \picantti{ij} & = \canpi^{ij}_{(5)\,\TT} + \dots\,.
\end{align}
\end{subequations}
\switchoffwide
At the mentioned stage we also need an order counting for the deviations between field momentum and
canonical field momentum due to $\pimati{ij}$ and the tracelessness violation of the field momentum $\pihati{ij}$.
As one can see in \Sec{sec:matmom} $\pihatmati{ij}$ starts at $\Order{\cInv{9}}$
and thus cannot contribute to the expansion of the Hamilton and momentum constraint later, while
$\pimati{ij}$ starts at $\Order{\cInv{5}}$.

In anticipation of the source expressions in \Sec{sec:sources} we will introduce the matter variables used there to discuss the
order counting of the field variables, because all fields depend on the source expression and thus the order counting of the matter variables.
The mass $m_a$, canonical matter momentum $\vmom{a}$, and spin variables $\vspin{a}$ are {\em formally} counted as 
$m_a \sim \Order{\cInv{2}}$, $\vmom{a} \sim \Order{\cInv{3}}$, and $\vspin{a} \sim \Order{\cInv{3}}$ for 
dimensional reasons only (remember that for {\em maximal} spins one would have $\vspin{a} \sim \Order{\cInv{4}}$ instead, see e.g. \cite[Appendix A]{Steinhoff:Wang:2009}).
This counting comes from the fact that after setting $c=\grav=1$ we require all quantities to be in units of length. 
Let us introduce symbols with a bar below them being the quantities in SI units and the other symbols the quantities 
in units of length, then it holds $m_a = \tfrac{\grav}{c^2} \underline{m}_a$ for the mass, $t = c \underline{t}$ for the time,
$\vmom{a} = \tfrac{\grav}{c^3} \underline{\vcanmom}_{a}$ for the linear momentum,
and similarly for the spin variables. Although we mentioned that $\grav$ has different units in $\dim\ne3$ than in $\dim=3$, 
we treat their $\cInv{1}$ order as in $\dim=3$ for simplicity.
So the order counting comes from the $c$ powers inserted to reconstruct the SI units.
It should be noted that these counting rules will in general not give correct \emph{absolute} orders in $c$ if the SI units of the
final expression are not taken into account. However, \emph{relative} orders are always meaningful, which is all that is relevant
for perturbative expansions. Further notice that different counting rules are obtained if one assumes that all quantities are
expressed in terms of mass units instead of length units when setting $c=\grav=1$, which is also often used in the literature.
\subsection{Hamilton Constraint}
The Hamilton constraint
\switchonwide
\begin{align}
 \frac{1}{16\pi\sqrt{\gamma} \grav} \left[
	\gamma \; \ricci{(\dim)}{}
	- \gamma_{ij} \gamma_{k l} \pi^{ik} \pi^{jl}
	+ \frac{1}{\dim-1} \left( \gamma_{ij} \pi^{ij} \right)^2 \right] = \src{}\,,
\end{align}
has to be expanded in a post-Newtonian manner. After inserting the metric and momentum decomposition, \eqref{eq:gammadecomp} and \eqref{eq:pidecomp},
and all field expansions \eqref{eq:fieldexpansions} it decomposes into
several Poisson equations for the post-Newtonian potential, namely
\begin{subequations}\label{eq:hamiltonconstraint}
\begin{align}
 -\frac{1}{16 \pi \grav} \Delta \phis{2} & =  \srcs{2}\label{subeq:phi2constr}\,,\eanl
 -\frac{1}{16 \pi \grav} \Delta \phis{4} & =  \srcs{4} 
	- \phibs{2} \srcs{2}\label{subeq:phi4constr}\,,\eanl
 -\frac{1}{16 \pi \grav} \Delta \phis{6} & =  \srcs{6} 
	- \phibs{2} \srcs{4} 
	+ (
		-\phibs{4} 
		+ \phibs{2}^2
	) \srcs{2}
 	- \frac{1}{16 \pi \grav}\biggl(
		- (\momls{3}{ij})^2 \neanl &
		+ 4 (\phibs{2} \htt_{ij})_{,ij}
	\biggr)\,,\eanl
 -\frac{1}{16 \pi \grav} \Delta \phis{8} & =  \srcs{8} 
	- \phibs{2} \srcs{6} 
	+ (
		-\phibs{4} 
		+ \phibs{2}^2
	)\srcs{4}
 	+ ( 
		- \phibs{6} 
		+ 2 \phibs{2} \phibs{4} \neanl &
		- \phibs{2}^3
	) \srcs{2}
	- \frac{1}{16 \pi \grav } \biggl(
		\frac{3\dim - 10}{\dim - 2} \phibs{2} (\momls{3}{ij})^2 
		- 2 \momls{3}{ij} \momls{5}{ij} 
		- 2 \momls{3}{ij} \pitti{ij} \neanl &
		+ \frac{4}{\dim - 2} \htt_{ij} \phibs{2}_{,i}\phibs{2}_{,j} 
		- \frac{1}{4} (\htt_{ij,k})^2 
		-\frac{16}{\dim - 2} (\phibs{2} \phibs{2}_{,j} \htt_{ij})_{,i} 
		+ 4 (\phibs{4} \htt_{ij})_{,ij} \neanl &
		+ \frac{1}{2} \Delta (\htt_{ij})^2 
		- \frac{1}{2} (\htt_{ik} \htt_{jk})_{,ij}
	\biggr)\,,\eanl
 -\frac{1}{16 \pi \grav} \Delta \phis{10} & = \srcs{10} 
	- \phibs{2} \srcs{8}
	+ (
		-\phibs{4}
		+ \phibs{2}^2
	)\srcs{6} 
	+ ( 
		- \phibs{6} 
		+ 2 \phibs{2} \phibs{4} \neanl &
		- \phibs{2}^3
	) \srcs{4} 
	+ \biggl(
		-\phibs{8} 
		+ 2\phibs{2} \phibs{6} 
		+ \phibs{4}^2 
		- 3 \phibs{4} \phibs{2}^2 
		+ \phibs{2}^4 \neanl &
		+ \frac{\dim - 4}{8(\dim - 1)} (\htt_{ij})^2
	\biggr)\srcs{2}
	- \frac{1}{16 \pi \grav} \biggl(
		\frac{3\dim - 10}{\dim - 2} \phibs{4} (\momls{3}{ij})^2 
		- 2 \htt_{ij} \momls{3}{ik} \momls{3}{jk} \neanl &
		- (\momls{5}{ij})^2 
		- 2\momls{3}{ij}\momls{7}{ij} 
		- (\pitti{ij})^2
		- 2 \frac{(\dim - 3)(3\dim - 10)}{(\dim - 2)^2} (\momls{3}{ij})^2 \phibs{2}^2 \neanl &
		+ \frac{8}{\dim - 2} \htt_{ij} \phibs{2}_{,i} \phibs{4}_{,j} 
		+ 2\frac{3\dim - 10}{\dim - 2} \momls{3}{ij} \momls{5}{ij} \phibs{2} 
		+ 2 \frac{3\dim - 10}{\dim - 2} \momls{3}{ij} \pitti{ij} \phibs{2} \neanl &
		- 4 \frac{\dim + 2}{(\dim - 2)^2} \htt_{ij} \phibs{2}_{,i} \phibs{2}_{,j} \phibs{2} 
		- \frac{1}{2} \htt_{ij,k} \htt_{ik,j} \phibs{2}
		- \frac{1}{4}\frac{\dim - 10}{\dim - 2} (\htt_{ij,k})^2 \phibs{2}
	\biggr) \neanl & 
	+ \td\,. \label{eq:ham10}
\end{align}
\end{subequations}
By virtue of \eqref{eq:HADM} the post-Newtonian Hamiltonians follow from an
integration of the right hand sides of these equations.
The last equation leads to the formal 3PN ADM Hamiltonian.
\subsection{Momentum Constraint}
Also the momentum constraint has to be expanded in a post-Newtonian manner. 
First of all one has to write the covariant divergence in a more explicit form.
For convenience it is also useful to write as much terms as possible in terms of
divergences of a traceless symmetric tensor, see \cite{Steinhoff:Wang:2009}.
Notice that we also did not remove the trace parts of the field momentum in
\begin{align}
 \momli{ij}_{\;,j} & = - 8 \pi \grav \;\src{i} + \biggl[
		\left(1 - \left(1+\bar{\phi}\right)^{4/(\dim-2)}\right)(\momli{ij} + \pitti{ij}) 
		+ V^k (\htt_{kj,i} + \htt_{ik,j} -  \htt_{ij,k}) \neanl &
		- \left(1 - \frac{2}{\dim}\right) V^{k}_{\;,k} \htt_{ij}
	\biggr]_{,j} 
	- \Delta(\htt_{ik} V^k) 
	+ \frac{1}{2} \htt_{\ell j,i} \pitti{\ell j} 
	- (\htt_{ik} \pitti{kj})_{,j} \neanl &
	+ \frac{1}{2} \htt_{\ell j,i} \pihati{\ell j} 
	- (\htt_{ik} \pihati{kj})_{,j} 
	- \frac{4}{\dim-2} \left(1+\bar{\phi}\right)^{(6-\dim)/(\dim-2)} \left(
	    \pihati{\ell i} \bar{\phi}_{,\ell} 
	    - \frac{1}{2} \pihati{\ell \ell} \bar{\phi}_{,i}
	\right)\,,
\end{align}
as the tracelessness condition may be violated for the non-canonical field momentum due to the spin.
The post-Newtonian expansion is necessary for the later integrations by parts and
to obtain the necessary field solutions for the momentum type fields. The expansion reads 
\begin{subequations}
\label{eq:momentumconstraint}
\begin{align}
 \momls{3}{ij}{}_{,j} & = - 8 \pi \grav \srcis{3}{i} \label{subeq:pi3constr}\,,\eanl
 \momls{5}{ij}{}_{,j} & = - 8 \pi \grav \srcis{5}{i} + \biggl[-\frac{4}{\dim-2} \momls{3}{ij} \phibs{2}\biggr]_{,j} \label{subeq:pi5constr} \,,\eanl
 \momls{7}{ij}{}_{,j} & = - 8 \pi \grav \srcis{7}{i} + \biggl[
	-\frac{4}{\dim-2} (\pitti{ij} + \momls{5}{ij}) \phibs{2} 
	+ \frac{2(\dim - 6)}{(\dim - 2)^2} \momls{3}{ij} \phibs{2}^2 
	- \frac{4}{\dim - 2} \momls{3}{ij} \phibs{4} \neanl &
	- \biggl(1-\frac{2}{\dim}\biggr) \htt_{ij} \vpots{3}{k}_{,k}
	+ \vpots{3}{k} \biggl(
		\htt_{jk,i} 
		+ \htt_{ik,j}
		- \htt_{ij,k}
	\biggr)
\biggr]_{,j} - \Delta (\htt_{ik} \vpots{3}{k}) \label{subeq:pi7constr}\,.
\end{align}
\end{subequations}
\switchoffwide
Notice that we did not insert the expansion for $\htt_{ij}$ and $\pitti{ij}$ since
this is only possible after their evolution were obtained and solved order by order later on.
\subsection{Integration by Parts}
Due to the complicated structure of $-\Delta \phis{10}/(16\pi\grav)$ in particular the appearance of $\phibs{8}$, $\phibs{6}$ and
$\momls{5}{ij}$, $\momls{7}{ij}$ it is necessary to simplify the integral over the right hand side of \eqref{eq:ham10}. Some of the 
mentioned fields are not even known in $\dim=3$ dimensions. The best way to remove them is to integrate by parts certain terms and 
afterwards use lower order Hamilton and momentum constraints.

In the used dimensional regularization one may always neglect boundary terms if
the integrands are not UV- and IR-divergent simultaneously. These more subtle terms
occur at 4PN point-mass calculations for the first time.
In the present calculation we always neglected boundary terms in the integrations by parts.

The parts of the Hamilton constraint coming from the expanded Ricci scalar in the conformal approximation have always a structure 
where a power of the post-Newtonian potential is coupled to a matter source of the Hamilton constraint. These terms can be simplified 
by inserting the lower order Hamilton constraint for the source and shift the emerging Laplacian to the coupled post-Newtonian potential 
via integrating by parts twice. This procedure can be used to eliminate $\phibs{8}$ and $\phibs{6}$ and the appropriate calculations 
are given by
\begin{align}
\dunderline{- \phibs{8} \srcs{2}} & = 
	-\phibs{2} \srcs{8} 
	+ \phibs{2}^2 \srcs{6} 
	+ (\phibs{4} \phibs{2} - \phibs{2}^3) \srcs{4} 
	+ (
		\phibs{6} \phibs{2} \neanl &
		- 2 \phibs{2}^2 \phibs{4} 
		+ \phibs{2}^4
	) \srcs{2} 
	- \frac{1}{16 \pi \grav} \biggl\{
		-\frac{3\dim - 10}{\dim - 2} \phibs{2}^2 (\momls{3}{ij})^2 \neanl &
		+ 2 \phibs{2} \momls{3}{ij} (\momls{5}{ij} + \pitti{ij}) 
		- \frac{4}{\dim - 2} \htt_{ij} \phibs{2}{}_{,i} \phibs{2}{}_{,j} \phibs{2}
		+ \frac{1}{4} \phibs{2} (\htt_{ij,k})^2 \neanl &
		+ \frac{16}{\dim - 2} (\phibs{2} \phibs{2}{}_{,i} \htt_{ij})_{,j} \phibs{2} 
		- 4 (\phibs{4} \htt_{ij})_{,ij} \phibs{2}
		- \frac{1}{2} \phibs{2} \Delta(\htt_{ij})^2 \neanl &
		+ \frac{1}{2} \phibs{2} (\htt_{ik} \htt_{jk})_{,ij}
	\biggr\} 
	+ \td\,,\label{eq:phi8h2PI}\\
\dunderline{-\phibs{6} \srcs{4}} & = 
	-\phibfo \srcs{6} 
	+\phibs{2} \phibfo \srcs{4} 
	+(
		\phibs{4} \phibfo 
		- \phibs{2}^2 \phibfo
	) \srcs{2} \neanl &
	- \frac{1}{16 \pi \grav} \biggl\{
		\phibfo (\momls{3}{ij})^2 
		- 4 \phibfo (\phibs{2} \htt_{ij})_{,ij}
	\biggr\} + \td\,, \label{eq:phi6src4PI} \\
\dunderline{3 \phibs{6} \phibs{2} \srcs{2}} & = \biggl\{
	-3\srcs{6} 
	+ 3\phibs{2} \srcs{4} 
	- (-3\phibs{4} + 3\phibs{2}^2)\srcs{2} \neanl &
	- \frac{1}{16 \pi \grav} \biggl[
		3(\momls{3}{ij})^2 
		- 12 \phibs{2}{}_{,ij} \htt_{ij}
	\biggr]
\biggr\}\phibft + \td\,.
\end{align}
The fields $\phibfo$ and $\phibft$ are the momentum dependent and the momentum independent part of $\phibs{4}$ given by 
$\phibfo = -16\pi\grav \tfrac{\dim-2}{4(\dim-1)}\Delta^{-1}\srcs{4}$ and $\phibfo = -16\pi\grav \tfrac{\dim-2}{4(\dim-1)}\Delta^{-1}(-\phibs{2}\srcs{2})$.
It is possible to simplify \eqref{eq:phi8h2PI} once more using
\begin{align}
 \dunderline{-\frac{1}{16\pi\grav}\left(-\frac{1}{2} \phibs{2} \Delta (\htt_{ij})^2\right)} & = -\frac{\dim - 2}{8(\dim - 1)} (\htt_{ij})^2 \srcs{2} + \td\,.
\end{align}

In the Hamilton constraint there are also several terms coupling different orders of the longitudinal field momentum. These terms can also be
simplified by inserting the decomposition \eqref{eq:momentumtildevpot}, removing the derivatives from the vector potential $V^i$ via an integration
by parts and inserting the lower order momentum constraints \eqref{eq:momentumconstraint} into the divergencies of the longitudinal field momentum.
This procedure is used in order to eliminate $\momls{5}{ij}$ and $\momls{7}{ij}$.
The respective integrations by parts are given by
\begin{align}
 \dunderline{-\frac{1}{16 \pi \grav} \left(-2 \momls{7}{ij} \momls{3}{ij}\right)} & = 
	2 \vpots{3}{i} \srcis{7}{i} 
	- \frac{1}{16 \pi \grav}\biggl\{
		-2 \htt_{ij} \biggl[
			-2\vpots{3}{j}_{,k} \momls{3}{ik} 
			+(\vpots{3}{k} \momls{3}{ij})_{,k} 
		\biggr] \neanl &
 		+ \frac{8}{\dim - 2} \momls{3}{ij} (\pitti{ij} + \momls{5}{ij}) \phibs{2} 
		- 4 \frac{\dim - 6}{(\dim - 2)^2} (\momls{3}{ij})^2 \phibs{2}^2 \neanl &
		+ \frac{8}{\dim - 2} (\momls{3}{ij})^2 \phibs{4}\biggr\} 
+ \td\,.
\end{align}
Last but not least the $(\momls{5}{ij})^2$ integration by parts is given by
\begin{align}
\dunderline{- \frac{1}{16 \pi \grav}\left(-( \momls{5}{ij} )^2\right)} & = 
	\vpots{5}{i} \srcis{5}{i} 
	- \frac{1}{16 \pi \grav} \biggl\{ 
		\frac{4}{\dim - 2} \momls{3}{ij} \momls{5}{ij}\phibs{2}
	\biggr\} + \td\,.
\end{align}
Although it is good to express field integrals in terms of sources and fields, the $\vpots{5}{i}$
potential is still very complicated.
Thus we try to express $(\momls{5}{ij})^2$ in yet another way.
From the transverse-traceless projection of an arbitrary second rank tensor field $A_{ij}$, namely
\begin{align}
 \TTproj{ij}{k\ell} A_{k\ell} & = \biggl(\delta_{i(k} \delta_{\ell)j} - \delta^{\text{LT}\,ij}_{k\ell} - \delta^{\text{Tr}\,ij}_{k\ell}\biggr)A_{k\ell}\,, \\
& = A_{ij} - \invlapl{1} \biggl((A_{j\ell})_{,\ell i} + (A_{i\ell})_{,\ell j} - \frac{1}{\dim - 1}\tproj{\dim-1}{ij} (A_{k\ell})_{,k\ell}\biggr) \neanl & \quad
- \frac{1}{\dim-1}\left(\delta_{ij} - \partial_i \partial_j \invlapl{1}\right)\delta_{k\ell} A_{k\ell}\,,\label{eq:ttprojection}
\end{align}
one can get the transverse-traceless projection ($\phibs{2} \momls{3}{k\ell}$ is traceless)
\begin{align}
 \TTproj{ij}{k\ell} (\phibs{2} \momls{3}{k\ell}) &= \phibs{2} \momls{3}{ij} + \frac{\dim-2}{4} \left\{\momls{5}{ij} + \momlones{5}{ij}\right\}\,, \label{eq:TTphipi}
\end{align}
via the momentum constraint \eqref{eq:momentumconstraint}.
Here
\begin{align}
 \momlones{5}{ij} &= \momlones{5}{i}_{,j} + \momlones{5}{j}_{,i} - \frac{1}{\dim-1}\tproj{\dim-1}{ij} \momlones{5}{k}_{,k} \label{eq:momentumtilde51}\,,
\end{align}
where
\begin{align}
 \momlones{5}{i} &= 8\pi\grav\invlapl{1}\srcis{5}{i}\label{eq:pitilde51}\,.
\end{align}
Notice that $\momlones{5}{i}$ is defined with a different sign than the usual vector potentials used throughout this article.
Now we can express $\momls{5}{ij}$ in terms of $\momlones{5}{ij}$, $\phibs{2}\momls{3}{ij}$ and the transverse-traceless projection \eqref{eq:TTphipi}.
From these considerations it follows that $(\momls{5}{ij})^2$ is given by
\begin{align}
 \dunderline{- \frac{1}{16 \pi \grav}\left(-( \momls{5}{ij} )^2\right)} & = 
  -\frac{1}{16\pi\grav} \biggl\{
     \frac{16}{(\dim-2)^2}\left[
	\left(\TTproj{ij}{k\ell}(\phibs{2}\momls{3}{k\ell})\right)^2
	-\phibs{2}^2(\momls{3}{ij})^2
     \right] \neanl &
     -\frac{8}{\dim-2}\phibs{2}\momls{3}{ij}\momlones{5}{ij}
     -(\momlones{5}{ij})^2
   \biggr\} + \td\,.
\end{align}
Some of the results above were partially checked by comparison with \cite{Steinhoff:Wang:2009}.

\subsection{The formal 3PN ADM Hamiltonian}
Performing the mentioned integration by parts leads to the following formal 3PN ADM Hamiltonian which can also be compared 
to \cite{Steinhoff:Wang:2009},
\begin{align}
  H_{\text{3PN}} & = \int \text{d}^\dim x \biggl[
	\srcs{10} 
	- 2 \phibs{2} \srcs{8} 
	+ (- \Sfour -4 \phibs{4} + 2 \phibs{2}^2)\srcs{6} 
 	+ (6 \phibs{2} \phibs{4} + \Sfour \phibs{2} \neanl &
		-2\phibs{2}^3) \srcs{4} 
	+ \biggl(
		4 \phibs{4}^2 
		+ \Sfour \phibs{4} 
		- 8 \phibs{4} \phibs{2}^2 
		- \Sfour \phibs{2}^2 
		+ 2\phibs{2}^4 \neanl &
		- \frac{1}{4(\dim - 1)} (\htt_{ij})^2
	\biggr)\srcs{2} 
	+ 2 \vpots{3}{i} \srcis{7}{i} 
	- \frac{1}{16 \pi \grav} \biggl(
		-(\momlones{5}{ij})^2 
		+ 2\frac{3\dim - 4}{\dim - 2} \phibs{4} (\momls{3}{ij})^2 \neanl &
		+ \Sfour (\momls{3}{ij})^2  
		- \frac{(3\dim-4)(3\dim-2)}{(\dim - 2)^2} (\momls{3}{ij})^2 \phibs{2}^2 
		- (\pitti{ij})^2 
		+ 8 \phibs{2} \momls{3}{ij} \pitti{ij} \neanl &
		- 2 \htt_{ij} \biggl[
			-2\vpots{3}{j}_{,k} \momls{3}{ik} 
			+(\vpots{3}{k} \momls{3}{ij})_{,k}
			+  \momls{3}{ik} \momls{3}{jk} 
			-4\frac{2\dim - 3}{\dim - 2} \phibs{2}_{,i} \phibs{4}_{,j} \neanl &
			-4\frac{3\dim - 4}{(\dim - 2)^2} \phibs{2} \phibs{2}_{,i} \phibs{2}_{,j} 
			-2\Sfour_{,i} \phibs{2}_{,j}
		\biggr]
		-8 \frac{\dim-1}{\dim-2} \phibs{2}\momlones{5}{ij}\momls{3}{ij} \neanl & 
		+16 \frac{2\dim-3}{(\dim-2)^2} \left(\TTproj{ij}{k\ell}(\phibs{2}\momls{3}{k\ell})\right)^2 
		+ \frac{2}{\dim - 2} (\htt_{ij,k})^2 \phibs{2}
	\biggr)
\biggr]\label{eq:H3PN}\,.
\end{align}
We changed all occurrences of $\phibfo$ to
$\Sfour = -2 \phibfo$ to gain a result which is comparable to \cite{Damour:Jaranowski:Schafer:2001}.
Note that, since we did not expand $\htt_{ij}$ and $\picantti{ij}$, we also need some formal 2PN terms which
may contribute to the 3PN kinetic energy or the 3PN interaction Hamiltonian after expanding $\htt_{ij}$,
\begin{align}
 H_{\text{2PN}} & = \int \text{d}^\dim x \biggl[\srcstt{8} + \dots
 - \frac{1}{16 \pi \grav } \biggl(\dots + \frac{4 (\dim - 1)}{\dim - 2} \phibs{2}{}_{,i} \phibs{2}{}_{,j} \htt_{ij} - \frac{1}{4} (\htt_{ij,k})^2 \biggr)\biggr]\label{eq:H2PNTT}\,.
\end{align}
Now we can split up the Hamiltonian into a 
kinetic part for the reduced canonical field variables [$\htt_{ij}$ and $\picantti{ij}$; after inserting \eqref{eq:piTTreplace}], 
an interaction part between these canonical fields and constraint fields, and a part independent of the canonical fields, i.e.,
\begin{align}
 	H_{\ADM} = H_{\ADM}^{\interaction} 
	+ H_{\ADM}^{\nonTT}
	+ \frac{1}{16\pi\grav}\int {\rm d}^\dim x \left[\frac{1}{4} (\htt_{ij,k})^2 + (\picantti{ij})^2\right] \,, \label{eq:Hsplit}
\end{align}
with
\begin{align}
H_{\ADM}^{\interaction} &= \frac{1}{16 \pi \grav} \int \text{d}^\dim x \biggl\{
	(B_{(4)ij}+\hat{B}_{(6)ij}) \htt_{ij} 
	- \frac{4 \pi \grav}{\dim - 1} (\htt_{ij})^2 \srcs{2} 
	+ \picantti{ij} C_{ij} \neanl &
	- \frac{2}{\dim - 2}\phibs{2} (\htt_{ij,k})^2
\biggr\} \,, \\
H_{\ADM}^{\nonTT} &= H_{\ADM} - H_{\ADM}^{\interaction} -  \frac{1}{16\pi\grav}\int {\rm d}^\dim x \left[\frac{1}{4} (\htt_{ij,k})^2 + (\picantti{ij})^2\right] \,,
\end{align}
and
\begin{align}
B_{(4)ij} &= 16\pi\grav \frac{\delta}{\delta \htt_{i j}} \int{ \text{d}^\dim x \, \srcs{8}  }
	- \frac{4(\dim-1)}{\dim-2} \phibs{2}{}_{, i} \phibs{2}{}_{, j} \,,\\
\hat{B}_{(6)ij} & = 16 \pi \grav \frac{\delta}{\delta \htt_{ij}} \int \text{d}^\dim x \biggl[
		\srcs{10}  
		-2\phibs{2} \srcs{8} 
		+ 2 \vpots{3}{k} \srcis{7}{k} 
	\biggr] 
	+ 2 \biggl[
		-2\vpots{3}{j}_{,k} \momls{3}{ik} \neanl & 
		+(\vpots{3}{k} \momls{3}{ij})_{,k} 
		+\momls{3}{ik} \momls{3}{jk} 
		-4\frac{2\dim - 3}{\dim - 2} \phibs{2}_{,i} \phibs{4}_{,j}
		+4\frac{3\dim - 4}{(\dim - 2)^2} \phibs{2} \phibs{2}_{,i} \phibs{2}_{,j} \neanl &
		-2\Sfour_{,i} \phibs{2}_{,j}\biggr]\label{eq:hatB6}\,,\\
 C_{ij} & = -2 \pimati{ij} - 8 \phibs{2} \momls{3}{ij}\label{eq:Cmatter}\,.
\end{align}
$H_{\ADM}$ consists of all Hamiltonians starting from the rest mass contribution up to $H_{3\text{PN}}$.
The $B_{(4)ij}$ part in $H_{\ADM}^{\interaction}$ comes from the formal 2PN Hamiltonian shown in \eqref{eq:H2PNTT}.
$H_{\ADM}^{\nonTT}$ can be obtained by removing all $\htt_{ij}$ and $\picantti{ij}$ parts in \eqref{eq:H2PNTT} and \eqref{eq:H3PN}
respectively.
Notice that the relevant source terms
in the expressions for $B_{(4)ij}$ and $\hat{B}_{(6)ij}$ are at most linear in $\htt_{ij}$. This allowed
us to single out these contributions by a functional derivative.

Now we arrived at a point mentioned in \Sec{sec:ADMHam}, namely where we are able to eliminate the constraint fields using lower
order Hamilton and momentum constraints, but where the dynamical field degrees of freedom are still in the Hamiltonian. 
The elimination of these and further simplifications of the calculation process are subject of the following section.

\section{Routhian and Application of Wave Equation}\label{sec:routhian_waveequation}
To obtain a fully reduced matter only Hamiltonian, we have to remove the dynamical degrees of freedom $\htt_{ij}$ and $\picantti{ij}$ by
solving the appropriate equations of motion and inserting their solutions into $H_{\ADM}$. However there are some subtleties in this procedure
which will be discussed in \Sec{subsec:routhian}.

From Hamilton's equations
\begin{align}
 \frac{1}{16\pi\grav} \pdiffq{\htt_{ij}}{t} & = \TTproj{ij}{k\ell} \frac{\delta H_{\ADM}}{\delta \picantti{k\ell}}\,,\\
 \frac{1}{16\pi\grav} \pdiffq{ \picantti{ij}}{t} & = -\TTproj{ij}{k\ell} \frac{\delta H_{\ADM}}{\delta \htt_{k\ell}}\,,
\end{align}
and the split of the ADM Hamiltonian \eqref{eq:Hsplit}
one gets the appropriate wave equations
\begin{align}
	\Box \htt_{i j} &= 16\pi\grav \TTproj{ij}{kl}
			\left[ 
				2 \frac{\delta H_{\ADM}^{\interaction}}{\delta \htt_{k l}}
				- \pdiffq{}{t} \frac{\delta H_{\ADM}^{\interaction}}{\delta \picantti{kl}} 
			\right] \,, \\
	\picantti{i j} &= \frac{1}{2} \left[ 
		\httdot_{i j}
		- 16\pi\grav \TTproj{ij}{kl} \frac{\delta H_{\ADM}^{\interaction}}{\delta \picantti{kl}} 
	\right] \,,
\end{align}
for the dynamical degrees of freedom of the gravitational field
(here $\Box = \Delta - \cInv{2} \partial_t^2$; remember that we use a $(\dim+1)$-metric with signature $\dim-1$). 
In order to receive the full wave equations one has to perform the variational derivative of the interaction Hamiltonian,
\begin{align}
 \Box \htt_{ij} & = \delta^{\text{TT}\,ij}_{k\ell} \biggl\{
	2 (B_{(4)k\ell}+\hat{B}_{(6)k\ell}) 
	- \frac{16\pi\grav}{\dim - 1} \srcs{2} \htt_{k\ell} 
	+ \frac{8}{\dim - 2} (\phibs{2} \htt_{k\ell,m})_{,m}
	- \pdiffq{}{t} C_{k\ell}\biggr\}\label{eq:boxhtt}\,,\\
 \picantti{ij} & = \frac{1}{2} \httdot_{ij} - \frac{1}{2}\delta^{\text{TT}\,ij}_{k\ell} C_{k\ell} \label{eq:pitt}\,.
\end{align}

\subsection{Near-Zone Expansion}\label{subsec:nzexpansion}
The wave equation
\begin{align}
 \Box h &= f\,, 
\end{align}
[which represents the components of \eqref{eq:boxhtt}] for a source $f$ has several solutions depending 
on the boundary conditions. In field theory mostly the retarded one is used. At the order considered in the present
article the time symmetric (i.e. conservative) solution is sufficient.%
\footnote{At the considered order $\htt_{(5)ij}$ can be neglected since the linear terms are not time symmetric and
the quadratic terms arise only due to $(\htt_{(5)ij,k})^2$ which is zero in $\dim=3$ because $\htt_{(5)ij}$ is only
a function of $t$ (there are no explicit terms in the source at $\cInv{5}$-order); $\htt_{(7)ij}$ is of too high order
to appear at formal 3PN order, see \cite{Jaranowski:Schafer:1997,Steinhoff:Wang:2009,Wang:Steinhoff:Zeng:Schafer:2011}. 
We therefore neglected $\htt_{(5)ij}$ and $\htt_{(7)ij}$ in \eqref{eq:httorders} and hence there are no contributions 
to \eqref{eq:H3PN} or \eqref{eq:H2PNTT}.}

The appropriate solution derives from a near-zone expansion, which is a formal expansion in $c^{-1}$.
(see e.g. \cite{Galtsov:2002,Cardoso:Dias:Lemos:2003,Blanchet:Damour:EspositoFarese:Iyer:2005,Blanchet:Damour:EspositoFarese:2004}.)
More precisely, the near-zone expansion is a series expansion in the small quantity $r/(c t)\ll 1$, which means that the distance from the field point to
the appropriate source point is small compared to the gravitational wavelength. So the near-zone expansion may be used
if retardation effects are negligible. Consider the Feynman propagator for a massless particle (which corresponds to the Green's function
of the wave equation)
\begin{align}
 G_{F}(\vct{x}, t) &= -\frac{1}{2\pi} \lim_{\varepsilon\to0}
 \int\text{d}k_0 \frac{1}{(2\pi)^\dim} \int\text{d}^\dim k \frac{e^{-i(k_0 x^0 - \scpm{\vct{k}}{\vct{x}})}}{k^2 - k_0^2 - i\varepsilon}\,.
\end{align}
For the reader's convenience we reintroduce the powers of $c$ in the following expressions. This means $k_0 = \omega/c$ and $x^0 = c t$.
As it was argued above and in e.g. \cite{Iwanenko:Sokolow:1953,Goldberger:2007} the condition $r/(ct) \ll 1$ corresponds to $k_0/k \ll 1$
which gives rise to the so-called potential gravitons in contrast to the radiation gravitons for which $k_0^2 \approx k^2$. For the radiation
gravitons the $i\varepsilon$ term becomes important, but for the potential ones it could be neglected and the Fourier amplitude 
in the propagator is expandable in $k_0/k$, namely
\begin{align}
 G_{F}(\vct{x}, t) &\stackrel{\text{NZ}}{=} -\frac{1}{2\pi c} 
 \int\text{d}\omega e^{-i \omega t}\sum_{n=0}^\infty \frac{\omega^{2n}}{c^{2n}} \frac{1}{(2\pi)^\dim} \int\text{d}^\dim k 
 \frac{e^{i\scpm{\vct{k}}{\vct{x}}}}{k^{2(n+1)}}\,.\label{eq:nzfourier}
\end{align}
This means that a near-zone expansion of the (time symmetric) Feynman propagator cannot contain any contributions leading to radiative losses.
The $\dim$-dimensional $k$-space integral in \eqref{eq:nzfourier} has to be performed by using
\begin{align}
 \int\text{d}^\dim k \frac{e^{i\scpm{\vct{k}}{\vct{x}}}}{k^{2\alpha}} &= 
 \frac{1}{(4\pi)^{\dim/2}} \frac{\Gamma\left(\frac{\dim}{2}-\alpha\right)}{\Gamma(\alpha)} \left(\frac{r^2}{4}\right)^{\alpha-\dim/2}\,,
\end{align}
which can be obtained by dimensional regularization (the source of the wave lies at the origin of the coordinate system). This result can
be reformulated in inverse Laplacians on a delta-type source by iterating \eqref{eq:invlapdelta} and \eqref{eq:invra} in the \descappendix,
\begin{align}
 \frac{1}{(2\pi)^\dim} \int\text{d}^\dim k  \frac{e^{i\scpm{\vct{k}}{\vct{x}}}}{k^{2(n+1)}} &= 
 -\frac{\Gamma\left(2+n-\frac{\dim}{2}\right)\Gamma\left(\frac{\dim}{2}-1-n\right)}{\Gamma\left(2-\frac{\dim}{2}\right)\Gamma\left(\frac{\dim}{2}-1\right)}
 (\Delta^{-1})^{n+1}\delta\,.
\end{align}
For $\dim\notin 2\mathbb{Z}$ (i.e. no odd dimensional spacetime) one can simplify the Gamma functions further by using the identity $\Gamma(1-z)\Gamma(z) = \pi/\sin(\pi z), z\notin\mathbb{Z}$ 
which leads to
\begin{align}
 \frac{1}{(2\pi)^\dim} \int\text{d}^\dim k  \frac{e^{i\scpm{\vct{k}}{\vct{x}}}}{k^{2(n+1)}} &= 
 -(-1)^n (\Delta^{-1})^{n+1}\delta\,,
\end{align}
and therefore after performing the $\omega$ Fourier transform to
\begin{align}
 G_{F}(\vct{x}, t) &\stackrel{\text{NZ}}{=} \sum_{n=0}^\infty \left(\frac{1}{c}\frac{\partial}{\partial t}\right)^{2n} (\Delta^{-1})^{n+1}\delta_{(\dim+1)}\,,
\end{align}
for the Feynman propagator. This result could immediately be used to write down
the near-zone expanded solution of the wave equation
\begin{align}
 h &= \Box^{-1} f \stackrel{\text{NZ}}{=} 
\Delta^{-1} \sum_{n=0}^{\infty}(\Delta^{-1})^{n}\left(\frac{1}{c}\frac{\partial}{\partial t}\right)^{2n}f \label{eq:nzexpansion}\,.
\end{align}
Notice that a near-zone expanded field in general does not converge at spatial infinity.

Now we are able to derive the solutions for the transverse-traceless part of the metric at a certain post-Newtonian order in the near-zone.

\subsection{Routhian}\label{subsec:routhian}
Before we can insert the wave equation \eqref{eq:boxhtt}
and its solution [see \eqref{eq:nzexpansion}, and for a more explicit form see \Sec{subsec:solwaveequation}], 
we have to transform the ADM Hamiltonian into a Routhian,
i.e. a Lagrangian in $\htt_{ij}$ and $\picantti{ij}$ and a Hamiltonian in the particle degrees of freedom, 
\begin{align}
R[\htt_{ij}, \httdot_{ij}] = H_{\ADM} - \frac{1}{16\pi \grav} \int \text{d}^\dim x \, \picantti{ij} \httdot_{ij}\,. 
\end{align}
This is necessary because otherwise the
equation of motion, e.g., for $\vx{a}$,
following from the Hamiltonian (for simplicity we omit the spin variables here)
\begin{align}
  H_{\ADM}(\vx{a}, \vmom{a}, \htt_{ij}(\vx{b}, \vmom{b}), \picantti{ij}(\vx{b}, \vmom{b}))\,,
\end{align}
using the Poisson brackets would be
\begin{align}
 \dot{\hat{z}}_a^k &= \pdiffq{H_{\ADM}}{\mom{a}{k}} + \frac{\delta H_\ADM}{\delta \htt_{ij}} \pdiffq{\htt_{ij}}{\mom{a}{k}} 
				+ \frac{\delta H_\ADM}{\delta \picantti{ij}} \pdiffq{\picantti{ij}}{\mom{a}{k}}\,.
\end{align}
This equation is obviously wrong, because $\htt_{ij}$ and $\picantti{ij}$ are dynamical degrees of freedom and may not
lead to additional terms in the equations of motion, also when their solutions are inserted.
On the other hand using $R(\vx{a}, \vmom{a}, \htt_{ij}(\vx{b}, \vmom{b}), \httdot_{ij}(\vx{b}, \dot{\hat{\vct{z}}}_b, \vmom{b}, \dot{\hat{\vct{p}}}_b))$ 
one has for the same equation of motion
\begin{align}
 \dot{\hat{z}}_a^k &= \pdiffq{R}{\mom{a}{k}} + \underbrace{\frac{\delta R}{\delta \htt_{ij}}}_{=0} \pdiffq{\htt_{ij}}{\mom{a}{k}}\,,
\end{align}
which has no additional terms coming from the chain rule, because they vanish due to $\htt_{ij}$ fulfills the equations of motion 
in the appropriate approximation. Hence one can obtain the equation of motion in the usual way and one does not have to keep track of the field insertions. This is analogous to the construction of a Fokker action, see, e.g., \cite{Damour:EspositoFarese:1995} and references therein.

A Fokker-like construction of a matter-only Lagrangian (or Routhian) can not account
for dissipative effects, see \cite{Galley:2012} for a discussion.
However, dissipative Hamiltonians in the ADM formalism can be constructed in the following way \cite{Jaranowski:Schafer:1997, Wang:Steinhoff:Zeng:Schafer:2011}:
The matter variables entering the solution of $\htt_{ij}$ and $\picantti{ij}$ are substituted by new (``primed'') variables
and thus the binary Hamiltonian now contains four types of canonical matter variables. This procedure prevents
occurrence of wrong contributions in the equations of motion, too. The primed variables will be treated as
explicitly time dependent and lead to an explicitly time dependent Hamiltonian. 
After calculating the equations of motion for canonical positions $\vx{a}$ and momenta $\vmom{a}$, the primed variables are 
again identified with the old ones which makes another regularization procedure necessary 
\cite{Jaranowski:Schafer:1997}. Another approach to construct an action principle for dissipative 
systems was suggested in \cite{Galley:2012} recently.

\subsection{Reduction of the Routhian using the Wave Equation}
The part of the Routhian labeled as TT contains $H_\ADM^\interaction$, the field kinetic part of the Hamiltonian and the part coming from 
the Legendre transform (and thus only terms coming from the transverse-traceless degrees of freedom),
\begin{align}
 R^{\TT} &= \frac{1}{16 \pi \grav} \int \text{d}^\dim x \biggl[
	\left(B_{(4)ij} + \hat{B}_{(6)ij}\right) \htt_{ij} 
	- \frac{4 \pi \grav}{\dim - 1} (\htt_{ij})^2 \srcs{2} 
	+ \picantti{ij} C_{ij} \neanl &
	- \frac{2}{\dim - 2}\phibs{2} (\htt_{ij,k})^2 
	+ \frac{1}{4} (\htt_{i j , k})^2 
	+ (\picantti{i j})^2 
	- \httdot_{ij} \picantti{ij}
\biggr]\,.
\end{align}
Inserting $\Box \htt_{ij}$ and $\picantti{ij}$ from \eqref{eq:boxhtt} and \eqref{eq:pitt}, or
\begin{align}
 \htt_{ij} (B_{(4)ij} + \hat{B}_{(6)ij}) & = \frac{1}{2} \htt_{ij} \Box \htt_{ij} 
	+ \frac{8\pi\grav}{\dim - 1} (\htt_{ij})^2 \srcs{2} 
	- \frac{4}{\dim - 2} \htt_{ij} (\phibs{2} \htt_{ij,k})_{,k} \neanl &
	+ \frac{1}{2} \htt_{ij} \pdiffq{}{t} C_{ij} 
	+ \td\,,
\end{align}
leads to
\begin{align}
 R^{\TT} &= \frac{1}{16 \pi \grav} \int \text{d}^\dim x \biggl[
	\frac{1}{4} \htt_{ij} \Box \htt_{ij} 
	+ \frac{4\pi\grav}{\dim - 1} (\htt_{ij})^2 \srcs{2}
	+ \frac{2}{\dim - 2}\phibs{2} (\htt_{ij,k})^2  \neanl &
 	+ \frac{1}{2} \pdiffq{}{t} \biggl[\htt_{ij}  C_{ij}\biggr] 
	- \frac{1}{4} C_{ij} \TTproj{ij}{k\ell} C_{k\ell} \biggr]\,,\label{eq:Routhian3PN}
\end{align}
where the last part will appear in the matter part of the final Routhian, 
because there is no $\htt_{ij}$ and no $\picantti{ij}$ or $\httdot_{ij}$ in $C_{ij}$.
Notice that we kept a total time derivative here. If we would drop it the $\dot{C}_{ij}$ terms would not cancel in the next step.
These terms are not impossible to handle, but it is advised to remove them to simplify the calculation.

\subsection{Insertion of the Near-Zone Wave Equation for Further Simplification}
Now we need to split up the first expression in the TT part of the Routhian \eqref{eq:Routhian3PN}. The near-zone
expansion of the transverse-traceless
part of the metric $\htt_{ij} = \htt_{(4)ij} + \htt_{(6)ij} + \dots$, which is explained in detail in \Sec{subsec:nzexpansion}, 
contributes only via $\htt_{(4)ij}$ and $\htt_{(6)ij}$ 
at 3PN level. This expansion leads to
\begin{align}
\htt_{ij} \Box \htt_{ij} & = \htt_{(4)\,ij} \Box \htt_{(4)\,ij} + 2 \htt_{(4)\,ij} \Box \htt_{(6)\,ij} + \td + \ttd\,,
\end{align}
where $\ttd$ denotes a total time derivative.
Notice that there is a difference between $\Box \htt_{(6)\,ij}$ and $(\Box \htt_{ij})_{(6)}$ in the near-zone expansion 
as time derivatives raise the formal order of a field in contrast to spatial derivatives. This
is a specific feature of the near-zone expansion, as in the far-zone time and space derivatives are equal in magnitude. 
These considerations lead to the difference in the following box operations,
\begin{align}
 \Box \htt_{(6)\,ij} & = \Delta \htt_{(6)\,ij} - \partial_t^2 \htt_{(6)\,ij} \label{eq:boxhtt6}\,,\\
 (\Box \htt_{ij})_{(6)} & = \Delta \htt_{(6)\,ij} - \partial_t^2 \htt_{(4)\,ij} \label{eq:box6htt}\,,
\end{align}
where $\partial_t^2 \htt_{(6)\,ij}$ is of formal $\cInv{8}$ order and hence the total time derivative can be neglected. From this it follows that
\begin{align}
 \htt_{ij} \Box \htt_{ij} & = \htt_{(4)\,ij} \Delta \htt_{(4)\,ij} - \htt_{(4)\,ij} \partial_t^2 \htt_{(4)\,ij} + 2 \htt_{(4)\,ij} \underbrace{\Delta \htt_{(6)\,ij}}_{(\Box \htt_{ij})_{(6)} + \partial_t^2 \htt_{(4)\,ij}} + \td\,,\label{eq:httboxhtt}
\end{align}
such that
\begin{align}
 (\htt_{ij} \Box \htt_{ij})_{(10)} & = 2 \htt_{(4)\,ij} (\Box \htt_{ij})_{(6)} + \htt_{(4)\,ij} \partial_t^2 \htt_{(4)\,ij} + \td\,,
\end{align}
where $(\Box \htt_{ij})_{(6)}$ is given by the $\cInv{6}$ part of Eq. \eqref{eq:boxhtt}.
After another integration by parts this immediately leads to
(we need only the leading order of $C_{ij}$ which is at $\cInv{5}$)
\begin{align}
R^{\TT}_{\text{3PN}} &= \frac{1}{16 \pi \grav} \int \text{d}^\dim x \biggl[
  \htt_{(4)ij}  \hat{B}_{(6)ij} 
  -\frac{1}{4} (\httdot_{(4)\,ij})^2 
  -\frac{4\pi\grav}{\dim - 1} (\htt_{(4)ij})^2 \srcs{2} \neanl &
  -\frac{2}{\dim - 2}\phibs{2} (\htt_{(4)ij,k})^2  
  +\frac{1}{2} \httdot_{(4)ij}  C_{(5)ij} 
  -\frac{1}{4} C_{(5)ij} \TTproj{ij}{k\ell} C_{(5)k\ell} 
\biggr]\,.
\end{align}
Rearranging some of the terms into pure matter and pure transverse-traceless parts, one obtains the full final 3PN Routhian,
\begin{subequations}
\label{eq:FinalRouthian3PN}
\begin{align}
R^{\matter}_{\text{3PN}} &=
\int \text{d}^\dim x \biggl[
  \srcsnontt{10} 
  -2\phibs{2} \srcsnontt{8} 
  + (- \Sfour -4 \phibs{4} + 2 \phibs{2}^2)\srcs{6} \neanl &
  + (6 \phibs{2} \phibs{4} + \Sfour \phibs{2} -2\phibs{2}^3) \srcs{4} 
  + \biggl(4 \phibs{4}^2 
    + \Sfour \phibs{4} 
    - 8 \phibs{4} \phibs{2}^2 
    - \Sfour \phibs{2}^2 \neanl &
    + 2\phibs{2}^4 
  \biggr)\srcs{2}
  +2 \vpots{3}{i} \srcisnontt{7}{i} 
  - \frac{1}{16 \pi \grav} \biggl(
    -(\momlones{5}{ij})^2
    +2\frac{3\dim - 4}{\dim - 2} \phibs{4} (\momls{3}{ij})^2 \neanl &
    + \Sfour (\momls{3}{ij})^2  
    - \frac{(3\dim - 4)(3\dim - 2)}{(\dim - 2)^2} (\momls{3}{ij})^2 \phibs{2}^2 
    + \biggl(\frac{4(\dim - 1)}{\dim -2} \TTproj{ij}{k\ell} (\phibs{2} \momls{3}{k\ell})\biggr)^2 \neanl &
    - 4\frac{2\dim - 3}{\dim - 2} \phibs{2} \momls{3}{ij}\momlones{5}{ij} 
  \biggr)
\biggr] \,, \label{eq:FinalMatterRouthian3PN}\\
R^{\TT}_{\text{3PN}} &= \frac{1}{16 \pi \grav} \int \text{d}^\dim x \biggl[
  \htt_{(4)ij}  \hat{B}_{(6)ij} 
  -\frac{1}{4} (\httdot_{(4)\,ij})^2 
  -\frac{4\pi\grav}{\dim - 1} (\htt_{(4)ij})^2 \srcs{2} \neanl &
  -\frac{2}{\dim - 2}\phibs{2} (\htt_{(4)ij,k})^2 
  -\httdot_{(4)ij} \pimatis{5}{ij}
  -4 \phibs{2} \momls{3}{ij} \httdot_{(4)ij} 
\biggr]\,,\label{eq:FinalTTRouthian3PN}
\end{align}
\end{subequations}
where $\hat{B}_{(6)ij}$ is given by \eqref{eq:hatB6}.
Note, that one does not need to calculate the $\htt_{(6)ij}$ field, which is not 
fully known in closed form. An explicit form of $\hat{B}_{(6)ij}$
with all sources inserted is derived in the next section, cf.\ \eqref{eq:B6matter}.

\section{Sources} \label{sec:sources}
The construction of the sources linear in spin follows along the lines of \cite{Steinhoff:Schafer:2009:2}.
This requires the introduction of a local Lorentz frame%
\footnote{Such frames were originally invented by \'{E}lie Cartan and named ``rep\`{e}re mobile'', which is French for ``moving frame''.
In $\dim=3$ it is also called ``triad'' or ``dreibein''; in $\dim=4$ ``tetrad'' or ``vierbein''. For arbitrary integer $\dim$ it is called ``vielbein''.}
 as we want flat space Poisson brackets for the spin. In particular
we also apply the Schwinger time gauge for the $(\dim+1)$-dimensional framefield
which effectively reduces it to a $\dim$-dimensional spatial framefield $\triad{i}{j}$. 

During the following calculations we neglect the $\pimati{ij}$ terms, which are far too high in their order, 
such that the modified source terms are given by \cite[Eqs. (6.33) and (6.34)]{Steinhoff:2011}
\begin{align}
 \src{} & = \sum_a \biggl[
  -\nmom{a} \delta_a +\frac{\mom{a}{j}\gamma^{ji}}{\nmom{a}}\hat{A}_a^{k\ell} \triad{m}{k} e^{(m)}_{\quad \ell,i}\dl{a}  \neanl &
  +\biggl\{
    \frac{1}{2} 
      \biggl(
	\frac{\triad{r}{\ell}\triad{s}{i}\mom{a}{j}}{\nmom{a}} 
	+ \gamma^{mn}\frac{\triad{r}{m}\triad{s}{i}\mom{a}{j}\mom{a}{n}\mom{a}{\ell}}{(\nmom{a})^2(m_a - \nmom{a})}
      \biggr) \gamma^{k\ell} \biggl(
	\gamma^{nj}\;\christoffel{(\dim)}{i}{nk} + \gamma^{in}\;\christoffel{(\dim)}{j}{nk}
      \biggr)\delta_a  \neanl &
      - \biggl(
	\frac{\mom{a}{\ell}}{m_a - \nmom{a}}\gamma^{ij}\gamma^{k\ell}\triad{r}{j}\triad{s}{k}\delta_a
      \biggr)_{,i}
  \biggr\}\spin{a}{r}{s}
 \biggr]\,,\label{eq:sourcehamilton}\\
 \src{i} & = \sum_a \biggl[
  \mom{a}{i}\delta_a
  - \hat{A}_a^{k\ell} \triad{m}{k} e^{(m)}_{\quad \ell,i}\dl{a} 
  + \frac{1}{2} \biggl(
    \gamma^{mk}\triad{r}{i}\triad{s}{k}\delta_a \neanl &
    - \frac{\mom{a}{\ell}\mom{a}{k}}{\nmom{a} (m_a - \nmom{a})} 
	(\gamma^{mk} \delta^{p}_{i} + \gamma^{mp} \delta^{k}_{i}) 
	    \gamma^{q\ell} \triad{r}{q}\triad{s}{p}\delta_a
    \biggr)_{,m}\spin{a}{r}{s}
  \biggr]\label{eq:sourcemomentum}\,,
\end{align}
where $m_a$ is the mass of the $a$th object, $\vmom{a}$ its canonical momentum, $\spin{a}{r}{s}$ its canonical spin,
$\nmom{a} = -\sqrt{m_a^2 + \gamma^{ij}\mom{a}{i}\mom{a}{j}}$, and $\dl{a} = \delta(\vct{x} - \vx{a})$ is the $\dim$-dimensional
Dirac delta located at $\vct{x}=\vx{a}$. Furthermore
\begin{align}
  \hat{A}_a^{k\ell} &= \gamma^{ik}\gamma^{j\ell} \left(\frac{1}{2}\hat{S}_{a\,ij} + \frac{m_a \mom{a}{(i}(nS_a)_{j)}}{\nmom{a} (m_a - \nmom{a})}\right)\label{eq:Ahat}\,,\\
  (nS_a)_i &= -\frac{\mom{a}{j}\gamma^{jk}\hat{S}_{a\,ki}}{m_a}\label{eq:nS}\,,
\end{align}
to linear order in spin. However the $\hat{A}_a^{k\ell}$-terms do not contribute to the expanded source expressions at the orders considered here.
The matter position and matter momentum variables are canonical conjugate to each other, namely
\begin{align}
 \{\xa{a}{i},\mom{b}{j}\} &= \delta_{ij} \delta_{ab}\,,
\end{align}
and the spin variables fulfill also canonical Poisson bracket relations, namely
\begin{align}
  \{\spin{a}{i}{j}, \spin{a}{k}{\ell}\} = \delta_{ik} \spin{a}{j}{\ell} - \delta_{i\ell} \spin{a}{j}{k} 
	- \delta_{jk} \spin{a}{i}{\ell} + \delta_{j\ell} \spin{a}{i}{k}\,,
\end{align}
where the canonical spin tensor $\spin{a}{i}{j}$ is related to the canonical spin vector $\hat{\vct{S}}_{a}$ via
$\spin{a}{i}{j} = \varepsilon_{ijk} \hat{S}_{a\,(k)}$ and $\varepsilon_{ijk}$ is the Levi-Civita symbol. The appropriate Poisson brackets for the canonical spin vector are given by
\begin{align}
  \{\hat{S}_{a\,(i)}, \hat{S}_{a\,(j)}\} = \varepsilon_{ijk}\hat{S}_{a\,(k)}\,.
\end{align}

\subsection{Framefield Expansion}
We choose to work within a symmetrical framefield
gauge $\triad{i}{j}=\triad{j}{i}$ \cite{Kibble:1963}, so the dreibein can be written
as a matrix square root of the metric, symbolically $\triad{i}{j} = \sqrt{\gamma_{ij}}$ or more explicit
\begin{align}
  \triad{i}{k} \triad{k}{j} = \gamma_{ij}\label{eq:dreibein}\,. 
\end{align}
Notice that $\gamma_{ij}$ is positive definite and we require the same for $\triad{i}{j}$
such that it is unique.
The second relation, \eqref{eq:dreibein}, can be inverted order by order, namely
\begin{subequations}
\begin{align}
 \triads{0}{i}{k} \triads{0}{k}{j} &= \gamma_{(0)\,ij} \stackrel{\ADMTT}{\Rightarrow} \triads{0}{i}{j} = \delta_{ij}\,,\label{eq:dreibeinzero}\\
 \triads{2}{i}{k} \triads{0}{k}{j} + \triads{0}{i}{k} \triads{2}{k}{j} &= \gamma_{(2)\,ij} \stackrel{\ADMTT}{\Rightarrow} \triads{2}{i}{j} = \frac{1}{2}\gamma_{(2)\,ij} = \frac{2}{\dim-2} \phibs{2} \delta_{ij}\,,\\
 \vdots \nonumber
\end{align}
\end{subequations}
and at the end of the day one gets
\begin{subequations}
\begin{align}
 \triads{0}{i}{j} & = \delta_{ij}\,, \\
 \triads{2}{i}{j} & = \frac{2}{\dim - 2} \phibs{2} \delta_{ij}\,, \\
 \triads{4}{i}{j} & = 
  \biggl(\frac{2}{\dim - 2} \phibs{4}
  - \frac{\dim - 4}{(\dim - 2)^2} \phibs{2}^2\biggr) \delta_{ij} 
  + \frac{1}{2} \htt_{ij} \,, \\
 \triads{6}{i}{j} & = 
  \biggl(\frac{2}{\dim - 2} \phibs{6} 
  - \frac{\dim - 4}{(\dim - 2)^2} \phibs{2} \phibs{4} 
  + \frac{2}{3} \frac{(\dim - 4)(\dim - 3)}{(\dim - 2)^3} \phibs{2}^3\biggr) \delta_{ij} 
  - \frac{1}{\dim - 2} \phibs{2} \htt_{ij}\,,
\end{align}
\end{subequations}
for the framefield perturbations.
The antisymmetric part of the framefield (which is zero in this gauge)
can be interpreted as rotational degrees of freedom in the choice of the local frame.
Such a rotation does not change the length of the spins.
Recall that an antisymmetric matrix is an infinitesimal generator of rotations and in $\dim$ dimensions
has $\tfrac{1}{2}\,\dim (\dim-1)$ independent entries. This is exactly the number of rotation planes in $\dim$ dimensions.

\subsection{Constraint Sources}
After performing the expansion of the sources \eqref{eq:sourcehamilton} and \eqref{eq:sourcemomentum} in powers of $\cInv{1}$ and also expanding the
metric-framefield relation \eqref{eq:dreibein}, we are able to write down the appropriate post-Newtonian contributions
to the source of the Hamilton constraint, namely
\begin{subequations}
\label{eq:PNSources}
\begin{align}
 \srcs{2} & = \sum_a m_a \dl{a}\,,\neanl
 \srcs{4} & = \sum_a \biggl[
	\frac{\vmom{a}^2}{2 m_a} \dl{a} 
	+ \frac{1}{2 m_a} \mom{a}{i} \spin{a}{i}{j} \dl{a}{}_{,j}
\biggr]\,,\eanl
 \srcs{6} & = \sum_a \biggl[
	-\frac{(\vmom{a}^2)^2}{8 m_a^3} \dl{a} 
	- \frac{2}{\dim-2} \frac{\vmom{a}^2}{m_a} \phibs{2}\dl{a} 
	+ \frac{2}{\dim - 2} \frac{\mom{a}{i}}{m_a}\spin{a}{i}{j} \phibs{2}_{,j}\dl{a} 
	- \frac{\vmom{a}^2}{8 m_a^3} \mom{a}{i}\spin{a}{i}{j} \dl{a}{}_{,j} \neanl & 
	- \frac{2}{\dim - 2} \frac{\mom{a}{i}}{m_a} \spin{a}{i}{j} (\phibs{2} \dl{a})_{,j}
\biggr]\,,\eanl
\srcs{8} & = \sum_a \biggl[
	\frac{(\vmom{a}^2)^3}{16 m_a^5} \dl{a} 
	+ \frac{1}{\dim - 2} \frac{(\vmom{a}^2)^2}{m_a^3}\phibs{2}\dl{a} 
	+ \frac{\dim + 2}{\dim - 2} \frac{\vmom{a}^2}{m_a} \phibs{2}^2 \dl{a} 
	- \frac{2}{\dim - 2} \frac{\vmom{a}^2}{m_a} \phibs{4} \dl{a} \neanl &
	- \frac{1}{2 m_a} \mom{a}{i} \mom{a}{j} \htt_{ij}\dl{a} 
	- \frac{1}{\dim-2} \frac{\vmom{a}^2}{m_a^3} \mom{a}{i}\spin{a}{i}{j} \phibs{2}_{,j}\dl{a} 
	- \frac{2(\dim+2)}{(\dim-2)^2} \frac{\mom{a}{i}}{m_a} \spin{a}{i}{j} \phibs{2}\phibs{2}_{,j} \dl{a} \neanl & 
	+ \frac{2}{\dim - 2} \frac{\mom{a}{i}}{m_a} \spin{a}{i}{j} \phibs{4}_{,j} \dl{a} 
	+ \frac{1}{2 m_a} \mom{a}{i} \spin{a}{j}{k} \htt_{ij,k} \dl{a}
\biggr] 
+\sum_a \partial_j \biggl[
	\frac{(\vmom{a}^2)^2}{16 m_a^5} \mom{a}{i}\spin{a}{i}{j}\dl{a} \neanl&
	+ \frac{1}{\dim-2}\frac{\vmom{a}^2}{m_a^3}\mom{a}{i}\spin{a}{i}{j}\phibs{2}\dl{a} 
	+ \frac{\dim+2}{(\dim-2)^2}\frac{\mom{a}{i}}{m_a}\spin{a}{i}{j}\phibs{2}^2\dl{a} 
	- \frac{2}{\dim-2}\frac{\mom{a}{i}}{m_a}\spin{a}{i}{j}\phibs{4}\dl{a} \neanl &
	+ \frac{1}{4 m_a} \mom{a}{i}\spin{a}{k}{i}\htt_{jk}\dl{a} 
	- \frac{1}{4 m_a} \mom{a}{i}\spin{a}{k}{j}\htt_{ik}\dl{a}
\biggr]\,,\eanl
\srcs{10} & = \sum_a\biggl[
	- \frac{5 (\vmom{a}^2)^4}{128 m_a^7}\dl{a}
	- \frac{3}{4(\dim-2)}\frac{(\vmom{a}^2)^3}{m_a^5} \phibs{2}\dl{a} 
	- \frac{\dim+6}{2(\dim-2)^2}\frac{(\vmom{a}^2)^2}{m_a^3}\phibs{2}^2\dl{a} \neanl &
	- \frac{2\dim(\dim+2)}{3(\dim-2)^3}\frac{\vmom{a}^2}{m_a}\phibs{2}^3\dl{a} 
	+ \frac{1}{\dim-2}\frac{(\vmom{a}^2)^2}{m_a^3}\phibs{4}\dl{a}
	+ \frac{2(\dim+2)}{(\dim-2)^2}\frac{\vmom{a}^2}{m_a}\phibs{2}\phibs{4}\dl{a} \neanl &
	- \frac{2}{\dim-2}\frac{\vmom{a}^2}{m_a}\phibs{6}\dl{a}
	+ \frac{\vmom{a}^2}{4 m_a^3}\mom{a}{i}\mom{a}{j}\htt_{ij}\dl{a} 
	+ \frac{4}{\dim-2}\frac{\mom{a}{i}\mom{a}{j}}{m_a}\htt_{ij}\phibs{2}\dl{a} \neanl &
	+ \frac{3}{4(\dim-2)}\frac{(\vmom{a}^2)^2}{m_a^5}\mom{a}{i}\spin{a}{i}{j}\phibs{2}_{,j}\dl{a} 
	+ \frac{\dim+6}{(\dim-2)^2}\frac{\vmom{a}^2}{m_a^3}\mom{a}{i}\spin{a}{i}{j}\phibs{2}\phibs{2}_{,j}\dl{a} \neanl &
	+ \frac{2\dim(\dim+2)}{(\dim-2)^3}\frac{\mom{a}{i}}{m_a}\spin{a}{i}{j}\phibs{2}^2\phibs{2}_{,j}\dl{a} 
	- \frac{1}{\dim-2}\frac{\vmom{a}^2}{m_a^3}\mom{a}{i}\spin{a}{i}{j}\phibs{4}_{,j}\dl{a} \neanl &
	- \frac{2(\dim+2)}{(\dim-2)^2}\frac{\mom{a}{i}}{m_a}\spin{a}{i}{j}(\phibs{2}\phibs{4})_{,j}\dl{a} 
	+ \frac{2}{\dim-2}\frac{\mom{a}{i}}{m_a}\spin{a}{i}{j}\phibs{6}_{,j}\dl{a} \neanl &
	- \frac{\vmom{a}^2}{4 m_a^3}\mom{a}{i}\spin{a}{j}{k}\htt_{ij,k}\dl{a} 
	- \frac{4}{\dim-2}\frac{\mom{a}{i}}{m_a}\spin{a}{j}{k}\htt_{ij,k}\phibs{2}\dl{a} \neanl &
	+ \frac{3}{\dim-2}\frac{\mom{a}{i}}{m_a}\htt_{ik}\spin{a}{j}{k}\phibs{2}_{,j}\dl{a}
	- \frac{1}{\dim-2}\frac{\mom{a}{i}}{m_a}\spin{a}{i}{k}\htt_{jk}\phibs{2}_{,j}\dl{a}
\biggr]
+\td\,.
\end{align}
\end{subequations}
In $\srcs{10}$ there is a $\phibs{6}$ term left. But all occurrences of $\phibs{6}$ can be cast into the form $-\frac{4}{\dim-2} \srcs{4}\phibs{6}$ 
which we integrate by parts using \eqref{eq:phi6src4PI}. Then $\phibs{6}$ disappears and gets substituted by
\begin{align}
 -\frac{4}{\dim-2} \srcs{4}\phibs{6} & = -\frac{4}{\dim-2}\biggl[
  \sum_a \biggl\{
    \frac{(\vmom{a}^2)^2}{16 m_a^3}\Sfour\dl{a}
    +\frac{\dim+2}{4(\dim-2)}\frac{\vmom{a}^2}{m_a}\phibs{2}\Sfour\dl{a} \neanl &
    +\frac{m_a}{2}(\phibs{4} -\phibs{2}^2 )\Sfour\dl{a}
    +\frac{\vmom{a}^2 \mom{a}{i}\spin{a}{i}{j}}{16 m_a^3} \Sfour\dl{a}{}_{,j} \neanl &
    +\frac{\dim+2}{4(\dim-2)} \frac{\mom{a}{i}\spin{a}{i}{j}}{m_a} \phibs{2}\Sfour\dl{a}{}_{,j} \biggr\} \neanl &
    -\frac{1}{16\pi\grav}\biggl\{\frac{1}{2}(\momls{3}{ij})^2\Sfour - 2\htt_{ij}\Sfour\phibs{2}_{,ij}\biggr\} 
  \biggr] +\td\label{eq:srcH4phi6}\,.
\end{align}
Furthermore we are also able to write down the sources for the momentum constraint in their full form, which are given by
\begin{subequations}
\label{eq:PNSourcesMom}
\begin{align}
 \srcis{3}{i} & = \sum_a \biggl[\mom{a}{i} \dl{a} + \frac{1}{2} (\spin{a}{i}{j} \dl{a})_{,j}\biggr]\,,\eanl
 \srcis{5}{i} & = \frac{1}{2} \sum_a \biggl[ -\frac{\mom{a}{k}}{2 m_a^2} (\mom{a}{j}\spin{a}{i}{k} + \mom{a}{i} \spin{a}{j}{k}) \dl{a}\biggr]_{,j}\,,\eanl
 \srcis{7}{i} & = \frac{1}{2} \sum_a \biggl[ \biggl(-\frac{1}{2}(\htt_{jk}\spin{a}{i}{k} + \htt_{ik}\spin{a}{j}{k}) + \frac{3 \vct{\canmom}_a^2}{8 m_a^4}\mom{a}{k}(\mom{a}{j}\spin{a}{i}{k} + \mom{a}{i} \spin{a}{j}{k}) \neanl &
+\frac{2}{\dim-2}\frac{\mom{a}{k}}{m_a^2}(\mom{a}{j}\spin{a}{i}{k} + \mom{a}{i} \spin{a}{j}{k})\phibs{2}\biggr) \dl{a}\biggr]_{,j}\,.
\end{align}
\end{subequations}
With the source expressions \eqref{eq:PNSources} including \eqref{eq:srcH4phi6} and \eqref{eq:PNSourcesMom} from above, we may express $B_{(4)ij}$ and $\hat{B}_{(6)ij}$ in terms of the matter variables.
They are given by
\begin{subequations}
\begin{align}
 B_{(4)ij} &= 16\pi\grav\sum_a \biggl[
  -\frac{\mom{a}{i}\mom{a}{j}}{2m_a}\dl{a}-\frac{\mom{a}{i}\spin{a}{j}{k}}{2m_a}\dl{a,k}
 \biggr] - \frac{4(\dim-1)}{\dim-2}\phibs{2}_{,i}\phibs{2}_{,j}\,,\label{eq:B4matter}\\
 \hat{B}_{(6)ij} & = 16\pi\grav \sum_a \biggl[
	\frac{\vmom{a}^2}{4 m_a^3} \mom{a}{i}\mom{a}{j}\dl{a} 
	+ \frac{\dim+2}{\dim-2} \frac{\mom{a}{i}\mom{a}{j}}{m_a}\phibs{2}\dl{a} \neanl &
	+ \frac{\vmom{a}^2}{4 m_a^3}\mom{a}{i}\spin{a}{j}{k}\dl{a,k}
	+ \frac{\dim+2}{\dim-2} \frac{\mom{a}{i}}{m_a}\spin{a}{j}{k}(\phibs{2}\dl{a})_{,k} \neanl &
	+ \frac{\dim+4}{2(\dim-2)} \frac{\mom{a}{i}}{m_a}\spin{a}{k}{j} \phibs{2}_{,k} \dl{a}
	- \frac{\dim}{2(\dim-2)} \frac{\mom{a}{k}}{m_a}\spin{a}{k}{j} \phibs{2}_{,i} \dl{a} \neanl &
	+ \frac{1}{2} \left(\vpots{3}{j}_{,k} + \vpots{3}{k}_{,j}\right) \spin{a}{k}{i} \dl{a}
\biggr] \neanl &
+ 2\momls{3}{ik}\left(\momls{3}{k}_{,j}-\momls{3}{j}_{,k}\right) 
+ \frac{\dim-2}{\dim-1}\momls{3}{ij}\momls{3}{k}_{,k}
+ 2\momls{3}{ij}_{,k}\vpots{3}{k} \neanl &
+ 8 \frac{2\dim-3}{\dim-2} \phibs{4}\phibs{2}_{,ij} 
+ 8 \frac{3\dim-4}{(\dim-2)^2} \phibs{2}\phibs{2}_{,i}\phibs{2}_{,j}
+ 4 \frac{\dim-4}{\dim-2} \Sfour \phibs{2}_{,ij}\label{eq:B6matter}\,.
\end{align}
\end{subequations}
We expressed $\hat{B}_{(6)ij}$ in a more convenient way now, since the $\momls{3}{i}$ vector potential has a 
much more simple structure than the $\vpots{3}{i}$ vector potential.
\subsection{Matter Correction to the Canonical Field Momentum} \label{sec:matmom}
Since we eliminated the transverse-traceless part of the canonical field momentum via using the relation between canonical
field momentum and velocity of the $\htt_{ij}$ field \eqref{eq:pitt}, there are terms containing matter parts 
of the field momentum left in the Routhian \eqref{eq:FinalTTRouthian3PN}. These can be calculated
from e.g. \cite[Eq. (2.34)]{Steinhoff:Wang:2009} where $\pimati{ij}$ is given by
\begin{align}
 \pimati{ij} & = 16 \pi \grav \sum_a \pia{a}{ij}\dl{a}\label{eq:pimatter}\,.
\end{align}
From \cite[Eqs. (3.33) and (3.34)]{Steinhoff:Wang:2009} one gets the closed form expression
\begin{align}
 \pia{a}{ij} & = \frac{1}{2} \gamma^{ik}\gamma^{j\ell} \frac{m_a \mom{a}{(k} nS_{a\,\ell)}}{\nmom{a} (m_a - \nmom{a})}\,.
\end{align}
The part containing the antisymmetric $\hat{A}^{[ij]}_a$ was neglected, because 
it is of order $\cInv{7}$ ($B^{ij}_{k\ell}$ starts at $\cInv{4}$) which is not necessary here. 
Here we only need $\pia{a}{ij}$ to the order $\cInv{5}$. A power counting (see beginning of subsection \ref{subsub:ordercounting}) 
tells us that we only have to take the leading order approximation of the above expression, reading
\begin{align}
 \pia{(5)a}{ij} & = \frac{\mom{a}{k}}{8 m_a^2} \left(\mom{a}{i}\spin{a}{k}{j} + \mom{a}{j}\spin{a}{k}{i}\right)\label{eq:pimatter5}\,.
\end{align}
This expression has a vanishing trace, which makes obvious that we can neglect $\pihati{ij}$, Eq.\ \eqref{eq:pitrfix}, at the considered order.
\section{Field Solutions and Integration}\label{sec:fieldsolutions}
After obtaining expressions for the sources \eqref{eq:PNSources}, \eqref{eq:PNSourcesMom} and the Routhian \eqref{eq:FinalRouthian3PN} 
which gives the Hamiltonian in the matter degrees of freedom after an integration, 
we need the fields to be inserted into the Routhian. These can be derived by solving the lower order constraint equations
in case of the post-Newtonian potential and the non-propagating parts of the field momentum (see \ref{subsubsec:solconstraints}). 
For the propagating degrees of freedom the wave equation has to be solved (see \ref{subsec:solwaveequation}).

\subsection{\texorpdfstring{$\dim$}{\dim}-dimensional Solutions of the Constraints}\label{subsubsec:solconstraints}
With $K = \frac{\Gamma\left(\frac{\dim}{2} - 1\right)}{\pi^{\frac{\dim}{2} - 1}} \grav$, the Hamilton constraint equations
\eqref{subeq:phi2constr}, \eqref{subeq:phi4constr}, the momentum constraint equation \eqref{subeq:pi3constr}, and the various
transformation formulas \eqref{eq:momentumtildepitilde}, \eqref{eq:momentumtildevpot}, and \eqref{eq:pitildevpot} relating
the longitudinal field momentum and its corresponding vector potentials, we find using the inverse Laplacians listed in the \descappendix 
\ref{sec:invlaptechnique} that
\begin{align}
 \phis{2} & = 4 K \sum_a \frac{m_a}{r_a^{\dim - 2}}\label{eq:phi2sol}\,,\\
 \phis{4} & = 4 K \sum_a \biggl[
    \frac{\vmom{a}^2}{2 m_a} \frac{1}{r_a^{\dim-2}} 
    + \frac{\mom{a}{i}\spin{a}{i}{j}}{2 m_a} \left(\frac{1}{r_a^{\dim-2}}\right)_{,j}
    - K\frac{\dim-2}{\dim-1} \sum_{b\ne a} \frac{m_a m_b}{r_{ab}^{\dim-2} r_b^{\dim-2}}
 \biggr]\label{eq:phi4sol}\,, \\
 \pipots{3}{i} & =  K \sum_a \biggl[
    2\frac{\mom{a}{i}}{r^{\dim - 2}_a}
    +\spin{a}{i}{j}\left(\frac{1}{r_a^{\dim-2}}\right)_{,j}
  \biggr]\label{eq:pitilde3sol}\,,\\
\vpots{3}{i} & = K \sum_a \left[
    2\frac{\mom{a}{i}}{r^{\dim - 2}_a} 
    - \frac{\dim - 2}{2(\dim - 1)(4 - \dim)} \mom{a}{j} \left(\frac{1}{r^{\dim - 4}_a}\right)_{,ij}
    + \spin{a}{i}{j}\left(\frac{1}{r_a^{\dim-2}}\right)_{,j}
   \right]\label{eq:vpot3sol}\,,\\ 
\momls{3}{ij} & =  K \sum_a \biggl[
  2\mom{a}{i} \left(\frac{1}{r^{\dim - 2}_a}\right)_{,j} 
  + 2\mom{a}{j} \left(\frac{1}{r^{\dim - 2}_a}\right)_{,i} 
  - \frac{\dim - 2}{(\dim - 1)(4 - \dim)} \mom{a}{k} \left(\frac{1}{r^{\dim - 4}_a}\right)_{,ijk} \neanl &
  - \frac{2}{\dim - 1} \delta_{ij} \mom{a}{k} \left(\frac{1}{r^{\dim - 2}_a}\right)_{,k}
  + \spin{a}{i}{k}\left(\frac{1}{r_a^{\dim-2}}\right)_{,kj} 
  + \spin{a}{j}{k}\left(\frac{1}{r_a^{\dim-2}}\right)_{,ki}
  \biggr]\label{eq:momentumtilde3sol}\,.
\end{align}
Remember that the momentum constraint can be solved for $\pitildei{i}$ with the help of \eqref{eq:pisolve}.
The more complicated fields like $\momls{5}{i}$ or
$\phis{6}$ were so far only found in $\dim=3$ dimensions \cite{Jaranowski:Schafer:1998}. Also the leading order of the transverse-traceless part
of the metric is only partially known in $\dim$ dimensions. We will discuss these issues in the following subsection.
\subsection{Solutions of the Wave Equation} \label{subsec:solwaveequation}
Consider now the wave equation \eqref{eq:boxhtt} for $\htt_{ij}$. There $\htt_{ij}$ is given in terms
of a post-Newtonian approximate source $S_{ij}$, namely
\begin{align}
 \Box\htt_{ij} &= \TTproj{ij}{k\ell} (S_{(4)k\ell} + S_{(6)k\ell})\,.
\end{align}
By taking into account the near-zone expansion of $\htt_{ij}$, \eqref{eq:nzexpansion}, one gets 
\begin{align}
 \htt_{(4)ij} &= \TTproj{ij}{k\ell} \Delta^{-1} S_{(4)k\ell}\,,\label{eq:htt4}\\
 \htt_{(6)ij} &= \TTproj{ij}{k\ell} \Delta^{-1} \left(S_{(6)k\ell} + \Delta^{-1} \partial_t^2 S_{(4)k\ell}\right)
  = \Delta^{-1}\left(\TTproj{ij}{k\ell}  S_{(6)k\ell} + \httddot_{(4)ij}\right)\,,\label{eq:htt6}
\end{align}
for the leading order and next-to-leading order expressions of $\htt_{ij}$ in the near-zone. 
Here the sources are given by
\begin{align}
 S_{(4)ij} &= 2 B_{(4)ij}\,,\\
 S_{(6)ij} &= 2 \hat{B}_{(6)ij} - \frac{16\pi\grav}{\dim - 1} \srcs{2} \htt_{(4)ij} 
	+ \frac{8}{\dim - 2} (\phibs{2} \htt_{(4)ij,k})_{,k}
	- \pdiffq{}{t} C_{(5)ij}\label{eq:sourcehtt6}\,,
\end{align}
where $B_{(4)ij}$ and $\hat{B}_{(6)ij}$ are given by \eqref{eq:B4matter} and \eqref{eq:B6matter}, and $C_{(5)ij}$ by
\eqref{eq:Cmatter} via \eqref{eq:pimatter} and \eqref{eq:pimatter5}.
Notice that we removed the post-Newtonian order-counting parameter $c$ in \eqref{eq:htt4} and \eqref{eq:htt6}. 
Fortunately there is no need to evaluate \eqref{eq:htt6} here. In fact the $\htt_{(4)ij}$ dependence of \eqref{eq:sourcehtt6}
renders the calculation almost impossible.

Then the solution of the wave equation at leading order and linear in spin is given by
\begin{align}
 \htt_{(4)ij} &= 4 K \sum_a \biggl[
    \frac{\mom{a}{i}\mom{a}{j}}{ m_a} \frac{1}{r_a^{\dim-2}} 
    - \frac{1}{4-\dim}\frac{\mom{a}{k}\mom{a}{(i}}{ m_a} \left(\frac{1}{r_a^{\dim-4}}\right)_{,j)k} \neanl &
    -\frac{1}{\dim-1}\delta_{ij} \biggl(\frac{\vmom{a}^2}{m_a} \frac{1}{r_a^{\dim-2}}
    -\frac{1}{2(4-\dim)}\frac{\mom{a}{k}\mom{a}{\ell}}{m_a}\left(\frac{1}{r_a^{\dim-4}}\right)_{,k\ell}\biggr) \neanl &
    +\frac{1}{2(\dim-1)(4-\dim)}\frac{\vmom{a}^2}{m_a}\left(\frac{1}{r_a^{\dim-4}}\right)_{,ij} \neanl &
    + \frac{\dim-2}{8(\dim-1)(\dim-4)(\dim-6)}\frac{\mom{a}{k}\mom{a}{\ell}}{m_a} \left(\frac{1}{r_a^{\dim-6}}\right)_{,ijk\ell} \neanl &
    +\frac{\mom{a}{k}\spin{a}{\ell}{m}}{m_a}\biggl\{
      \left(\delta_{k(i}\delta_{j)\ell}\partial_m 
      -\frac{1}{\dim-1}\delta_{ij}\delta_{k\ell}\partial_m\right)\frac{1}{r_a^{\dim-2}} \neanl &
      +\frac{1}{2(4-\dim)}\left(\frac{1}{\dim-1}\delta_{k\ell}\partial_i \partial_j \partial_m 
      - \delta_{\ell(i}\partial_{j)}\partial_k\partial_m\right)\frac{1}{r_a^{\dim-4}}
    \biggr\}
\biggr] + \htt_{(4\,0)ij}\label{eq:htt4sol}\,,
\end{align}
where $\htt_{(4\,0)ij}$ is the momentum (and spin-) independent part of the transverse-traceless part of the metric which is
generated by the TT-projection of $\Delta^{-1}(\phibs{2}{}_{,i}\phibs{2}{}_{,j})$ and is only known 
in $\dim=3$ see \cite[Eq. (A20)]{Jaranowski:Schafer:1998}, namely
\begin{align}
 \htt_{(4\,0)ij} &= \gravthree^2 \sum_a \sum_{b\ne a} m_a m_b \biggl\{
    -\frac{4}{s_{ab}}\left(\frac{1}{r_{ab}} + \frac{1}{s_{ab}}\right)\nxa{ab}{i}\nxa{ab}{j}
    +\frac{1}{4}\left(\frac{r_a + r_b}{r_{ab}^3} + \frac{12}{s_{ab}^2}\right)\nxa{a}{i}\nxa{b}{j} \neanl &
    +2\left(\frac{2}{s_{ab}^2} - \frac{1}{r_{ab}^2}\right)(\nxa{a}{i}\nxa{ab}{j}+\nxa{a}{j}\nxa{ab}{i}) \neanl &
    +\biggl[
      \frac{5}{8 r_{ab} r_a}
      -\frac{1}{8 r_{ab}^3} \left(\frac{r_b^2}{r_a} + 3 r_a\right) 
      - \frac{1}{s_{ab}} \left(\frac{1}{r_a} + \frac{1}{s_{ab}}\right)
    \biggr] \nxa{a}{i}\nxa{a}{j} \neanl &
    +\biggl[
      \frac{5 r_a}{8 r_{ab}^3} \left(\frac{r_a}{r_b} - 1\right)
      -\frac{17}{8 r_{ab} r_a}
      +\frac{1}{2 r_a r_b}
      +\frac{1}{s_{ab}} \left(\frac{1}{r_a} + \frac{4}{r_{ab}}\right)
    \biggr] \delta_{ij}
 \biggr\}\,.\label{eq:htt40}
\end{align}
Both solutions were also obtained by using the inverse Laplacians in the \descappendix \ref{sec:invlaptechnique}.
Obviously, most of the parts of $\htt_{(6)ij}$ are of the same type as $\htt_{(4\,0)ij}$ (see \eqref{eq:htt6} and \eqref{eq:sourcehtt6}). 
That is the reason why we eliminated it from the integrands.

\subsection{Distributional Contributions}\label{subsec:distcontrib}
As long as the Riesz-kernel method is not used (where a Dirac delta is substituted by the so-called Riesz-kernel)
one has to take care of delta parts when differentiating certain functions. Consider for example
the field $\phis{2} = 4 K\sum_a \tfrac{m_a}{r_a^{\dim-2}}$ and differentiate it two times. Using the ordinary derivative
it would give
\begin{align}
 \partial_i^{\text{ord}} \partial_j^{\text{ord}} \phis{2} & = 0\,.
\end{align}
But as we already know from the constraint equations the second derivative of $\phis{2}$ should be
\begin{align}
 \partial_i \partial_j \phis{2} & = -16\pi\grav\frac{1}{\dim} \delta_{ij} \sum_a m_a \dl{a}\,.
\end{align}
Fortunately there is a result from the theory of distributions \cite{Jaranowski:Schafer:1998} which 
defines the so-called distributional derivative
\begin{align}
 \partial_i f &= \partial_i^{\text{ord}} f
  + \frac{(-1)^k}{k!} \frac{\partial^k \delta(\vct{x})}{\partial x^{i_1} \dots \partial x^{i_k}} \oint_\Sigma \text{d}\Omega_{\dim-1}\, n^i f x^{i_1} \dots x^{i_k}\,.
\end{align}
Here $f$ is a positive homogeneous function of degree $\lambda$ (i.e. $f(a\vct{x}) = a^\lambda f(\vct{x})$ for $a\ge0$) and
$k := -\lambda + 1 - \dim$ is a non-negative integer. This means $f$ must decay with an exponent linear in the dimension 
$\dim$ which does not apply to fields generated by a Riesz-kernel type source (see \descappendix \ref{subsec:rieszkernel}).
There are not only distributional contributions from the field derivatives, but from the fields themselves (some parts of
the higher order field momenta).

\subsection{Ultraviolet-Analysis}\label{subsec:UVana}
As for gauge theories in quantum field theory, dimensional regularization should be
used in classical general relativity \cite{Damour:Jaranowski:Schafer:2001}.
Therefore first all integrals must be evaluated in generic $\dim$ dimensions and then the limit
$\dim \rightarrow 3$ is calculated. However, certain integrals are very difficult to solve for
generic $\dim$. In practice one therefore evaluates the integrals in $\dim=3$ first and then determines
possible additional contributions that arise from dimensional regularization. That is, one
analyses the $\dim$-dependence of the integrals close to the singular sources, i.e., in the
UV. (Only close to singularities regularization is important.) This is the purpose of the
present section. Other necessary integration techniques are provided in Appendix \ref{sec:integrationtechniques}.

The UV-analysis in generic dimension $\dim$ is a
necessary ingredient to correctly derive the Hamiltonians at formal 3PN level.
This includes the 3PN point-mass Hamiltonian (see \cite{Damour:Jaranowski:Schafer:2001}) and
the NNLO spin-orbit and spin(1)-spin(2) Hamiltonians considered in the present article.
It would also be necessary for the yet unknown NNLO spin(1)-spin(1) Hamiltonian.

For integrals only obtained for $\dim=3$ one has no control on poles in $1/(\dim-3)$.
There are two different problems with such poles: First the poles do not appear in pure $\dim=3$ calculations and thus lead to ambiguous results after
integrations by parts in integrands containing such poles (in one representation there are poles, in another maybe not). This comes from the
fact that some of the pole terms can also give finite contributions which must be added to the $\dim=3$ result. 
Second the poles have to cancel each other in order to extract a finite result from the $\dim$ dimensional integration in the limit $\dim\to3$ (or one must be able to absorb all 
poles through a renormalization procedure as in \cite{Blanchet:Damour:EspositoFarese:2004}).
Both problems are well-known and also discussed in \cite{Damour:Jaranowski:Schafer:2001}.
In the following we will provide some more technical details on how to perform the UV-analysis depending on the structure of the integrand.

All integrals involving $\htt_{(4\,0)\,ij}$, $\TTproj{ij}{k\ell}(\phibs{2}\momls{3}{k\ell})$ and the high order 
potentials such as $\phibs{6}$ or $\momls{5}{i}$ are not available in $\dim$ dimensions and were only calculated in $\dim=3$ dimensions here.
In all other integrals the limit $\dim\to3$
is straightforward, although integrations in $\dim$ dimensions sometimes involve around one
million terms on which the limit must be performed.  
In case of the TT-projection of $\phibs{2}\momls{3}{ij}$, the fields are available in $\dim$ dimensions. Hence, one can split up this part of the 
Hamiltonian in one-particle TT-projections (which can be performed in $\dim$ dimensions) and two-particle TT-projections (which can only be evaluated for $\dim=3$ with the presented methods).
For the latter ones must still perform the UV-analysis.

The term UV-analysis in this context refers to the short-range behavior of the integrand around a specific point. This will become
more clear during the following explanation.
Let us now consider the decay of the integrand $f(r_a, r_b, \vnxa{a}, \vnxa{b})$ around the source $a$.
First of all the integral is split up into a ball integral around one of the sources, say for
example source particle $a$, and an integral over the whole $\mathbb{R}^\dim$ without this ball,
\begin{align}
 \int \text{d}^\dim x\, f(r_a, r_b, \vnxa{a}, \vnxa{b}) &= \int_{B_{\ell_a}(\vx{a})} \text{d}^\dim x\, f(r_a, r_b, \vnxa{a}, \vnxa{b}) 
 + \int_{\mathbb{R}^\dim \setminus B_{\ell_a}(\vx{a})} \text{d}^\dim x\, f(r_a, r_b, \vnxa{a}, \vnxa{b})\,,
\end{align}
where $0<\ell_a\ll r_{ab}$. The variables $r_b$ and $\vnxa{b}$ of the other source ($b\neq a$) are expressed in terms of $r_a$, $r_{ab}$ and $\vnxa{ab}$,
\begin{align}
 r_b & = |\vct{x}-\vx{b}| = |\vct{x}-\vx{a} + \vx{a} - \vx{b}|
 =\sqrt{r_a^2 + r_{ab}^2 + 2 r_a r_{ab} \scpm{\vnxa{a}}{\vnxa{ab}}} \,,\label{eq:rbinsertion}\\
 \vnxa{b} & = \frac{r_a}{r_b} \vnxa{a} + \frac{r_{ab}}{r_{b}} \vnxa{ab}\,, 
\end{align}
such that all $\vct{x}$-dependent expressions come from $r_{a}$ and $\vnxa{a}$ type variables.
Next we concentrate on the ball integral around $a$,
\begin{align}
 \int_{B_{\ell_a}(\vct{x}_a)} \text{d}^\dim x f(r_{a}, \vnxa{a}) & = \int \text{d}\Omega_{a,\dim-1} \int_{0}^{\ell_a}\text{d}r_a\, r_a^{\dim-1} f(r_{a}, \vnxa{a})\,.\label{eq:ballintegration}
\end{align}
Now the integrand is expanded in $r_a$ (leaving $\vnxa{a}$ untouched).
This is possible because $a$ and $b$ are well separated, and the ball contains only a small neighborhood of the source $a$.
Then the integrand takes the form of a polynomial in $r_a$ and one can pick out the terms
contributing poles at $\dim=3$ (the ones with an exponent giving $-3$ for $\dim=3$ on $r_a$).
The next step is to count the number of $\vnxa{a}$-vectors in each term and remove terms with an odd number of these vectors. This is due to the
averaging procedure coming from the angular integration in \eqref{eq:ballintegration} using the formulas \eqref{eq:averageOmega} in the
\descappendix.
Consider for example an integrand $f(r_a, \vnxa{a}) = C(\dim) r_a^{6-3\dim} \scpm{\vmom{a}}{\vnxa{a}}\scpm{\vmom{b}}{\vnxa{a}}$ (this integrand
is of the form qualified as dangerous in \cite[Eq. (3.1)]{Damour:Jaranowski:Schafer:2001}) then
\eqref{eq:ballintegration} gives
\begin{align}
 \int_{B_{\ell_a}(\vct{x}_a)} \text{d}^\dim x f(r_{a}, \vnxa{a}) & = 
C(\dim) \int \text{d}\Omega_{a,\dim-1} \scpm{\vmom{a}}{\vnxa{a}}\scpm{\vmom{b}}{\vnxa{a}} \int_{0}^{\ell_a}\text{d}r_a\, r_a^{\dim-1}r_a^{6-3\dim} \neanl & 
 = C(\dim) \mom{a}{i}\mom{b}{j}\int \text{d}\Omega_{a,\dim-1} \nxa{a}{i}\nxa{a}{j} \int_{0}^{\ell_a}\text{d}r_a\, r_a^{5 - 2\dim}\,.
\end{align}
Using \eqref{eq:angavgtwo} and usual integration rules we obtain
\begin{align}
 \int_{B_{\ell_a}(\vct{x}_a)} \text{d}^\dim x f(r_{a}, \vnxa{a}) & = 
\frac{C(\dim)}{\dim}\Omega_{a,\dim-1} \scpm{\vmom{a}}{\vmom{b}} \frac{\ell_a^{2(3-\dim)}}{2(3-\dim)}\,.
\end{align}
The last integration step was performed by means of an analytic continuation from $\dim<3$.
At this stage there are three possibilities: The first one is that $C(\dim)$ contains several factors of $\dim-3$ which cancel the pole in the last factor and even lead to a vanishing limit when $\dim \rightarrow 3$.
Then the potentially dangerous term is actually not dangerous at all. The second possibility is that $C(\dim) \sim \dim-3$ which would also lead to a
cancellation of the pole but would give a finite contribution which has to be taken into account to get the correct Hamiltonian.
Last but not least $C(\dim)$ could have such a structure that a pole remains and so this term has to be renormalized or canceled by another
pole to give a physically meaningful result.

The procedure mentioned above is only valid if there is no TT-projection (and in particular no $\htt_{(4\,0)ij}$) appearing. 
The analysis of the $\htt_{(4\,0)\,ij}$ type integrals works as follows (during this discussion we talk about a two-particle system 
where we consider $r_1$ as expansion point):
The $\htt_{(4\,0)\,ij}$ part is given in terms of inverse Laplacians and field variables in $\dim$ dimensions by
\begin{align}
 \htt_{(4\,0)\,ij} & = - \frac{8(\dim-1)}{\dim-2} \TTproj{ij}{kl}\Delta^{-1} (\phibs{2}{}_{, k} \phibs{2}{}_{, l})\,,
\end{align}
see \eqref{eq:htt40} for the explicit solution in $\dim=3$.
Now one can insert the $\phibs{2}$ field, given by
\begin{align}
 \phibs{2} = \frac{\dim-2}{4(\dim-1)} (m_1 u_1 + m_2 u_2)\,,
\end{align}
where $u_a = - 16\pi \grav\Delta^{-1} \dl{a}\sim r_a^{2-\dim}$. After interchanging TT-projector and inverse Laplacian, one gets 
$\TTproj{ij}{kl} (u_{a,k} u_{a,l})=0$. One can see this by using $u_{a,k}u_{a,\ell} \sim \nxa{a}{k}\nxa{a}{\ell} r_{a}^{2-2\dim}$ 
which can be rewritten using 
\begin{align}
\partial_i \partial_j r_{a}^{4-2\dim} &= -2(\dim-2)(\delta_{ij}-2(\dim-1)\nxa{a}{i}\nxa{a}{j})r_{a}^{2-2\dim}\,,
\end{align}
as $u_{a,k}u_{a,\ell} \sim \frac{1}{2(\dim-1)}(\delta_{ij}r_{a}^{2-2\dim} + \frac{1}{2(\dim-2)}\partial_i \partial_j r_{a}^{4-2\dim})$. 
This will obviously be projected to zero by the TT-projector. Thus, after TT-projection there is only one part left, namely
\begin{align}
 \htt_{(4\,0)\,ij} & = - 2 \frac{\dim-2}{2(\dim-1)} m_1 m_2 \Delta^{-1} \TTproj{ij}{kl}  (u_{1, k} u_{2, l})\,.
\end{align}
Under the TT-projector one can integrate by parts because $\partial_k \TTproj{ij}{kl} = 0$ and obtains
\begin{align}
 \htt_{(4\,0)\,ij} & = 2 \frac{\dim-2}{2(\dim-1)} m_1 m_2 \Delta^{-1} \TTproj{ij}{kl}  (u_{1} \partial_{k} \partial_{l} u_{2})\,.
\end{align}
Notice that the derivative should act on the $u_2$ term because all quantities will be Taylor expanded in $r_1$ around $\vct{x} = \vx{1}$
and $u_1$ is already proportional to an $r_1$-power. At this stage of processing the $\htt_{(4\,0)\,ij}$
terms one is able to use the same analysis procedure as mentioned above:
Check whether there are powers of $r_1$ with exponents smaller than $-3$ for $\dim=3$ and expand $\partial_{k} \partial_{l} u_{2}$ around $r_1$ to an 
order sufficient to reach the critical value in the exponent of $r_1$.
The idea is the expansion of the $r_2$ variable in the TT-projector such that the Taylor expansion consists only of
multiple inverse Laplacians on powers of $r_1$, which can be calculated using \eqref{eq:invra} from the \descappendix. 
Also bear in mind that the inverse Laplacian introduces additional powers of $r_1$ via the Green's function. The final 
ball integration can now be performed as discussed above.
The same technique can also be used to perform the UV-analysis of terms involving 
$\TTproj{ij}{kl}(\phibs{2}\momls{3}{kl})$.

To check our UV-analysis code we first reproduced the pole coefficients given in \cite[Table 1]{Damour:Jaranowski:Schafer:2001}.
The {\em finite} UV-contribution to the 3PN point-mass Hamiltonian in our representation \eqref{eq:FinalRouthian3PN} is given by
\begin{align}
 \Delta H^{\text{3PN,UV}}_{\text{PM}}(\dim) &= \frac{2 \Lambda^{6-2\dim} (\dim-2)(\dim+1)(96-40\dim-28\dim^2+\dim^3) \pi^{3-3\dim/2} \Gamma\left(\frac{\dim}{2}\right)^3}{3(\dim-4)(\dim-1)^4(\dim+2)} \times \neanl &
 \frac{(\grav)^3}{\rel^\dim}  m_1 m_2 \left(\dim\scpm{\vnunit}{\vmom{1}}^2 - \vmom{1}^2 + \dim\scpm{\vnunit}{\vmom{2}}^2 - \vmom{2}^2\right)\,,
 \label{eq:UVcontribPM}
\end{align}
where $\Lambda$ is a UV-cutoff scale which does not contribute in the limit $\dim\to3$. A similar analysis for the 2PN point-mass Hamiltonian
gave no contribution. We found no net contribution to the spin-dependent Hamiltonians, though poles and finite parts appeared in intermediate expressions.
That is, Hadamard regularization would have been sufficient to obtain the correct
linear-in-spin Hamiltonians presented in  the present work. The same situation was
also found for the harmonic-gauge calculation of the equations of motion in 
\cite{Marsat:Bohe:Faye:Blanchet:2012,Bohe:Marsat:Faye:Blanchet:2012}.
\section{Results}\label{sec:results}

After discussing the several simplifications given above to reduce the integral of the formal 3PN Routhian to a form
which can be handled appropriately, we continue by giving a short description of the integrands showing up at this order.
The integrands can be divided into three different types:
\begin{itemize}
 \item the delta-type $\int {\rm d}^\dim x f(\vct{x}) \dl{1}$, 
 \item the Riesz-type $\int {\rm d}^\dim x\, n^{i_1}_1 \dots n^{i_k}_1 n^{j_1}_2 \dots n^{j_\ell}_2 r_1^\alpha r_2^\beta$,
 \item and the generalized Riesz-type $\int {\rm d}^3 x\, n^{i_1}_1 \dots n^{i_k}_1 n^{j_1}_2 \dots n^{j_\ell}_2 r_1^\alpha r_2^\beta s_{12}^\gamma$.
\end{itemize}
The solution of these three types of integrals will be shown in \descappendix \ref{sec:integrationtechniques}. 

We used our {\sc Mathematica} code to perform an integration of \eqref{eq:FinalRouthian3PN} directly
with all fields inserted up to linear order in spin, neglecting spin(1)$^2$ and spin(2)$^2$
terms afterwards. From this we obtained the fully reduced matter Hamiltonians for point masses,
for the spin-orbit- and the spin(1)-spin(2)-interaction.
There were no $1/(\dim-3)$ poles for the linear-in-spin Hamiltonians giving rise to finite parts
or being singular for $\dim\to3$. Such poles only appeared in some intermediate
steps in the UV-analysis (see \Sec{subsec:UVana}) but finally identically canceled. For
the point-mass parts there appeared a finite contribution given in \eqref{eq:UVcontribPM}. This
together with our integration result reproduced the result from the literature exactly. 

Now we are able to write down the NNLO linear-in-spin fully reduced Hamiltonians in terms of matter variables only, by 
using \eqref{eq:FinalRouthian3PN} for the Routhian, inserting the sources \eqref{eq:PNSources}, 
$\hat{B}_{(6)ij}$ \eqref{eq:B6matter}, $\pimati{ij}$ \eqref{eq:pimatter5}, the solution for 
the constraint fields \eqref{eq:phi2sol}--\eqref{eq:momentumtilde3sol}, \eqref{eq:momentumtilde51}, \eqref{eq:pitilde51}, 
and the propagating degrees of freedom \eqref{eq:htt4sol} and \eqref{eq:htt40}. 
Notice that certain field derivatives
may lead to distributional contributions mentioned in \ref{subsec:distcontrib}. The integrals
can (for a binary) be solved by using the techniques given in \descappendix \ref{sec:integrationtechniques}.
Both Hamiltonians are valid for any compact objects like black holes or neutron stars.
Their center-of-mass frame versions are given in \ref{subsec:comhamiltonians} where the gyromagnetic ratios in the spin-orbit case
are also given in \cite{Nagar:2011,Barausse:Buonanno:2011}.

\subsection{Next-to-next-to-leading Order Spin-Orbit Hamiltonian}

The spin-orbit Hamiltonian given in this subsection is the higher order gravitational analogue of the
interaction of an electron's spin interacting with the electron's orbital angular momentum in the case
of an e.g. hydrogen atom. In quantum electrodynamics this interaction is responsible for the 
fine structure in the spectrum. Here the spin is obviously no quantum mechanical quantity, it only characterizes
the rotation (i.e. its direction and magnitude) of a gravitating mass in the gravitational field of another mass.
Notice that the fine structure constant $\alpha$ in electromagnetic theory is substituted by Newton's gravitational
constant $\gravthree$ here.
The result for the NNLO spin-orbit Hamiltonian reads
\begin{align}
 H^{\text{NNLO}}_{\text{SO}} & = \frac{\gravthree}{\rel^2} \biggl[
	\biggl(
		\frac{7 m_2 (\vmom{1}^2)^2}{16 m_1^5} 
		+ \frac{9 \scpm{\vnun}{\vmom{1}}\scpm{\vnun}{\vmom{2}}\vmom{1}^2}{16 m_1^4} 
		+ \frac{3 \vmom{1}^2 \scpm{\vnun}{\vmom{2}}^2}{4 m_1^3 m_2}\nonumber\\
&		+ \frac{45 \scpm{\vnun}{\vmom{1}}\scpm{\vnun}{\vmom{2}}^3}{16 m_1^2 m_2^2}
		+ \frac{9 \vmom{1}^2 \scpm{\vmom{1}}{\vmom{2}}}{16 m_1^4}
		- \frac{3 \scpm{\vnun}{\vmom{2}}^2 \scpm{\vmom{1}}{\vmom{2}}}{16 m_1^2 m_2^2}\nonumber\\
&		- \frac{3 (\vmom{1}^2) (\vmom{2}^2)}{16 m_1^3 m_2}
		- \frac{15 \scpm{\vnun}{\vmom{1}}\scpm{\vnun}{\vmom{2}} \vmom{2}^2}{16 m_1^2 m_2^2}
		+ \frac{3 \scpm{\vnun}{\vmom{2}}^2 \vmom{2}^2}{4 m_1 m_2^3}\nonumber\\
&		- \frac{3 \scpm{\vmom{1}}{\vmom{2}} \vmom{2}^2}{16 m_1^2 m_2^2}
		- \frac{3 (\vmom{2}^2)^2}{16 m_1 m_2^3}
	\biggr)((\vnun \times \vmom{1})\vspin{1})
	+\biggl(
		- \frac{3 \scpm{\vnun}{\vmom{1}}\scpm{\vnun}{\vmom{2}}\vmom{1}^2}{2 m_1^3 m_2}\nonumber\\
&		- \frac{15 \scpm{\vnun}{\vmom{1}}^2\scpm{\vnun}{\vmom{2}}^2}{4 m_1^2 m_2^2}
		+ \frac{3 \vmom{1}^2 \scpm{\vnun}{\vmom{2}}^2}{4 m_1^2 m_2^2}
		- \frac{\vmom{1}^2 \scpm{\vmom{1}}{\vmom{2}}}{2 m_1^3 m_2}
		+ \frac{\scpm{\vmom{1}}{\vmom{2}}^2}{2 m_1^2 m_2^2}\nonumber\\
&		+ \frac{3 \scpm{\vnun}{\vmom{1}}^2 \vmom{2}^2}{4 m_1^2 m_2^2}
		- \frac{(\vmom{1}^2) (\vmom{2}^2)}{4 m_1^2 m_2^2}
		- \frac{3 \scpm{\vnun}{\vmom{1}}\scpm{\vnun}{\vmom{2}}\vmom{2}^2}{2 m_1 m_2^3}\nonumber\\
&		- \frac{\scpm{\vmom{1}}{\vmom{2}} \vmom{2}^2}{2 m_1 m_2^3}
	\biggr)((\vnun \times \vmom{2})\vspin{1})
	+\biggl(
		- \frac{9 \scpm{\vnun}{\vmom{1}} \vmom{1}^2}{16 m_1^4}
		+ \frac{\vmom{1}^2 \scpm{\vnun}{\vmom{2}}}{m_1^3 m_2}\nonumber\\
&		+ \frac{27 \scpm{\vnun}{\vmom{1}}\scpm{\vnun}{\vmom{2}}^2}{16 m_1^2 m_2^2}
		- \frac{\scpm{\vnun}{\vmom{2}}\scpm{\vmom{1}}{\vmom{2}}}{8 m_1^2 m_2^2}
		- \frac{5 \scpm{\vnun}{\vmom{1}} \vmom{2}^2}{16 m_1^2 m_2^2}\nonumber\\
&		+ \frac{\scpm{\vnun}{\vmom{2}}\vmom{2}^2}{m_1 m_2^3}
	\biggr)((\vmom{1} \times \vmom{2})\vspin{1})
\biggr] \nonumber\\
&+ \frac{\gravthree^2}{\rel^3} \biggl[
	\biggl(
		-\frac{3 m_2 \scpm{\vnun}{\vmom{1}}^2}{2 m_1^2}
		+\left(
			-\frac{3 m_2}{2 m_1^2}
			+\frac{27 m_2^2}{8 m_1^3}
		\right) \vmom{1}^2
		+\left(
			\frac{177}{16 m_1}
			+\frac{11}{m_2}
		\right) \scpm{\vnun}{\vmom{2}}^2\nonumber\\
&		+\left(
			\frac{11}{2 m_1}
			+\frac{9 m_2}{2 m_1^2}
		\right) \scpm{\vnun}{\vmom{1}} \scpm{\vnun}{\vmom{2}}
		+\left(
			\frac{23}{4 m_1}
			+\frac{9 m_2}{2 m_1^2}
		\right) \scpm{\vmom{1}}{\vmom{2}}\nonumber\\
&		-\left(
			\frac{159}{16 m_1}
			+\frac{37}{8 m_2}
		\right) \vmom{2}^2
	\biggr)((\vnun \times \vmom{1})\vspin{1})
	+\biggl(
		\frac{4 \scpm{\vnun}{\vmom{1}}^2}{m_1}
		+\frac{13 \vmom{1}^2}{2 m_1}\nonumber\\
&		+\frac{5 \scpm{\vnun}{\vmom{2}}^2}{m_2}
		+\frac{53 \vmom{2}^2}{8 m_2}
		- \left(
			\frac{211}{8 m_1}
			+\frac{22}{m_2}
		\right) \scpm{\vnun}{\vmom{1}} \scpm{\vnun}{\vmom{2}}\nonumber\\
&		-\left(
			\frac{47}{8 m_1}
			+\frac{5}{m_2}
		\right)\scpm{\vmom{1}}{\vmom{2}}
	\biggr)((\vnun \times \vmom{2})\vspin{1})
	+\biggl(
		-\left(
			\frac{8}{m_1}
			+\frac{9 m_2}{2 m_1^2}
		\right)\scpm{\vnun}{\vmom{1}}\nonumber\\
&		+\left(
			\frac{59}{4 m_1}
			+\frac{27}{2 m_2}
		\right)\scpm{\vnun}{\vmom{2}}
	\biggr)((\vmom{1} \times \vmom{2})\vspin{1})
\biggr]\nonumber\\
&+\frac{\gravthree^3}{\rel^4} \biggl[
	\left(
		\frac{181 m_1 m_2}{16}
		+ \frac{95 m_2^2}{4}
		+ \frac{75 m_2^3}{8 m_1}
	\right) ((\vnun \times \vmom{1})\vspin{1})\nonumber\\
&	- \left(
		\frac{21 m_1^2}{2}
		+ \frac{473 m_1 m_2}{16}
		+ \frac{63 m_2^2}{4}
	\right)((\vnun \times \vmom{2})\vspin{1})
\biggr]
 + (1\leftrightarrow2)\label{eq:HNNLOSO}\,.
\end{align}
This Hamiltonian is formally at 3PN but for maximally rotating objects the post-Newtonian
order goes up to 3.5PN. Recently in \cite{Marsat:Bohe:Faye:Blanchet:2012,Bohe:Marsat:Faye:Blanchet:2012} the NNLO spin-orbit contributions
to the acceleration and spin-precession in harmonic gauge were calculated and agreement with the equations
of motion following from our Hamiltonian was found.\footnote{In \cite{Hartung:Steinhoff:2011:1}
there is a typo in the term $-\frac{\gravthree}{\rel^2}\frac{15 \scpm{\vnun}{\vmom{1}} \vmom{2}^2}{16 m_1^2 m_2^2} ((\vmom{1} \times \vmom{2})\vspin{1})$. 
The coefficient has to be $-\frac{5}{16}$ instead of $-\frac{15}{16}$. The result given here is correct. Thanks to S. Marsat for pointing this out.}
From a combinatorial point of view there are 66 algebraically 
different possible contributions to the Hamiltonian for each object 
(written in terms of the canonical spin tensor), but 24 of them do not 
appear in the canonical representation used here.

\subsection{Next-to-next-to-leading Order Spin(1)-Spin(2) Hamiltonian}

Also the spin(1)-spin(2) Hamiltonian has an electromagnetic
counterpart. It is the gravitational analogue to e.g. the coupling
between electron spin and spin of the atomic nucleus, responsible for the
hyperfine structure in the electromagnetic spectrum. Of course in our case the 
spin(1)-spin(2) interaction leads to the modulation of gravitational
waves but does not lead to a hyperfine structure in the emitted atomic electromagnetic spectrum.
The result for the NNLO spin(1)-spin(2) Hamiltonian reads
\begin{align}
  H^{\text{NNLO}}_{\text{SS}} & =  \frac{\gravthree}{\rel^3}\biggl[
	\frac{((\vmom{1} \times \vmom{2})\,\vspin{1})((\vmom{1} \times \vmom{2})\,\vspin{2})}{16 m_1^2 m_2^2}
	-\frac{9 ((\vmom{1} \times \vmom{2})\,\vspin{1})((\vnun \times \vmom{2})\,\vspin{2})\scpm{\vnun}{\vmom{1}}}{8 m_1^2 m_2^2} \nlqn
	-\frac{3 ((\vnun \times \vmom{2})\,\vspin{1})((\vmom{1} \times \vmom{2})\,\vspin{2})\scpm{\vnun}{\vmom{1}}}{2 m_1^2 m_2^2} \nlqn
	+((\vnun \times \vmom{1})\,\vspin{1})((\vnun \times \vmom{1})\,\vspin{2})\biggl(
		\frac{9 \vmom{1}^2}{8 m_1^4}
		+ \frac{15 \scpm{\vnun}{\vmom{2}}^2}{4 m_1^2 m_2^2}
		- \frac{3 \vmom{2}^2}{4 m_1^2 m_2^2}
	\biggr) \nlqn
	+((\vnun \times \vmom{2})\,\vspin{1})((\vnun \times \vmom{1})\,\vspin{2})\biggl(
		-\frac{3 \vmom{1}^2}{2 m_1^3 m_2}
		+\frac{3 \scpm{\vmom{1}}{\vmom{2}}}{4 m_1^2 m_2^2} \nlqn
		-\frac{15 \scpm{\vnun}{\vmom{1}}\scpm{\vnun}{\vmom{2}}}{4 m_1^2 m_2^2}
	\biggr) 
	+((\vnun \times \vmom{1})\,\vspin{1})((\vnun \times \vmom{2})\,\vspin{2})\biggl(
		\frac{3 \vmom{1}^2}{16 m_1^3 m_2} \nlqn
		-\frac{3 \scpm{\vmom{1}}{\vmom{2}}}{16 m_1^2 m_2^2} 
		-\frac{15 \scpm{\vnun}{\vmom{1}}\scpm{\vnun}{\vmom{2}}}{16 m_1^2 m_2^2}
	\biggr) \nlqn
	+ \scpm{\vmom{1}}{\vspin{1}}\scpm{\vmom{1}}{\vspin{2}}\biggl(
		\frac{3 \scpm{\vnun}{\vmom{2}}^2}{4 m_1^2 m_2^2} 
		- \frac{\vmom{2}^2}{4 m_1^2 m_2^2}
	\biggr) \nlqn
	+ \scpm{\vmom{1}}{\vspin{1}}\scpm{\vmom{2}}{\vspin{2}}\biggl(
		-\frac{\vmom{1}^2}{4 m_1^3 m_2}
		+\frac{\scpm{\vmom{1}}{\vmom{2}}}{4 m_1^2 m_2^2}
	\biggr) \nlqn
	+ \scpm{\vmom{2}}{\vspin{1}}\scpm{\vmom{1}}{\vspin{2}}\biggl(
		\frac{5\vmom{1}^2}{16 m_1^3 m_2}
		-\frac{3\scpm{\vmom{1}}{\vmom{2}}}{16 m_1^2 m_2^2}
		-\frac{9\scpm{\vnun}{\vmom{1}}\scpm{\vnun}{\vmom{2}}}{16 m_1^2 m_2^2}
	\biggr) \nlqn
	+ \scpm{\vnun}{\vspin{1}}\scpm{\vmom{1}}{\vspin{2}}\biggl(
		\frac{9 \scpm{\vnun}{\vmom{1}} \vmom{1}^2}{8 m_1^4}
		-\frac{3 \scpm{\vnun}{\vmom{2}} \vmom{1}^2}{4 m_1^3 m_2}
		-\frac{3 \scpm{\vnun}{\vmom{2}} \vmom{2}^2}{4 m_1 m_2^3}
	\biggr) \nlqn
	+ \scpm{\vmom{1}}{\vspin{1}}\scpm{\vnun}{\vspin{2}}\biggl(
		-\frac{3 \scpm{\vnun}{\vmom{2}} \vmom{1}^2}{4 m_1^3 m_2}
		-\frac{15 \scpm{\vnun}{\vmom{1}}\scpm{\vnun}{\vmom{2}}^2}{4 m_1^2 m_2^2}
		+\frac{3 \scpm{\vnun}{\vmom{1}} \vmom{2}^2}{4 m_1^2 m_2^2} \nlqn
		-\frac{3 \scpm{\vnun}{\vmom{2}} \vmom{2}^2}{4 m_1 m_2^3}
	\biggr)
	+ \scpm{\vnun}{\vspin{1}}\scpm{\vnun}{\vspin{2}}\biggl(
		-\frac{3 \scpm{\vmom{1}}{\vmom{2}}^2{}}{8 m_1^2 m_2^2} 
		+\frac{105 \scpm{\vnun}{\vmom{1}}^2 \scpm{\vnun}{\vmom{2}}^2}{16 m_1^2 m_2^2} \nlqn
		-\frac{15 \scpm{\vnun}{\vmom{2}}^2 \vmom{1}^2}{8 m_1^2 m_2^2} 
		+\frac{3 \vmom{1}^2\scpm{\vmom{1}}{\vmom{2}}}{4 m_1^3 m_2}
		+\frac{3 \vmom{1}^2 \vmom{2}^2}{16 m_1^2 m_2^2}
		+\frac{15 \vmom{1}^2 \scpm{\vnun}{\vmom{1}}\scpm{\vnun}{\vmom{2}}}{4 m_1^3 m_2}
	\biggr) \nlqn
	+ \scpm{\vspin{1}}{\vspin{2}}\biggl(
		\frac{\scpm{\vmom{1}}{\vmom{2}}^2}{16 m_1^2 m_2^2}
		-\frac{9 \scpm{\vnun}{\vmom{1}}^2 \vmom{1}^2}{8 m_1^4}
		-\frac{5 \scpm{\vmom{1}}{\vmom{2}} \vmom{1}^2}{16 m_1^3 m_2} 
		-\frac{3 \scpm{\vnun}{\vmom{2}}^2\vmom{1}^2}{8 m_1^2 m_2^2} \nlqn
		-\frac{15 \scpm{\vnun}{\vmom{1}}^2 \scpm{\vnun}{\vmom{2}}^2}{16 m_1^2 m_2^2} 
		+\frac{3 \vmom{1}^2 \vmom{2}^2}{16 m_1^2 m_2^2}
		+\frac{3 \vmom{1}^2 \scpm{\vnun}{\vmom{1}}\scpm{\vnun}{\vmom{2}}}{4 m_1^3 m_2} \nlqn
		+\frac{9 \scpm{\vmom{1}}{\vmom{2}}\scpm{\vnun}{\vmom{1}}\scpm{\vnun}{\vmom{2}}}{16 m_1^2 m_2^2}
	\biggr)
\biggr]  \nln
 + \frac{\gravthree^2}{\rel^4}\biggl[
	((\vnun \times \vmom{1})\,\vspin{1})((\vnun \times \vmom{1})\,\vspin{2})\biggl(
		\frac{12}{m_1}
		+\frac{9 m_2}{m_1^2}
	\biggr) \nlqn
	-\frac{81}{4 m_1}((\vnun \times \vmom{2})\,\vspin{1})((\vnun \times \vmom{1})\,\vspin{2})
	-\frac{27}{4 m_1}((\vnun \times \vmom{1})\,\vspin{1})((\vnun \times \vmom{2})\,\vspin{2})
	\nlqn
	-\frac{5}{2 m_1}\scpm{\vmom{1}}{\vspin{1}}\scpm{\vmom{2}}{\vspin{2}}
	+\frac{29}{8 m_1}\scpm{\vmom{2}}{\vspin{1}}\scpm{\vmom{1}}{\vspin{2}}
	-\frac{21}{8 m_1}\scpm{\vmom{1}}{\vspin{1}}\scpm{\vmom{1}}{\vspin{2}}
	\nlqn
	+\scpm{\vnun}{\vspin{1}}\scpm{\vmom{1}}{\vspin{2}}\biggl\{
		\left(\frac{33}{2 m_1} + \frac{9 m_2}{m_1^2}\right)\scpm{\vnun}{\vmom{1}}
		-\left(\frac{14}{m_1} + \frac{29}{2 m_2}\right)\scpm{\vnun}{\vmom{2}}
	\biggr\} \nlqn
	+\scpm{\vmom{1}}{\vspin{1}}\scpm{\vnun}{\vspin{2}}\biggl\{
		\frac{4}{m_1}\scpm{\vnun}{\vmom{1}}
		-\left(\frac{11}{m_1} + \frac{11}{m_2}\right)\scpm{\vnun}{\vmom{2}}
	\biggr\} \nlqn
	+\scpm{\vnun}{\vspin{1}}\scpm{\vnun}{\vspin{2}}\biggl\{
		-\frac{12}{m_1} \scpm{\vnun}{\vmom{1}}^2
		-\frac{10}{m_1} \vmom{1}^2
		+\frac{37}{4 m_1}
			\scpm{\vmom{1}}{\vmom{2}} \nlqn
		+\frac{255}{4 m_1}
			\scpm{\vnun}{\vmom{1}}\scpm{\vnun}{\vmom{2}}
	\biggr\} 
	+\scpm{\vspin{1}}{\vspin{2}}\biggl\{
		-\left(\frac{25}{2 m_1} + \frac{9 m_2}{m_1^2}\right) \scpm{\vnun}{\vmom{1}}^2
		+ \frac{49}{8 m_1} \vmom{1}^2 \nlqn
		+ \frac{35}{4 m_1}
			\scpm{\vnun}{\vmom{1}}\scpm{\vnun}{\vmom{2}} 
		- \frac{43}{8 m_1}
			\scpm{\vmom{1}}{\vmom{2}}
	\biggr\}
\biggr]\nln
+\frac{\gravthree^3}{\rel^5}\biggl[
	-\scpm{\vspin{1}}{\vspin{2}}\left(
		\frac{63}{4} m_1^2
		+\frac{145}{8} m_1 m_2
	\right) 
	+ \scpm{\vnun}{\vspin{1}}\scpm{\vnun}{\vspin{2}}\left(
		\frac{105}{4} m_1^2
		+\frac{289}{8} m_1 m_2
	\right)
\biggr] \nln
 + (1\leftrightarrow2)\label{eq:HNNLOS1S2}\,,
\end{align}
This Hamiltonian is formally also at 3PN but for maximally rotating objects the post-Newtonian
order goes up to 4PN.
Notice that from a combinatorial point of view there are 167 algebraically 
different possible contributions to the Hamiltonian for all objects
(written in terms of the canonical spin tensor), but 75 of them do not 
appear in the canonical representation used here. 

\subsection{Hamiltonians in Center-Of-Mass Frame}\label{subsec:comhamiltonians}
For later computations of, e.g., the mentioned orbital parametrizations of a binary system
it is convenient to provide the Hamiltonians in the center-of-mass frame ($\vmom{1} = -\vmom{2} = \vmom{}$).
In this frame in dimensionless quantities (see e.g. \cite{Tessmer:Hartung:Schafer:2010,Tessmer:Hartung:Schafer:2012} for rescaling)
they are given by
\begin{subequations}
\label{eq:Hredfull}
\begin{align}
H_{\text{COM SO}}^{\text{NNLO}}
&= \frac{1}{4\rel^5} \left[
    21\sqrt{1-4\eta}(\eta+1)\scpm{\vang}{\vct{\Delta}}
    +\frac{1}{2} (-2\eta^2+33\eta+42)\scpm{\vang}{\vct{\Sigma}}
  \right] \neanl&
  + \frac{\eta}{32 \rel^4} \biggl[
    -\sqrt{1-4\eta}\left((256+45\eta)\scpm{\vnunit}{\vmom{}}^2 + (314+39\eta)\vmom{}^2\right)\scpm{\vang}{\vct{\Delta}} \neanl & \quad
    +\left((-256+275\eta)\scpm{\vnunit}{\vmom{}}^2 + (-206+73\eta)\vmom{}^2\right)\scpm{\vang}{\vct{\Sigma}}
  \biggr] \neanl&
  + \frac{\eta}{32 \rel^3} \biggl[
    \sqrt{1-4\eta}\bigl(
	15\scpm{\vnunit}{\vmom{}}^4
	+ 3(9\eta-4) \scpm{\vnunit}{\vmom{}}^2 \vmom{}^2 \neanl & \quad\quad
	+ 2(22\eta-9) (\vmom{}^2)^2
    \bigr)\scpm{\vang}{\vct{\Delta}} 
    -\bigl(
	15(2\eta-1)\scpm{\vnunit}{\vmom{}}^4 \neanl &\quad\quad
	+3 (6\eta^2 - 11 \eta + 4) \scpm{\vnunit}{\vmom{}}^2 \vmom{}^2
	+2 (5\eta^2 - 3\eta + 2) (\vmom{}^2)^2
    \bigr)\scpm{\vang}{\vct{\Sigma}}
  \biggr]\,,\\
H_{\text{COM SS}}^{\text{NNLO}}
&=
\eta\biggl\{
  \frac{1}{4\rel^5} \left[
    (79\eta + 105) \scpm{\vnunit}{\vspin{1}}\scpm{\vnunit}{\vspin{2}}
    -(63 + 19\eta) \scpm{\vspin{1}}{\vspin{2}}
  \right] \neanl &\quad
    +\frac{1}{\rel^4} \biggl[
      -\left(
	\frac{303}{4}\eta \scpm{\vnunit}{\vmom{}}^2 
	+ \left(\frac{125}{4}\eta+9\right) \vmom{}^2
      \right) \scpm{\vnunit}{\vspin{1}}\scpm{\vnunit}{\vspin{2}} \neanl & \quad\quad
      \left(
      	-\left(18+\frac{25}{4}\eta\right) \scpm{\vnunit}{\vmom{}}^2 
	+ \left(9 + \frac{47}{2}\eta\right) \vmom{}^2
      \right) \scpm{\vspin{1}}{\vspin{2}} \neanl & \quad\quad
      - \frac{9}{4} (7\eta+4) \scpm{\vmom{}}{\vspin{1}}\scpm{\vmom{}}{\vspin{2}} \neanl & \quad\quad
      +\left(34\eta+\frac{27}{2}\right)\scpm{\vnunit}{\vmom{}}
	(\scpm{\vmom{}}{\vspin{1}}\scpm{\vnunit}{\vspin{2}} + \scpm{\vnunit}{\vspin{1}}\scpm{\vmom{}}{\vspin{2}}) \neanl & \quad\quad
      +\frac{3}{2}\sqrt{1-4\eta}(\eta+3)\scpm{\vnunit}{\vmom{}}
	(\scpm{\vmom{}}{\vspin{1}}\scpm{\vnunit}{\vspin{2}} - \scpm{\vnunit}{\vspin{1}}\scpm{\vmom{}}{\vspin{2}})
    \biggr] \neanl &\quad
   +\frac{1}{\rel^3} \biggl[
      \frac{1}{8}\biggl(
	105 \eta^2 \scpm{\vnunit}{\vmom{}}^4
	+15 \eta (3\eta - 2) \scpm{\vnunit}{\vmom{}}^2 \vmom{}^2 \neanl &\quad\quad\quad 
	+\frac{3}{2} (10\eta^2 + 13 \eta - 6) (\vmom{}^2)^2
      \biggr)\scpm{\vnunit}{\vspin{1}}\scpm{\vnunit}{\vspin{2}} \neanl &\quad\quad
      +\frac{1}{8}\biggl(
	-3 (8\eta^2 - 37 \eta + 12) \scpm{\vnunit}{\vmom{}}^2 \vmom{}^2 
	+ (7\eta^2 - 23 \eta +9) (\vmom{}^2)^2
      \biggr)\scpm{\vspin{1}}{\vspin{2}} \neanl &\quad\quad
      +\frac{1}{4}\biggl(
	9 \eta^2 \scpm{\vnunit}{\vmom{}}^2 
	+ \frac{1}{2} (4\eta^2 + 25 \eta - 9) \vmom{}^2
      \biggr)\scpm{\vmom{}}{\vspin{1}}\scpm{\vmom{}}{\vspin{2}} \neanl &\quad\quad
      -\frac{3}{8}\biggl(
	+ 15 \eta^2 \scpm{\vnunit}{\vmom{}}^2 
	+ \frac{1}{2} (10\eta^2 + 21 \eta - 9) \vmom{}^2
      \biggr)\scpm{\vnunit}{\vmom{}}\neanl&
	\quad\quad\quad\quad\times(\scpm{\vmom{}}{\vspin{1}}\scpm{\vnunit}{\vspin{2}} + \scpm{\vnunit}{\vspin{1}}\scpm{\vmom{}}{\vspin{2}}) \neanl &\quad\quad
      +\frac{9}{16}\sqrt{1-4\eta}(1 - 2\eta)\scpm{\vnunit}{\vmom{}}(\scpm{\vmom{}}{\vspin{1}}\scpm{\vnunit}{\vspin{2}} - \scpm{\vnunit}{\vspin{1}}\scpm{\vmom{}}{\vspin{2}})
   \biggr]
\biggr\}\,.
\end{align}
\end{subequations}
There $\vct{\Delta} = \vspin{1}-\vspin{2}$ and $\vct{\Sigma} = \vspin{1}+\vspin{2}$ the differences and sums of 
the spin vectors and $\vang$ is the orbital angular momentum $\vang = \rel \vnunit \times \vmom{}$.


\section{Kinematical Consistency: The Approximate Poincar\'{e} Algebra} \label{sec:kinematicalconsistency}
For a space-time which is asymptotically flat the Poincar\'{e} algebra must be fulfilled at spacial infinity, e.g. \cite{Beig:OMurchadha:1987}.
The generators of the Poincar\'{e} algebra can be expressed in terms of the canonical variables describing
the physical system, i.e. matter variables like linear momenta, position variables or spins. Also propagating
field degrees of freedom enter the generators of the Poincar\'{e} algebra. Throughout this section we set
$\dim=3$. The relations between the generators
are given by
\begin{subequations}
\begin{align}
 \{P_i, H\} &= 0\,,\quad\{J_i, H\} = 0\label{eq:PAJH}\,,\\
 \{J_i, P_j\} &= \epsilon_{ijk} P_k\,,\quad\{J_i, J_j\} = \epsilon_{ijk} J_k\,,\\
 \{J_i, G_j\} &= \epsilon_{ijk} G_k\,,\\
 \{G_i, H\} &= P_i\,,\label{eq:PAGHeqP}\\
 \{G_i, P_j\} &= \cInv{2} \delta_{ij} H\,,\label{eq:PAGPeqH}\\
 \{G_i, G_j\} &= -\cInv{2} \epsilon_{ijk} J_k\,,\label{eq:PAGGeqJ}
\end{align}
\end{subequations}
where $\vct{P}$ is the total linear momentum, $J^{ij}$ is the total angular momentum tensor and $J_i = \tfrac{1}{2}\epsilon_{ijk} J^{jk}$ 
the associated dual vector, $\vct{G}$ is the center-of-mass vector and $H$ the Hamiltonian of the physical system.
Total linear momentum $\vct{P}$ and total angular momentum
$J^{ij} = - J^{ji}$ are given by
\begin{align}
 \vct{P} = \sum_a \vmom{a}\,, \quad
 J^{ij} = \sum_a \left[\hat{z}^i_a \mom{a}{j} - \hat{z}^j_a \mom{a}{i} + \spin{a}{i}{j}\right]\,,
\end{align}
see also, e.g., \cite{Damour:Jaranowski:Schafer:2000, Damour:Jaranowski:Schafer:2008:1}.
For the contributions of the propagating field degrees of freedom see, e.g., \cite{Steinhoff:Wang:2009,Steinhoff:2011}.
However, these contributions can be dropped within $\vct{P}$ and $J^{ij}$ here as we are considering the \emph{conservative} 
matter-only Hamiltonian instead of the ADM Hamiltonian (the latter still depends on the canonical field variables).

\subsection{General Considerations for a Center-Of-Mass Vector Ansatz}
As in \cite{Damour:Jaranowski:Schafer:2000, Damour:Jaranowski:Schafer:2008:1,Hartung:Steinhoff:2011:1} we
use an ansatz for the center-of-mass
vectors $\vct{G}$ at next-to-next-to-leading order
(at lower orders it is also possible to directly calculate $\vct{G}$ from certain integrals).
For constructing the center-of-mass vectors one has to consider the irreducible algebraic quantities which can be
generated from $\spin{a}{i}{j}$, $\mom{a}{i}$ and $\nxa{ab}{i}$. Since the Newtonian center-of-mass vector
\begin{align}
\vct{G}_\text{N} &= \sum_a m_a \vx{a}\,,
\end{align}
is at $\cInv{2}$ and the Newtonian Hamiltonian 
\begin{align}
H_\text{N} &= \sum_a \frac{\vmom{a}^2}{2 m_a} - \sum_a \sum_{b\ne a} \frac{\gravthree m_a m_b}{2 r_{ab}}\,,
\end{align}
is at $\cInv{4}$ the higher order corrections to the center-of-mass vector are also one post-Newtonian order below the appropriate Hamiltonian.
Thus the momentum and $\gravthree$ powers there are also reduced. 

Let us demonstrate these considerations at point-mass level: The Newtonian Hamiltonian has only $p^2$ terms at $\gravthree^0$ [which could be
$\vmom{1}^2$, $\vmom{2}^2$, $\scpm{\vnun}{\vmom{1}}^2$, $\scpm{\vnun}{\vmom{2}}^2$, and $\scpm{\vmom{1}}{\vmom{2}}$]\footnote{Although only $\vmom{1}^2$ and $\vmom{2}^2$ are appearing at the Newtonian order. This discussion should only provide some idea of appearing momentum powers at certain post-Newtonian orders.}
and $p^0$ terms at $\gravthree^1$ (which is only one term). At 1PN there appear $p^4$ terms at $\gravthree^0$, $p^2$ terms at $\gravthree^1$ and $p^0$
at $\gravthree^2$. At 3PN there are $p^8$ terms at $\gravthree^0$ and $p^0$ terms at $\gravthree^4$. 
The center-of-mass vectors belonging to the Hamiltonians above have the following momentum powers emerging there:
The Newtonian center-of-mass vector mentioned above has $p^0$ at $\gravthree^0$, the 1PN one contains $p^2$ at $\gravthree^0$ and $p^0$ at $\gravthree^1$ 
and at 3PN level it contains $p^6$ at $\gravthree^0$ and $p^0$ at $\gravthree^3$.

Now we will discuss how to construct the linear-in-spin corrections for the center-of-mass vectors. Symbolically they can be written in the form
\begin{align}
 \vct{G}_{\text{SO}} &= \text{SO-scalar}\cdot\text{PM-vector} + \text{PM-scalar}\cdot\text{SO-vector}\,,\label{eq:GSO}\\
 \vct{G}_{\text{SS}} &= S_1 S_2\text{-scalar}\cdot\text{PM-vector} + \text{SO-scalar}\cdot\text{SO-vector} + \text{PM-scalar}\cdot S_1 S_2\text{-vector}\label{eq:GSS}\,.
\end{align}
Notice that we are formally working in generic dimension, where a
spin-\emph{vector} can not be defined. As the Poincare-Algebra must also hold
in generic dimensions, it must be possible to construct the center-of-mass
vector in terms of the spin-\emph{tensor}.
This is fortunate, as identities such as (6.1) in \cite{Marsat:Bohe:Faye:Blanchet:2012} 
would complicate the situation if one is forced to work with a spin vector in $\dim=3$.

Let us now summarize the vector quantities which can be built at certain spin-levels 
and may be used to construct the center-of-mass vectors.
\begin{table}[htpb]
\begin{tabular}{c|c}
 Vector & Irreducible Quantities  \\
\hline 
 point-mass (PM) & $\vx{a}$, $\vmom{a}$, $\vnxa{ab}$ \\
 spin-orbit (SO) (for $\spin{a}{i}{j}$) & $\nxa{ab}{j}\spin{a}{i}{j}$,$\mom{a}{j}\spin{a}{i}{j}$, $\mom{b}{j}\spin{a}{i}{j}$ \\
 spin($a$)-spin($b$) ($S_a S_b$) & $\nxa{ab}{k}\spin{a}{k}{j}\spin{b}{i}{j}$, $\mom{a}{k}\spin{a}{k}{j}\spin{b}{i}{j}$, $\mom{b}{k}\spin{a}{k}{j}\spin{b}{i}{j}$ \\
\end{tabular}
\caption{Vector quantities at certain spin levels}
\label{tab:vectors}
\end{table}

The mentioned vectors must perhaps be multiplied by scalar quantities, see \eqref{eq:GSO} and \eqref{eq:GSS}.
If the number of momentum variables in the spin-orbit or spin(1)-spin(2) scalars given in the following Table \ref{tab:scalars} is not sufficient for 
the appropriate $\gravthree$-order they have to be filled up by the point-mass scalars, namely the linear momentum powers.
\begin{table}[htpb]
\begin{tabular}{c|c}
 Scalar & Irreducible Quantities \\
 \hline
 PM & linear momentum powers $p^n$ \\
 SO & $(\nxa{ab}{i}\mom{a}{j}\spin{a}{i}{j})$, $(\nxa{ab}{i}\mom{b}{j}\spin{a}{i}{j})$, $(\mom{a}{i}\mom{b}{j}\spin{a}{i}{j})$ \\
 $S_1 S_2$ & $(\mom{1}{i}\mom{2}{j}\spin{1}{i}{j})(\mom{1}{i}\mom{2}{j}\spin{2}{i}{j})$,
	      $(\nun{i}\mom{1}{j}\spin{1}{i}{j})(\nun{i}\mom{1}{j}\spin{2}{i}{j})$,\\
	     & $(\nun{i}\mom{1}{j}\spin{1}{i}{j})(\nun{i}\mom{2}{j}\spin{2}{i}{j})$,
	     $(\mom{1}{i}\mom{2}{j}\spin{1}{i}{j})(\mom{1}{i}\mom{2}{j}\spin{2}{i}{j})$,\\
	      &$(\nun{i}\mom{1}{j}\spin{1}{i}{j})(\nun{i}\mom{1}{j}\spin{2}{i}{j})$,
	      $(\nun{i}\mom{1}{j}\spin{1}{i}{j})(\nun{i}\mom{2}{j}\spin{2}{i}{j})$,\\
	      &$(\nun{i}\mom{2}{j}\spin{1}{i}{j})(\nun{i}\mom{1}{j}\spin{2}{i}{j})$,
	      $(\nun{i}\mom{2}{j}\spin{1}{i}{j})(\nun{i}\mom{2}{j}\spin{2}{i}{j})$,\\
	      &$(\spin{1}{i}{j}\spin{2}{i}{j})$,
		$(\nun{i}\nun{j}\spin{1}{i}{k}\spin{2}{j}{k})$,\\
		&$(\nun{i}\mom{1}{j}\spin{1}{i}{k}\spin{2}{j}{k})$,
		$(\nun{i}\mom{2}{j}\spin{1}{i}{k}\spin{2}{j}{k})$, \\
	      &$(\mom{1}{i}\nun{j}\spin{1}{i}{k}\spin{2}{j}{k})$, 
		$(\mom{2}{i}\nun{j}\spin{1}{i}{k}\spin{2}{j}{k})$,\\
		&$(\mom{1}{i}\mom{2}{j}\spin{1}{i}{k}\spin{2}{j}{k})$,
		$(\mom{2}{i}\mom{1}{j}\spin{1}{i}{k}\spin{2}{j}{k})$, \\
		&$(\mom{1}{i}\mom{2}{j}\spin{1}{i}{j})(\nun{i}\mom{1}{j}\spin{2}{i}{j})$,
		$(\mom{1}{i}\mom{2}{j}\spin{1}{i}{j})(\nun{i}\mom{2}{j}\spin{2}{i}{j})$,\\
		&$(\nun{i}\mom{1}{j}\spin{1}{i}{j})(\mom{1}{i}\mom{2}{j}\spin{2}{i}{j})$, 
	      $(\nun{i}\mom{2}{j}\spin{1}{i}{j})(\mom{1}{i}\mom{2}{j}\spin{2}{i}{j})$ \\
\end{tabular}
\caption{Scalar quantities at certain spin levels}
\label{tab:scalars}
\end{table}

Also important is that every spin is counted like a linear momentum because they have the same $\cInv{1}$-order, see \ref{subsub:ordercounting}. 
This means the formal 3PN spin-orbit and spin(1)-spin(2) Hamiltonians have only contributions up to $\gravthree^3$ ($\gravthree^4$ 
contributions cannot contain any spins since they are momentum-independent for point-masses).
Notice that the Hamiltonians can only be constructed from the irreducible scalar quantities given above. This is demanded by the
Poincar\'e algebra, namely \eqref{eq:PAJH} ($H$ should be invariant under translations and rotations and thus is a scalar).
The $\vx{a}$ contribution in the center-of-mass vector can be fixed by \eqref{eq:PAGPeqH} using the lower order Hamiltonian.
This Hamiltonian can always be written in the form $H = \sum_a h_a$, where the $h_a$ are
translation invariant, $\{h_a, \vct{P}\} = 0$.
(In the post-Newtonian approximation of general relativity
all Hamiltonians have such a structure that the $h_a$ are translational invariant.)
If we make an ansatz for the center-of-mass vector of the form
\begin{align}
\vct{G} &= \sum_a h_a \vx{a} + \vct{Y}\label{eq:Gansatz}\,,
\end{align}
we see that ($\cInv{1} = 1$)
\begin{align}
 \left\{\sum_a h_a \xa{a}{i} + Y^i, \sum_b \mom{b}{j}\right\} &= \sum_a \biggl[\underbrace{\{h_a,P^j\}}_{=0}\xa{a}{i} 
  +  h_a \delta_{ij}\biggr] + \underbrace{\{Y^i,P^j\}}_{\stackrel{!}{=}0} = \sum_a h_a \delta_{ij} = H \delta_{ij} \label{eq:GPeqHdemand}\,.
\end{align}
Equation \eqref{eq:GPeqHdemand} demands that $\{Y^i,P^j\}=0$ and so
$\vct{Y}$ must be translational invariant. We have shown that the part of the
center-of-mass vector which is not translation invariant, i.e., $\sum_a h_a \vx{a}$, can be
read off from the Hamiltonian.

From these consideration it follows that in the spin-orbit case the center-of-mass vector consists of 52 algebraic independent
quantities for one object and in the spin(1)-spin(2) case there are 86 algebraic independent quantities for both objects. 
Notice that up to the formal 3PN level and linear order in spin all center-of-mass vectors can be fixed uniquely by using the Poincar\'e algebra.

\subsection{Next-to-next-to-leading Order linear-in-spin Center-Of-Mass Vectors}
Now we take the ansatz for the center-of-mass vector \eqref{eq:Gansatz} where $\vct{Y}$ has to be constructed
from the irreducible quantities given in Tables \ref{tab:vectors} and \ref{tab:scalars} with the decompositions \eqref{eq:GSO} and \eqref{eq:GSS}, but without
$\vx{a}$ vectors and put them into \eqref{eq:PAGHeqP} with the Hamiltonians \eqref{eq:HNNLOSO} and \eqref{eq:HNNLOS1S2} for the spin-orbit
and spin(1)-spin(2) case. From this all unknown coefficients mentioned above could be fixed uniquely.
The center-of-mass vector contributions given here implements the change in the binding energy of the system
due to the NNLO linear-in-spin interaction Hamiltonians. This results also in context of the energy-mass equivalence
in a modified gravitating mass and thus in a correction to the Newtonian center-of-mass vector 
$\vct{G}_{\text{N}} = \sum_a m_a \vx{a}$, which does not take any interactions into account.
The correction to the center-of-mass vector from NNLO spin-orbit interactions finally results as
\begin{align}
  \vct{G}^{\text{NNLO}}_{\text{SO}} & = 
 \frac{(\vmom{1}^2)^2}{16 m_1^5} (\vmom{1}\times\vspin{1})\nonumber\\
&+ (\vmom{2}\times\vspin{1})
 	\biggl[
 		\frac{\gravthree}{\rel}
 			\biggl(
 				- \frac{3 \scpm{\vnun}{\vmom{1}}\scpm{\vnun}{\vmom{2}}}{8 m_1 m_2}
				- \frac{\scpm{\vmom{1}}{\vmom{2}}}{8 m_1 m_2}
			\biggr)
		+\frac{\gravthree^2}{\rel^2} 
			\biggl(
				-\frac{47 m_1}{16} 
				-\frac{21 m_2}{8}
			\biggr)
	\biggr]\nonumber\\
&+ (\vmom{1}\times\vspin{1})
	\biggl[
		\frac{\gravthree}{\rel}
			\biggl(
				\frac{9 m_2 \vmom{1}^2}{16 m_1^3}
				- \frac{5 \vmom{2}^2}{8 m_1 m_2}
			\biggr)
		+\frac{\gravthree^2}{\rel^2} 
			\biggl(
				\frac{57 m_2}{16} 
				+\frac{15 m_2^2}{8 m_1}
			\biggr)
	\biggr]\nonumber\\
&+ (\vnun\times\vspin{1})
	\biggl[
		\frac{\gravthree}{\rel}
			\biggl(
				\frac{9\scpm{\vnun}{\vmom{1}}\scpm{\vnun}{\vmom{2}}^2}{16 m_1 m_2}
				+ \frac{\scpm{\vnun}{\vmom{2}}\scpm{\vmom{1}}{\vmom{2}}}{8 m_1 m_2}
				+ \frac{\scpm{\vnun}{\vmom{1}}\vmom{2}^2}{16 m_1 m_2}
			\biggr)\nonumber\\
&\quad		+\frac{\gravthree^2}{\rel^2}
			\biggl(
				-\frac{5 m_2}{8} \scpm{\vnun}{\vmom{1}}
				+\left\{\frac{13 m_1}{8} + \frac{11 m_2}{4}\right\} \scpm{\vnun}{\vmom{2}}
			\biggr)
	\biggr]\nonumber\\
&- \frac{\gravthree}{\rel} \vmom{1} \frac{\scpm{\vnun}{\vmom{2}}((\vnun \times \vmom{2})\vspin{1})}{2 m_1 m_2}\nonumber\\
&+ \frac{\gravthree}{\rel} \vmom{2}
	\biggl(
		-\frac{\scpm{\vnun}{\vmom{2}}((\vnun \times \vmom{1})\vspin{1})}{8 m_1 m_2}
		+\frac{\scpm{\vnun}{\vmom{1}}((\vnun \times \vmom{2})\vspin{1})}{2 m_1 m_2}\nonumber\\
&\quad		-\frac{((\vmom{1} \times \vmom{2})\vspin{1})}{8 m_1 m_2}
	\biggr)\nonumber\\
&+ \vnun
	\biggl[
		\frac{\gravthree}{\rel}
			\biggl(
				\biggl\{
					\frac{m_2 \vmom{1}^2}{16 m_1^3}
					+ \frac{15 \scpm{\vnun}{\vmom{2}}^2}{16 m_1 m_2}
					- \frac{3 \vmom{2}^2}{16 m_1 m_2}
				\biggr\}((\vnun \times \vmom{1})\vspin{1})\nonumber\\
&\quad\quad				+\biggl\{
					-\frac{3 \scpm{\vnun}{\vmom{1}}\scpm{\vnun}{\vmom{2}}}{2 m_1 m_2}
					-\frac{\scpm{\vmom{1}}{\vmom{2}}}{2 m_1 m_2}
				\biggr\}((\vnun \times \vmom{2})\vspin{1})\nonumber\\
&\quad\quad			+\frac{13 \scpm{\vnun}{\vmom{2}}}{8 m_1 m_2}((\vmom{1} \times \vmom{2})\vspin{1})
			\biggr)\nonumber\\
&\quad		+\frac{\gravthree^2}{\rel^2}
			\biggl(
				\left\{\frac{m_2}{2} + \frac{5 m_2^2}{4 m_1}\right\} ((\vnun \times \vmom{1})\vspin{1})
				+\left\{-2 m_1 - 5 m_2\right\} ((\vnun \times \vmom{2})\vspin{1})
			\biggr)
	\biggr]\nonumber\\
&+ \frac{\hat{\vct{z}}_1}{\rel}
	\biggl[
		\frac{\gravthree}{\rel}
			\biggl(
				\biggl\{
					\frac{3 \scpm{\vnun}{\vmom{1}} \scpm{\vnun}{\vmom{2}}}{m_1 m_2}
					+\frac{\scpm{\vmom{1}}{\vmom{2}}}{m_1 m_2}
				\biggr\} ((\vnun \times \vmom{2})\vspin{1})\nonumber\\
&\quad\quad			+\biggl\{
					\frac{3 \scpm{\vnun}{\vmom{1}}}{4 m_1^2}
					-\frac{2 \scpm{\vnun}{\vmom{2}}}{m_1 m_2}
				\biggr\}((\vmom{1} \times \vmom{2})\vspin{1})\nonumber\\
&\quad\quad			+\biggl\{
					- \frac{5 m_2 \vmom{1}^2}{8 m_1^3}
					- \frac{3 \scpm{\vnun}{\vmom{1}} \scpm{\vnun}{\vmom{2}}}{4 m_1^2}
					- \frac{3 \scpm{\vnun}{\vmom{2}}^2}{2 m_1 m_2}\nonumber\\
&\quad\quad\quad			- \frac{3 \scpm{\vmom{1}}{\vmom{2}}}{4 m_1^2}
					+ \frac{3 \vmom{2}^2}{4 m_1 m_2}
				\biggr\} ((\vnun \times \vmom{1})\vspin{1})
			\biggr)\nonumber\\
&		+\frac{\gravthree^2}{\rel^2}
			\biggl(
				\left\{
					-\frac{11 m_2}{2}
					-\frac{5 m_2^2}{m_1}
				\right\} ((\vnun \times \vmom{1})\vspin{1})
				+\left\{
					6 m_1
					+ \frac{15 m_2}{2}
				\right\} ((\vnun \times \vmom{2})\vspin{2})
			\biggr)
	\biggr]\nonumber\\
& + (1\leftrightarrow2)\,,\label{eq:GNNLOSO}
\end{align}
and the NNLO spin(1)-spin(2) part reads
\begin{align}
  \vct{G}^{\text{NNLO}}_{\text{SS}} & = 
	\frac{\gravthree^2}{\rel^3} ((\vnun \times \vspin{2})\times \vspin{1}) \biggl(\frac{17}{8} m_1 + m_2\biggr) \nln
	+ \frac{\gravthree}{\rel^2}\biggl[ \vmom{1} \biggl(
		-\frac{\scpm{\vnun}{\vspin{1}}\scpm{\vmom{2}}{\vspin{2}}}{4 m_1 m_2}
		+\frac{3 \scpm{\vnun}{\vspin{1}} \scpm{\vnun}{\vspin{2}} \scpm{\vnun}{\vmom{2}}}{4 m_1 m_2}
	\biggr) \nlqn
	+ 
	(\vnun \times \vspin{1}) \biggl(
		-\frac{((\vmom{1}\times\vmom{2})\,\vspin{2})}{4 m_1 m_2}
		+\frac{3((\vnun\times\vmom{1})\,\vspin{2})\scpm{\vnun}{\vmom{2}}}{4 m_1 m_2}
	\biggr) \nlqn
	-
	(\vmom{1} \times \vspin{1}) \frac{((\vnun \times \vmom{2})\,\vspin{2})}{8 m_1 m_2}
	-
	(\vmom{2} \times \vspin{1}) \frac{((\vnun \times \vmom{1})\,\vspin{2})}{4 m_1 m_2} \nlqn
	-
	((\vmom{1} \times \vspin{2})\times \vspin{1}) \frac{\scpm{\vnun}{\vmom{2}}}{4 m_1 m_2} 
	-
	((\vmom{2} \times \vspin{2})\times \vspin{1}) \frac{\scpm{\vnun}{\vmom{1}}}{4 m_1 m_2} \biggr]\nln
	+\frac{\vx{1}}{\rel} \biggl(
		\frac{2 \gravthree^2 (2 m_1 + m_2)}{\rel^3}\biggl[
			\scpm{\vspin{1}}{\vspin{2}}
			-2\scpm{\vnun}{\vspin{1}}\scpm{\vnun}{\vspin{2}}
		\biggr] \nlqn
		+\frac{\gravthree}{\rel^2}\biggl[
			-\frac{3 ((\vnun\times\vmom{1})\,\vspin{1})((\vnun\times\vmom{1})\,\vspin{2})}{2 m_1^2} \nlqn
			+\frac{3 ((\vnun\times\vmom{2})\,\vspin{1})((\vnun\times\vmom{1})\,\vspin{2})}{2 m_1 m_2}
			+\frac{3 ((\vnun\times\vmom{1})\,\vspin{1})((\vnun\times\vmom{2})\,\vspin{2})}{8 m_1 m_2} \nlqn
			-\frac{\scpm{\vmom{2}}{\vspin{1}}\scpm{\vmom{1}}{\vspin{2}}}{8 m_1 m_2}
			+\frac{\scpm{\vmom{1}}{\vspin{1}}\scpm{\vmom{2}}{\vspin{2}}}{4 m_1 m_2}
			+\frac{3 \scpm{\vmom{2}}{\vspin{1}}\scpm{\vnun}{\vspin{2}}\scpm{\vnun}{\vmom{1}}}{2 m_1 m_2} \nlqn
			-\frac{3 \scpm{\vnun}{\vspin{1}}\scpm{\vmom{1}}{\vspin{2}}\scpm{\vnun}{\vmom{1}}}{2 m_1^2}
			+\frac{3 \scpm{\vnun}{\vspin{1}}\scpm{\vmom{2}}{\vspin{2}}\scpm{\vnun}{\vmom{1}}}{4 m_1 m_2} \nlqn
			+\frac{3 \scpm{\vmom{1}}{\vspin{1}}\scpm{\vnun}{\vspin{2}}\scpm{\vnun}{\vmom{2}}}{4 m_1 m_2} \nlqn
			-\scpm{\vnun}{\vspin{1}}\scpm{\vnun}{\vspin{2}}\biggl\{
				\frac{15 \scpm{\vnun}{\vmom{1}} \scpm{\vnun}{\vmom{2}}}{4 m_1 m_2}
				+\frac{3 \scpm{\vmom{1}}{\vmom{2}}}{4 m_1 m_2}
			\biggr\} \nlqn
			+\scpm{\vspin{1}}{\vspin{2}}\biggl\{
				\frac{3 \scpm{\vnun}{\vmom{1}}^2}{2 m_1^2}
				-\frac{3 \scpm{\vnun}{\vmom{1}}\scpm{\vnun}{\vmom{2}}}{4 m_1 m_2}
				+\frac{\scpm{\vmom{1}}{\vmom{2}}}{8 m_1 m_2}
			\biggr\}
		\biggr]
	\biggr) + (1\leftrightarrow2)\,.\label{eq:GNNLOSS}
\end{align}
From this the boost vector $\vct{K} = \vct{G} - t \vct{P}$ can be obtained, which explicitly depends on time $t$.

Notice that in \eqref{eq:GNNLOSO} (in contrast to \eqref{eq:GNNLOSS}) there appears a one-particle term without $\gravthree$ factor.
It comes from the displacement of the center-of-mass due to the rotation and the resulting special relativistic
Lorentz contractions of different parts of the object (which has to have a finite size). 
Further discussions about this issue can be found in \cite{Steinhoff:2011} which are clarified graphically in particular in Fig. 1 therein.
In \eqref{eq:GNNLOSS} there is no term without $\gravthree$ factor because all interactions between two spins are
transmitted by the gravitational field in general relativity.

\section{Test-Spin near Kerr Black Hole} \label{sec:testspin}
In the last section we checked whether our results are compatible with kinematical restrictions of the
Poincar\'e algebra. Here we derive an exact 
test-spin Hamiltonian to compare our results \eqref{eq:HNNLOSO} and \eqref{eq:HNNLOS1S2} with.
In the following subsections we restrict ourselves to $\dim=3$, because in the
test-spin case there are only delta integrals to be evaluated.
A partial check of our Hamiltonians against the test-spin case is contained in \cite{Steinhoff:Puetzfeld:2012} for the case of aligned spins.

There are various approaches to calculate the motion of a test-spin near a Kerr black hole (see, e.g., \cite{Barausse:Racine:Buonanno:2009} 
and references therein). Since in the counting used in \cite{Barausse:Racine:Buonanno:2009} 
the NNLO spin(1)-spin(2) interaction is at 4PN and therefore not considered therein, 
one needs to calculate the spin(1)-spin(2) contributions from 
\begin{align}
 H_{\text{Testspin}} &= -\hat{A}^{k\ell} \triad{m}{k} e^{(m)}_{\quad \ell,0} + \int \text{d}^3 x\,[\src{}\lapse - \src{i} \shiftup{i}]
 \label{eq:testspinhamiltonian}\,,
\end{align}
where $\lapse$, $\shift{i}$, the framefield $\triad{m}{k}$, and the implicit appearing metric 
provide the exterior gravitational field, and $\src{}$ and $\src{i}$ represent
the test-spin moving in it \cite{Steinhoff:2011}. Note that the framefield in the first term has to be evaluated at the
position of the test-spin.

Since all spin dependencies of the metric are at least quadratic in the Kerr spin, and the only contribution which is
linear in Kerr spin comes from the shift vector, the three-dimensional part of the metric and the lapse are
identical with the appropriate components of the isotropic Schwarzschild metric which also comes from
the spinless limit of the Kerr metric. The shift is given by the expressions in \cite[Eq. (54)]{Hergt:Schafer:2008:2}.

The metric components generated by particle `$1$' are given by
\begin{align}
 g_{00} &= -\left(\frac{1-\frac{m_1}{2 r_1}}{1+\frac{m_1}{2 r_1}}\right)^2\,,\\
 g_{ij} &= \left(1+\frac{m_1}{2 r_1}\right)^4 \delta_{ij}\,,\\
 g_{0i} &= \frac{2 m_1\, \nxa{1}{k}\kerrspin{1}{i}{k}}{r_1^2 \left(1 + \frac{m_1}{2 r_1}\right)^2}\,.
\end{align}
It is well-known that the metric components can be rewritten into a three-dimensional metric on the spatial hypersurface, lapse $\lapse$, 
and shift $\shiftup{i} = \gamma^{ij} \shift{j}$, using \eqref{eq:metricdecomp1} and \eqref{eq:metricdecomp2}.
So one can immediately see that
\begin{align}
 \gamma_{ij} = g_{ij} &= \left(1+\frac{m_1}{2 r_1}\right)^4 \delta_{ij}\label{eq:gammaiso}\,,\\
 \gamma^{ij} &= \left(1+\frac{m_1}{2 r_1}\right)^{-4} \delta_{ij}\label{eq:gammainviso}\,,\\
 \lapse &= \frac{1-\frac{m_1}{2 r_1}}{1+\frac{m_1}{2 r_1}}\,,
\end{align}
with the squares of the shift neglected, because they are quadratic in the Kerr spin, 
\begin{align}
 \shift{i} &= \frac{2 m_1\, \nxa{1}{k}\kerrspin{1}{i}{k}}{r_1^2 \left(1 + \frac{m_1}{2 r_1}\right)^2}
	    = -2 m_1 \kerrspin{1}{i}{k} \left(\frac{1}{r_1 \left(1 + \frac{m_1}{2 r_1}\right)}\right)_{,k}\,,
\end{align}
see \cite{Hergt:Schafer:2008:2}. Here $\kerrspin{1}{i}{j} = \spin{1}{i}{j}/m_1$ is the Kerr spin belonging to the black hole located
at position `1'. Note that to linear order in spin 
\begin{align}
  \pi^{ij} = \frac{6 m_1 \nxa{1}{k} \nxa{1}{(i} \kerrspin{1}{j)}{k}}{r_1^3 \left(1 + \frac{m_1}{2 r_1}\right)^4}\,,
\end{align}
(calculated from the inverse metric and the three-dimensional Christoffel symbols, see \cite[Eq. (65)]{Hergt:Schafer:2008:2}) 
fulfills the ADM gauge condition, so no further coordinate shift from quasi isotropic coordinates to another coordinate system 
is necessary. Due to the symmetric 
framefield gauge $\triad{i}{j} = \sqrt{\gamma_{ij}}$, the framefield is 
given by \eqref{eq:dreibein}
\begin{align}
 \triad{i}{j} &= \left(1+\frac{m_1}{2 r_1}\right)^2 \delta_{ij}\label{eq:dreibeiniso}\,.
\end{align}

For the test-spin in a Kerr field it is sufficient to calculate \eqref{eq:testspinhamiltonian}
where the sources are given by \eqref{eq:sourcehamilton} and \eqref{eq:sourcemomentum}.
Since the metric and the framefield are proportional to $\delta_{ij}$ many terms vanish in the source. The only terms remaining
are
\begin{align}
 \src{} &= \sum_a \biggl[-\nmom{a} \dl{a} 
	-\frac{1}{2}\frac{\hat{S}_{a\,li}\mom{a}{j}}{\nmom{a}} \gamma^{kl}\gamma^{ij}_{\quad,k}\dl{a}
	-\left(
		\frac{\mom{a}{l}}{m_a - \nmom{a}}\gamma^{ij}\gamma^{kl}\hat{S}_{a\,jk}\dl{a}
	\right)_{,i}
\biggr]\,,\\
 \src{i} &= \sum_a \biggl[\mom{a}{i}\dl{a}
  + \frac{1}{2}\biggl(
    \gamma^{jk} \hat{S}_{a\,ik} \dl{a}
    +\gamma^{jk}\gamma^{\ell p} \frac{2 \mom{a}{\ell}\mom{a}{(i}\hat{S}_{a\,k)p}}{\nmom{a}(m_a - \nmom{a})} \dl{a}
  \biggr)_{,j}
\biggl]\,.
\end{align}

Inserting sources, metric components and framefield, and
evaluating them at the testspin location
gives the exact result
\begin{align}
 H^{\text{Kerr}}_{\text{Testspin}} & \stackrel{\Order{a^1}}{=} 
 \frac{1-\frac{m_1}{2 \rel}}{1 + \frac{m_1}{2 \rel}} \sqrt{m_2^2 + \frac{\vmom{2}^2}{\left(1+\frac{m_1}{2 \rel}\right)^4}} \neanl &
 - \frac{m_1 
		\trpm{\vnxa{12}}{\vmom{2}}{\vspin{2}}
	      }{\rel^2 \left(1+\frac{m_1}{2 \rel}\right)^{6}} \left[
    \frac{1-\frac{m_1}{2 \rel}}{\sqrt{m_2^2 + \frac{\vmom{2}^2}{\left(1 + \frac{m_1}{2 \rel}\right)^4}}}
    +\frac{1}{m_2 + \sqrt{m_2^2 + \frac{\vmom{2}^2}{\left(1 + \frac{m_1}{2 \rel}\right)^4}}}
 \right] \neanl &
 + \frac{2 m_1 
		  \trpm{\vnxa{12}}{\vmom{2}}{\vkerrspin{1}}
		}{\rel^2 \left(1 + \frac{m_1}{2 \rel}\right)^6}
 + \frac{m_1}{\rel^3 \left(1 + \frac{m_1}{2 \rel}\right)^7}\Biggl[
     -\left(1 - \frac{5 m_1}{2 \rel}\right)\scpm{\vkerrspin{1}}{\vspin{2}} \neanl &
     -3 \left(1 - \frac{m_1}{2 \rel}\right)\biggl\{
	-\scpm{\vnxa{12}}{\vkerrspin{1}}\scpm{\vnxa{12}}{\vspin{2}} \neanl &
     + \frac{
	\trpm{\vnxa{12}}{\vmom{2}}{\vkerrspin{1}}\trpm{\vnxa{12}}{\vmom{2}}{\vspin{2}}
     - \scpm{\vnxa{12}}{\vmom{2}}^2 \scpm{\vkerrspin{1}}{\vspin{2}}
     + \scpm{\vnxa{12}}{\vmom{2}} \scpm{\vmom{2}}{\vkerrspin{1}}\scpm{\vnxa{12}}{\vspin{2}}}
    {\left(m_2 + \sqrt{m_2^2 + \frac{\vmom{2}^2}{\left(1 + \frac{m_1}{2 \rel}\right)^4}}\right)
    \sqrt{m_2^2 + \frac{\vmom{2}^2}{\left(1 + \frac{m_1}{2 \rel}\right)^4}}\left(1 + \frac{m_1}{2 \rel}\right)^4}\biggr\}
 \Biggr]\,,
\end{align}
which leads after a post-Newtonian expansion (the post-Newtonian order given in the subscript is a formal one)
\begin{align}
 H^{\text{Kerr}}_{\text{Testspin},\le 3\text{PN}} &\stackrel{\Order{a^1}}{\approx} m_2 c^2 
  + \left(\frac{\vmom{2}^2}{2 m_2} - \frac{m_1 m_2}{\rel}\right)
  + \cInv{2} \biggl(
    -\frac{(\vmom{2}^2)^2}{8 m_2^3} 
    - \frac{3 m_1 \vmom{2}^2}{2 m_2 \rel} \neanl &
    +\frac{m_1}{\rel^2} \biggl[
      \frac{m_1 m_2}{2}
	-2 \trpm{\vnxa{12}}{\vmom{2}}{\vkerrspin{1}}
      -\frac{3 \trpm{\vnxa{12}}{\vmom{2}}{\vspin{2}}}{2 m_2}
    \biggr] \neanl &
    + \frac{m_1}{\rel^3} 
	\left[-\scpm{\vkerrspin{1}}{\vspin{2}}+3 \scpm{\vnxa{12}}{\vkerrspin{1}}\scpm{\vnxa{12}}{\vspin{2}}\right]
  \biggr) \neanl & 
  + \cInv{4}\biggl(\frac{(\vmom{2}^2)^3}{16 m_2^5}
      +\frac{5 m_1 (\vmom{2}^2)^2}{8 m_2^3 \rel}
      +\frac{m_1 \vmom{2}^2}{m_2 \rel^2} \biggl[
	\frac{5 m_1 }{2}
	+\frac{5 \trpm{\vnxa{12}}{\vmom{2}}{\vspin{2}}}{8 m_2^2}
      \biggr] \neanl &
      + \frac{m_1}{\rel^3} \biggl[
	- \frac{m_1^2 m_2}{4}
	+ m_1 \biggl(6 \trpm{\vnxa{12}}{\vmom{2}}{\vkerrspin{1}} + \frac{5 \trpm{\vnxa{12}}{\vmom{2}}{\vspin{2}}}{m_2}\biggr) \neanl &
	- \frac{3 \trpm{\vnxa{12}}{\vmom{2}}{\vkerrspin{1}}\trpm{\vnxa{12}}{\vmom{2}}{\vspin{2}}}{2 m_2^2} 
	+ \frac{3 \scpm{\vnxa{12}}{\vmom{2}}^2 \scpm{\vkerrspin{1}}{\vspin{2}}}{2 m_2^2} \neanl &
	- \frac{3 \scpm{\vnxa{12}}{\vmom{2}}\scpm{\vmom{2}}{\vkerrspin{1}}\scpm{\vnxa{12}}{\vspin{2}}}{2 m_2^2} 
      \biggr] 
	+ \frac{6 m_1^2}{\rel^4} \left[\scpm{\vkerrspin{1}}{\vspin{2}}-2 \scpm{\vnxa{12}}{\vkerrspin{1}}\scpm{\vnxa{12}}{\vspin{2}}\right]
    \biggr) \neanl &
  + \cInv{6}\biggl(-\frac{5(\vmom{2}^2)^4}{128 m_2^7}
      -\frac{7 m_1 (\vmom{2}^2)^3}{16 m_2^5 \rel}
      -\frac{m_1 (\vmom{2}^2)^2}{m_2^3 \rel^2} \biggl[
	\frac{27 m_1}{16} 
	+ \frac{7 \trpm{\vnxa{12}}{\vmom{2}}{\vspin{2}}}{16 m_2^2}
      \biggr] \neanl &
      +\frac{m_1 \vmom{2}^2}{m_2 \rel^3} \biggl[
	-\frac{25 m_1^2}{8}
	-\frac{27 m_1 \trpm{\vnxa{12}}{\vmom{2}}{\vspin{2}}}{8 m_2^2}
	+\frac{9 \trpm{\vnxa{12}}{\vmom{2}}{\vkerrspin{1}}\trpm{\vnxa{12}}{\vmom{2}}{\vspin{2}}}{8 m_2^3} \neanl &
	+\frac{9 \scpm{\vnxa{12}}{\vmom{2}}\scpm{\vmom{2}}{\vkerrspin{1}}\scpm{\vnxa{12}}{\vspin{2}}}{8 m_2^3}
	-\frac{9 \scpm{\vnxa{12}}{\vmom{2}}^2 \scpm{\vkerrspin{1}}{\vspin{2}}}{8 m_2^3}
      \biggr]
      + \frac{m_1^2}{\rel^4} \biggl[
	\frac{m_1^2 m_2}{8} \neanl &
	+m_1 \biggl(
	  -\frac{21 \trpm{\vnxa{12}}{\vmom{2}}{\vkerrspin{1}}}{2}
	  -\frac{75 \trpm{\vnxa{12}}{\vmom{2}}{\vspin{2}}}{8 m_2}
	\biggr) \neanl &
	+\frac{9 \trpm{\vnxa{12}}{\vmom{2}}{\vkerrspin{1}}\trpm{\vnxa{12}}{\vmom{2}}{\vspin{2}}}{m_2^2} 
	+\frac{9 \scpm{\vnxa{12}}{\vmom{2}}\scpm{\vmom{2}}{\vkerrspin{1}}\scpm{\vnxa{12}}{\vspin{2}}}{m_2^2} \neanl &
	-\frac{9 \scpm{\vnxa{12}}{\vmom{2}}^2 \scpm{\vkerrspin{1}}{\vspin{2}}}{m_2^2}
      \biggr] 
      + \frac{m_1^3}{\rel^5} \biggl[
	-\frac{63}{4}\scpm{\vkerrspin{1}}{\vspin{2}}
	+\frac{105}{4} \scpm{\vnxa{12}}{\vkerrspin{1}}\scpm{\vnxa{12}}{\vspin{2}}      
      \biggr]
    \biggr)\,, 
\end{align}
in full agreement with the test-spin limit of the full point-mass Hamiltonian, the spin-orbit Hamiltonian
and the spin(1)-spin(2) Hamiltonian up to and including the formal 3PN order. For later checks we provide
also the expressions at formal 4PN order. The test-spin limit for the next-to-next-to-next-to-leading order 
(NNNLO) spin-orbit interaction is given by
\begin{align}
 H^{\text{Kerr}}_{\text{Testspin},4\text{PN}, \text{SO}} &= \biggl(
  \frac{45}{128} \frac{m_1 (\vmom{2}^2)^3}{ m_2^7 \rel^2}
  + \frac{13}{4}\frac{m_1^2 (\vmom{2}^2)^2}{m_2^5 \rel^3}
  + \frac{315}{32} \frac{m_1^3 \vmom{2}^2}{m_2^3 \rel^4}
  + \frac{105}{8} \frac{m_1^4}{m_2 \rel^5}\biggr) \trpm{\vnxa{12}}{\vmom{2}}{\vspin{2}} \neanl &
  + 14\frac{m_1^4}{\rel^5} \trpm{\vnxa{12}}{\vmom{2}}{\vkerrspin{1}}\,,
\end{align}
and for the NNNLO spin(1)-spin(2) interaction by
\begin{align}
 H^{\text{Kerr}}_{\text{Testspin},4\text{PN}, \text{SS}} &= \biggl(
  -\frac{15}{16}\frac{m_1 (\vmom{2}^2)^2}{m_2^6 \rel^3}
  -9 \frac{m_1^2 \vmom{2}^2}{m_2^4 \rel^4}
  -\frac{231}{8} \frac{m_1^3}{m_2^2 \rel^5}
  \biggr)[
    \trpm{\vnxa{12}}{\vmom{2}}{\vkerrspin{1}}\trpm{\vnxa{12}}{\vmom{2}}{\vspin{2}} \neanl &
    +\scpm{\vnxa{12}}{\vmom{2}}\scpm{\vmom{2}}{\vkerrspin{1}}\scpm{\vnxa{12}}{\vspin{2}} 
    -\scpm{\vnxa{12}}{\vmom{2}}^2 \scpm{\vkerrspin{1}}{\vspin{2}}
   ] \neanl &
  + \frac{14 m_1^4}{\rel^6} \biggl(
	2 \scpm{\vkerrspin{1}}{\vspin{2}}
	-3 \scpm{\vnxa{12}}{\vkerrspin{1}}\scpm{\vnxa{12}}{\vspin{2}}      
      \biggr)\,. 
\end{align}
Notice that the formal 3PN spin(1)-spin(2) test-spin contributions were not given in
\cite{Barausse:Racine:Buonanno:2009} (in their counting rules they would be at 4PN level). 
Further notice that there are no contributions coming from the first term in \eqref{eq:testspinhamiltonian} in the spin-orbit, and
the spin(1)-spin(2) case, since in isotropic Schwarzschild coordinates it vanishes identically, see 
\eqref{eq:Ahat}, \eqref{eq:nS}, \eqref{eq:gammaiso}, \eqref{eq:gammainviso}, and \eqref{eq:dreibeiniso}.

There are two possible further checks which use different approaches.
As mentioned in \cite{Hergt:Steinhoff:Schafer:2011}
a comparison of the effective field theory NNLO spin(1)-spin(2) potential in \cite{Levi:2011} to our
NNLO spin(1)-spin(2) Hamiltonian would be a very strong check of both results since the EFT results are
completely independent from the ADM formalism. 
Also a confirming check would be the derivation of both NNLO Hamiltonians 
using the spin-precession method shown in \cite{Damour:Jaranowski:Schafer:2008:1}.
Due to the complicated structure of these comparisons they will be postponed to later publications.
A very recent check of the NNLO spin-orbit Hamiltonian and the resulting equations of motion was
performed in \cite{Marsat:Bohe:Faye:Blanchet:2012, Bohe:Marsat:Faye:Blanchet:2012} in harmonic gauge.
Furthermore in \cite{Bohe:Marsat:Faye:Blanchet:2012} the near-zone metric was determined which is
an important step towards the mentioned template calculations.

\section{Conclusions and Outlook}\label{sec:conclusions}
We have derived the next-to-next-to-leading order spin-orbit and spin(1)-spin(2) Hamiltonians
for binary systems. The spin-orbit Hamiltonian completes the knowledge of binary black hole dynamics up to
and including 3.5PN order if the objects are rapidly rotating. For neutron stars also the leading order
cubic-in-spin Hamiltonians are needed as the results in \cite{Hergt:Schafer:2008:2, Hergt:Schafer:2008} are valid for
black holes only and tidal deformation effects become very important
\cite{Damour:Nagar:2009, Vines:Flanagan:2010, Bini:Damour:Faye:2012}.
The Hamiltonians were checked using two methods.
The fulfillment of the global approximate Poincar\'e algebra was a major criterion for the correctness of the
derived Hamiltonians in the extended ADM formalism. During this check the center-of-mass vectors could be determined
uniquely from an ansatz. Since the approximate Poincar\'e algebra is not sensitive to the static part of the
spin(1)-spin(2) Hamiltonian and fixed only the difference of the two coefficients at the highest order in $\gravthree$ of the spin-orbit Hamiltonian
we performed further checks. The most simple test is a linear-in-spin approximation of the Hamiltonian of a test-spin 
moving near a stationary Kerr black hole. We rederived the test-spin Hamiltonian from \cite{Barausse:Racine:Buonanno:2009} in a different 
manner in \Sec{sec:testspin} (avoiding the use of Dirac brackets).
A comparison was straightforward as the same gauge was used.
A more elaborate test is the recalculation of both NNLO Hamiltonians
via the spin-precession frequency method in \cite{Damour:Jaranowski:Schafer:2008:1} 
and will be part of a further publication. Also a comparison of the NNLO spin(1)-spin(2)
Hamiltonian with the NNLO spin(1)-spin(2) potential given in \cite{Levi:2011} will be part of a further publication and would be a
very strong check, because the derivation of this potential is completely independent from the ADM formalism.
The most important confirmation to date is the independent derivation of the NNLO spin-orbit equations of motion in harmonic gauge
\cite{Marsat:Bohe:Faye:Blanchet:2012, Bohe:Marsat:Faye:Blanchet:2012}.

The results given in this article complete the knowledge of the post-Newtonian approximate dynamics for binary black holes up
to 3.5PN. For general compact objects like neutron stars the leading order cubic-in-spin Hamiltonians are still unknown.
The NNLO spin(1)-spin(2) Hamiltonian is at 4PN, if both objects are rapidly rotating, but there are still some tasks left
to get the full post-Newtonian approximate dynamics up to and including 4PN. For general compact objects the leading order
quartic-in-spin Hamiltonians are also unknown, they are only known for black holes.\footnote{In \cite{Steinhoff:Puetzfeld:2012} the
authors argued that the spin(1)$^4$ Hamiltonians derived in \cite{Hergt:Schafer:2008} are incomplete.} Furthermore the NNLO spin(1)$^2$ Hamiltonian
-- which is also at 4PN if the object is maximally rotating and maybe stronger than NNLO spin(1)-spin(2) -- is completely unknown.
Last but not least the $\gravthree^3$ up to the $\gravthree^5$ corrections to the 4PN point-mass Hamiltonian are also still unknown 
\cite{Jaranowski:Schafer:2012}.

To get reasonable results for the templates the far-zone radiation field also must be calculated at higher order in the post-Newtonian
approximation and also at higher orders in spin. The energy and angular momentum loss are also not known at a post-Newtonian order
corresponding to next-to-next-to-leading order linear-in-spin. If radiation and fluxes are known at such high orders also a 
parameterization is necessary. These three major ingredients are needed for the analytical description of gravitational wave templates which are very 
sensitive to higher order post-Newtonian and spin corrections. Analytical results are still important because
for spinning binaries the parameter space (masses and spin-directions of the components) is very large and
numerical simulations are so time consuming that they cannot be used to cover the whole parameter space.%
\footnote{In \cite{Hinder:2010} they estimate that the simulation of non-spinning binaries for eight orbits
for mass ratio 1:1 consumes ca. $200\,000$ and for 1:10 up to two million CPU hours.}

\ifnotadp
\ifnotprd
\paragraph*{Acknowledgments}
\fi
\ifprd
\acknowledgments
\fi
\ack
\setlength{\bibsep}{0pt}
\bibliographystyle{utphys}
\input{Technical3PNPaper_refs}
\fi

\ifadp
\begin{acknowledgement}
\ack
\end{acknowledgement}
\input{refs_adp}

\fi

\newpage

\appendix

\section{Integration Techniques}\label{sec:integrationtechniques}
In this appendix we recapitulate the techniques needed to solve the integrals that appear during our
calculation. Most integration techniques are needed for $\dim=3$ only, but whenever possible we provide 
results for generic $\dim$ here as they may be used for checks. 
The short-range part of dimensional regularization using an UV-analysis is explained in
\Sec{subsec:UVana}.

\subsection{Inverse Laplacians}\label{sec:invlaptechnique}
Inverse Laplacians are necessary to obtain solutions for the fields in the constraints, $\phi$ and $\pitildei{i}$.
As one can see in \eqref{eq:hamiltonconstraint}, \eqref{eq:momentumconstraint}, and \eqref{eq:pisolve}, the constraint equations always reduce to Poisson-like equations.
On the right-hand-side of the equations there is always some source expression appearing. If the source only consists of Dirac deltas or
their derivatives the inverse Laplacians can be calculated very easily,
\begin{align}
 \Delta^{-1} \dl{a} &= -\frac{\Gamma\left(\frac{\dim}{2}-1\right)}{4\pi^{\dim/2}} \frac{1}{r_a^{\dim-2}}\,.\label{eq:invlapdelta}
\end{align}
Also for some arbitrary power of $r_a$ one can calculate the inverse Laplacian immediately, namely
\begin{align}
 \Delta^{-1} r_a^{\alpha} &= \frac{r_a^{\alpha+2}}{(\alpha+\dim)(\alpha+2)}\label{eq:invra}\,.
\end{align}
The generalization to the application of multiple inverse Laplacians is straightforward. More interesting is the calculation of
inverse Laplacians for powers of $r_a$ and $r_b$ which is a special case of the three-particle integral originally derived in
\cite[Eq. (7)]{Boos:Davydychev:1987} and also later given in \cite[Eq. above (B14)]{Chu:2009}.
The regular solution of $\Delta^{-1} \tfrac{1}{r_a r_b}$ in $\dim=3$ is well-known and given by
\begin{align}
 \Delta^{-1} \frac{1}{r_a r_b} &= \ln s_{ab} = \ln(r_a + r_b +r_{ab})\label{eq:lnS}\,.
\end{align}
It was first found in \cite{Fock:1960}, also used in \cite{Jaranowski:Schafer:1997} 
and finally rederived in \cite[Eq. (B14)]{Chu:2009}. 
(Notice that in \cite[Appendix C]{Blanchet:Damour:EspositoFarese:2004} a $\dim$-dimensional
generalization of \eqref{eq:lnS} was given, but its extension to higher inverse Laplacians,
i.e. calculation of $\Delta^{-n}(r_a^{2-\dim} r_b^{2-\dim})$, is highly non-trivial.)
Such solutions are for example necessary for the derivation of $\htt_{(4)ij}$ [see \eqref{eq:htt4}].
The inverse Laplacians for more general powers of $r_a$ and $r_b$ namely $r_a^m r_b^n$ ($n,m\ge-1$ and $n,m$ odd) in $\dim=3$ can be found by using the ansatz
\begin{align}
 \Delta^{-1} r_a^m r_b^n &= W_1^{m,n}(r_a,r_b,r_{ab}) + W_2^{m,n}(r_a,r_b,r_{ab})\ln s_{ab}\,,
\end{align}
where $W_1^{m,n}$ and $W_2^{m,n}$ are polynomials of $(n+m+2)$-th degree in $r_a$, $r_b$, and $r_{ab}$
which consist altogether of $2(\tfrac{m+n}{2} + 2)(m+n+3)$ unknown coefficients. These coefficients
have to be fixed by certain consistency conditions
\begin{align}
 \Delta \Delta^{-1} = 1\,,\quad
 \Delta_a \Delta^{-1} - \Delta^{-1} \Delta_a = 0\,,\quad
 \Delta_b \Delta^{-1} - \Delta^{-1} \Delta_b = 0\,,
\end{align}
where $\Delta_a = \partial_i^{(a)}\partial_i^{(a)}$ denotes the Laplacian with respect
to an object coordinate $\vx{a}$. These considerations can be generalized for higher inverse Laplacians. 
In \cite{Jaranowski:Schafer:1998} this technique was extensively used. 
(Generalizations of the mentioned method are discussed in \cite[Sect. V.A.]{Itoh:2004} where
the inverse Laplacians are denoted as superpotentials referring to a non-compact source. 
See also \cite[Sects. 5.1., 5.2.]{Futamase:Itoh:2007} and references therein.)

Before one of the inverse Laplacians discussed above can be applied, one has to get rid of possible
$\vnxa{1}$ and $\vnxa{2}$ vectors.
A way to eliminate these $\vnxa{}$-vectors is to rewrite them as derivatives with respect to the
particle coordinates, which are then commuted with the inverse Laplacian, see e.g. \cite{Jaranowski:1997} and 
\cite[Appendix C, Eqs. (C6)-(C8)]{Steinhoff:Schafer:Hergt:2008}.

\subsection{Averaging Procedures}\label{subsec:avgprocedures}

The averaging over the angle variables (necessary for Hadamard's finite part procedure and the local UV-analysis) 
is given by
\begin{subequations}
\label{eq:averageOmega}
\begin{align}
 \int \text{d}\Omega_{a,\dim-1} &= \Omega_{a,\dim-1}\,,\\
 \int \text{d}\Omega_{a,\dim-1} \nxa{a}{i_1}\nxa{a}{i_2} & =  \frac{1}{\dim}\Omega_{a,\dim-1}\delta_{i_1 i_2}\label{eq:angavgtwo}\,,\\
 \int \text{d}\Omega_{a,\dim-1} \nxa{a}{i_1}\nxa{a}{i_2}\nxa{a}{i_3}\nxa{a}{i_4} & = \frac{1}{\dim(\dim+2)}\Omega_{a,\dim-1}\left(\delta_{i_1 i_2}\delta_{i_3 i_4}+\delta_{i_1 i_3}\delta_{i_2 i_4}+\delta_{i_1 i_4}\delta_{i_2 i_3}\right)\,,\\
 &\vdots\,\nonumber\\
 \int \text{d}\Omega_{a,\dim-1} \nxa{a}{i_1}\dots\nxa{a}{i_{2k}} & = \frac{(\dim-2)!!}{(\dim+2(k-1))!!}\Omega_{a,\dim-1}\left(\delta_{\{i_1 i_2}\dots\delta_{i_{2k-1} i_{2k}\}}\right)\label{eq:genangavg}\,,
\end{align}
\end{subequations}
where \eqref{eq:genangavg} for $\dim=3$ coincides with Eq. (A28b) in \cite{Blanchet:Damour:1986}. 
$\Omega_{a,\dim-1}$ is the surface of the $\dim$-dimensional unit sphere around point $a$.
We use the $A_{\{i_1 i_2 \dots i_\ell\}}$ notation in the same manner like \cite[Appendix A5]{Blanchet:Damour:1986}. This means
$A_{\{i_1 i_2 \dots i_\ell\}} = \sum_{\sigma \in \mathfrak{S}} A_{\sigma(i_1)\dots\sigma(i_\ell)}$ where $\mathfrak{S}$ is the {\em smallest} set
of permutations $(1,\dots,\ell)$ making $A_{\{i_1 i_2 \dots i_\ell\}}$ fully symmetrical in $i_1,\dots,i_\ell$.
The factors in front of the deltas come from demanding that the trace over pairwise unit vectors should be one and 
the integral has the value of the surface of the unit sphere after tracing all unit vectors. An integral over an odd number 
of unit vectors is zero.

Another necessary averaging procedure is the averaging in a subspace perpendicular to an axis given by a certain vector (this
is necessary for reducing integrals of the Riesz-type and the generalized Riesz-type which may have a certain tensor structure
to a linear combination of scalar integrals of the appropriate type). 

A $\dim$-dimensional vector space can be decomposed into a line characterized by a certain vector and a $\dim-1$-dimensional subspace $S$ being
perpendicular to the vector. Let this vector be $\vct{a}$ with $\vct{a}^2 = 1$. Then the averaging of a tensor product of the vector $\vct{m}$ 
lying in the $\dim-1$-dimensional subspace (its components 
parametrized by angular variables) should give only contributions if the number of vectors is even, namely
\begin{subequations}
\begin{align}
 \langle 1 \rangle_S &= 1\,,\\
 \langle m_i \rangle_S &= 0\,,\\
 \langle m_i m_j \rangle_S &= \alpha \Aproj_{ij}\,,\label{eq:m2avg}\\
 & \vdots \nonumber
\end{align}
\end{subequations}
where $\langle\dots\rangle_S$ denotes the $\dim-1$-dimensional averaging and $\Aproj_{ij}$ the projector onto the subspace. The projector
fulfills the well-known identities
\begin{subequations}
\begin{align}
 \Aproj_{ij} & = \Aproj_{ji}\,,\\
 a^{i} \Aproj_{ij} & = 0\,,\\
 \Aproj_{ij} \Aproj_{jk} & = \Aproj_{ik}\,,
\end{align}
\end{subequations}
and is given by $\Aproj_{ij} = \delta_{ij} - a^i a^j$.
The identity in \eqref{eq:m2avg} is motivated by realizing that $\langle m_i m_j \rangle_S$ fulfills the first two projector identities. This can be rigorously shown under the conditions that $\vct{m}$ lies in the subspace perpendicular to $\vct{a}$ and $|\vct{m}|$ does not depend on the direction of $\vct{m}$.
If one contracts \eqref{eq:m2avg} with $\delta_{ij}$, then $\vct{m}^2$ can be pulled out of the averaging
brackets (as it is required to be constant in $S$) and thus
\begin{align}
 \delta_{ij} \langle m_i m_j \rangle_S &= \vct{m}^2 =\alpha \Aproj_{ii}\,,
\end{align}
from which one can conclude that ($\Aproj_{ii} = \dim-1$)
\begin{align}
 \alpha &= \frac{\vct{m}^2}{\dim-1}\,.
\end{align}
One can generalize the above result to
\begin{align}
 \langle m^{i_1} \dots m^{i_{2n}} \rangle_S & = 
	 \frac{(\vct{m}^2)^{n} (\dim-3)!!}{(\dim-3 + 2n)!!}(\Aproj^{\{i_1 i_2}\dots\Aproj^{i_{2n-1} i_{2n}\}}).
\end{align}

\subsection{Finite Part Integration}\label{subsec:finitepart}
In the Hamiltonian density there appear several integrals of the delta-type
$\int \text{d}^{\dim}x\,f(\vct{x})\dl{a}$. The function $f(\vct{x})$ is often singular at
$\vct{x} = \vx{a}$. In $\dim=3$ these integrals are often evaluated by the well-known
Hadamard's partie-finie (finite part) method (see e.g. \cite{Jaranowski:1997,Jaranowski:Schafer:1997,Jaranowski:Schafer:1998}). 
Consider a function $f(\vct{x})$ which is
well-defined in a neighborhood of $\vct{x}=\vx{a}$ and singular at this point.
Then it is possible to expand this function in a Laurent series around this point using
the auxiliary functions $f_\vct{n}$,
\begin{align}
\label{eq:PFLaurent}
 f_{\vct{n}}(\varepsilon) := f(\vx{a} + \varepsilon \vct{n}) = \sum_{\alpha = -N}^{\infty} a_{\alpha}(\vct{n})\varepsilon^\alpha\,.
\end{align}
One defines the zero order coefficient of the Laurent expansion averaged over the $\vct{n}$-vectors as the regularized value
of $f$ at $\vx{a}$, namely
\begin{align}
\label{eq:PFa0}
 f_{\text{reg}}(\vx{a}) := \frac{1}{4\pi} \int \text{d}\Omega_{2} a_0(\vct{n})\,.
\end{align}
See \eqref{eq:averageOmega} for the appropriate averaging formulas. Notice that according to
\cite{Jaranowski:Schafer:1999,Damour:Jaranowski:Schafer:2000:2} the finite part integration is ambiguous
at 3PN because in general
\begin{align}
 (f_1 f_2)_{\text{reg}}(\vx{a}) \ne (f_1)_{\text{reg}}(\vx{a})(f_2)_{\text{reg}}(\vx{a})\,,
\end{align}
consider e.g.\ $(\phis{2}^4)(\vx{a})$ vs. $(\phis{2}(\vx{a}))^4$ (see also \cite[Appendix 2]{Infeld:Plebanski:1960} for
a discussion of this ``tweedling of products'' property). We will discuss
how to avoid this issue in \Sec{subsec:contactterms}.

\subsection{The Riesz Kernel}\label{subsec:rieszkernel}
Because of the difficulties arising from using Dirac delta distributions as sources of fields in
a non-linear theory like General Relativity (e.g.\ the product of distributions is not well-defined),
reformulations of the delta distribution in terms of a function of finite width are also useful, e.g.\ the Riesz kernel.
This kernel tends to a Dirac delta when the regulator (the finite width) tends to zero. The Riesz kernel is given by
\begin{align}
 \label{eq:Rieszkernel}
\delta^{\epsilon_a}(\vct{x}-\vct{z}_a) &= \frac{\Gamma(\frac{\dim-\epsilon_a}{2})}{\pi^{\dim/2}2^{\epsilon_a}\Gamma(\frac{\epsilon_a}{2})}
	r_a^{\epsilon_a-\dim} \,.
\end{align}
Thus, in order to get rid of the singularities appearing due to the usage of Dirac deltas, one can 
replace them by the Riesz kernel and taking the limit $\epsilon_a \rightarrow 0$
for the Riesz kernel regulators after calculating the Hamiltonian.
Another advantage of the Riesz kernel is that there will be no distributional contributions to some of the
field derivatives mentioned in \ref{subsec:distcontrib}.
Yet it is important to use Riesz kernels in generic dimension to avoid ambiguous results (as they appear in the finite part method).
Unfortunately some of the integrals (in particular inverse Laplacians) for the solutions of the constraints or wave equation have not been solved to date
if a Riesz kernel type source is used. Therefore we used the Riesz kernel method only
when squares of delta distributions appeared or as a check for the other methods. 
Another problem of the Riesz kernel is that it breaks the general covariance of the theory explicitly via
violating the contracted Bianchi identities $\nabla_\mu T^{\mu\nu} = 0$ and one hopes
that after the process $\epsilon_a \to 0$ the covariance is restored. This can be checked by using the
Poincar\'{e} algebra, see \Sec{sec:kinematicalconsistency}. See \cite[above and below Eq. (94)]{Schafer:2009}
for a more detailed discussion of this issue.

\subsection{Treating Contact Terms}\label{subsec:contactterms}
Integration of the delta-type integrands is a little subtle. It was mentioned in \cite{Damour:Jaranowski:Schafer:2001} that by analysis of Fourier representations of the integrals
\emph{within dimensional regularization} one can show that
\begin{align}
 \int \text{d}^d x\,f_1 f_2 f_3 \delta_a &= (f_1)_\text{reg}(\vx{a}) (f_2)_\text{reg}(\vx{a}) (f_3)_\text{reg}(\vx{a})\,,
\end{align}
up to and including the formal 3PN order.
(But to our knowledge no general proof was given yet.)
We evaluated delta-type integrals with the help of this formula and used the
$\dim$-dimensional Riesz kernel to calculate the regularized values of the functions
at the source point appearing on the right hand side. We found that for all
cases the regularized values of the fields agreed with results from the usual finite part
method in $\dim=3$, though this does not hold for products of the fields like
$(\phis{2}^4)(\vx{a})$.
Further, the only field leading to nonvanishing finite parts in \eqref{eq:FinalRouthian3PN} is $\htt_{(4\,0)ij}$ 
and its derivatives.

\subsection{Reduction to Scalar Integrals}\label{sec:apppsc}
Besides the delta-type integrals discussed in the previous subsection (or after insertion of
Riesz kernels), the only other type of
integrals at formal 3PN level is of the form
\begin{align}
 \int \text{d}^3 x\,\nxa{1}{i_1}\cdots\nxa{1}{i_k}\cdot\nxa{2}{j_1}\cdots\nxa{2}{j_\ell}\cdot r_1^\alpha r_2^\beta s_{12}^\gamma\,.\label{eq:nintform}
\end{align}
The $s_{12}$ is introduced from derivatives of $\ln s_{12}$, which in turn arise from certain inverse
Laplacians.
This subsection provides a formalism to reduce these integrands with complicated tensor structure to a linear 
combination of pure scalar integrands of the generalized Riesz-type (discussed in the next subsection). It is only possible to rewrite the vectors as
derivatives with respect to the particle coordinates applied to a function of the type $r_1^\alpha r_2^\beta s_{12}^\gamma$
(as for the inverse Laplacians) for the case $\gamma=0,1,2\dots$, because the $s_{12}$ also
depends on the particle coordinate. For integer $\gamma=n\ge0$
\begin{align}
 s_{12}^n = (r_1 + r_2 + r_{12})^n &= \sum_{k=0}^n \sum^k_{\ell=0} \binom{n}{k} \binom{k}{\ell} r_1^\ell r_2^{k-\ell} r^{n-k}_{12}\,,\label{eq:sn}
\end{align}
and therefore the mentioned functions can be reduced to a linear combination of products of the form
$r_1^\alpha r_2^\beta r_{12}^\gamma$. This product structure is crucial to rewrite the vectors in \eqref{eq:nintform} 
into particle derivatives.
Thus one has to use another method to eliminate the vectors from the integrand for $\gamma<0$ where the functions cannot
reduced to a product form.
In \cite{Jaranowski:Schafer:1998} 
a method is mentioned which can be used to get rid of angle integrations resulting from the vectors by using an
averaging procedure of the integrand in prolate spheroidal coordinates.
We will present this procedure here in a slightly different way.

First of all one has to get rid of, e.g., the $\vnxa{2}$ vectors by using the identity
\begin{align}
 \vnxa{2} & = \frac{r_1}{r_2} \vnxa{1} + \frac{\rel}{r_2} \vnxa{12}\,.
\end{align}
Afterwards one still has to eliminate $\vnxa{1}$ vectors from the integrand.
By using $\vnxa{1} = (\vct{x} - \vx{1})/r_1$
and the orthogonal decomposition of $\vnxa{1}$ with respect to $\vnxa{12}$ one gets
\begin{align}
 \vnxa{1} = \scpm{\vnxa{1}}{\vnxa{12}}\, \vnxa{12} + \vnxa{1}^\perp \,, \qquad
 \nxa{1}{\perp\,i} = \Nproj^{ij} \nxa{1}{j} \,,
 \label{eq:orthogonaln1decomp}
\end{align}
where
\begin{align}
\scpm{\vnxa{1}}{\vnxa{12}} &\bydefd A = \frac{1}{2 r_1 r_{12}} (r_2^2 - r_1^2 -r_{12}^2) \,,\label{eq:Aexpression} \\
| \vnxa{1}^\perp | &\bydefd B = \frac{1}{2 r_1 \rel} \sqrt{\left[(r_1 + r_2)^2 - \rel^2\right]\left[\rel^2 - (r_1 - r_2)^2\right]} \,.\label{eq:Bexpression}
\end{align}
and $\Nproj^{ij} = \delta_{ij} - \nxa{12}{i} \nxa{12}{j}$ is the projector onto the subspace perpendicular to $\vnxa{12}$.
Equation \eqref{eq:Aexpression} can be derived by considering $r_2^2 = |\vct{x} - \vx{2}|^2 = |\vct{x} - \vx{1} + \vx{1} - \vx{2}|^2$ and expressing
it in terms of $r_1$, $r_{12}$ and $A$ itself. $B$ is given by $B^2 = n_1^i \Nproj^{ij} n_1^j = 1-A^2$ where we used the projector condition, $\nxa{12}{i}\Nproj^{ij}=0$, \eqref{eq:orthogonaln1decomp}, and \eqref{eq:Aexpression}.
Now a tensor product of such vectors can also be decomposed via \eqref{eq:orthogonaln1decomp} as
\begin{align}
 \int \text{d}^\dim x\, T_{i_1 \dots i_k} n_1^{i_1}\dots n_1^{i_k} & = 
 \int \text{d}^\dim x\, T_{i_1 \dots i_k} (n_1^\perp + A n_{12})^{i_1}\dots (n_1^\perp + A n_{12})^{i_k}\,,
\end{align}
where $T_{i_1 \dots i_k}$ collects all constant tensors like linear momenta, spin tensors, and $\vnunit$ vectors.
Notice that neither $A$ nor $B$ depend on the direction of $\vnxa{1}^\perp$. Furthermore the integration extends over the whole $\mathbb{R}^\dim$ 
such that one can rewrite the integrals of the type 
\begin{align}
 & \int \text{d}^\dim x\,S_{i_1\dots i_k} (r_1, r_2, r_{12})\, n_1^{\perp\,i_1} \dots n_1^{\perp\,i_k}\,,
\end{align}
in a more convenient way after a coordinate transform to prolate spheroidal coordinates. 
The Cartesian coordinates in terms of the prolate spheroidal coordinates $(\xi,\eta,\phi_1,\dots,\phi_{\dim-2})$ are given by 
(generalization of the coordinates given in \cite[pp. 752]{Abramowitz:Stegun:1964})
\begin{subequations}
\begin{align}
 x_1 & = f\xi\eta\,,\\
 x_2 & = f\sqrt{\left(\xi^2-1\right)\left(1-\eta^2\right)} \cos\phi_1\,,\\
 x_3 & = f\sqrt{\left(\xi^2-1\right)\left(1-\eta^2\right)} \sin\phi_1 \cos \phi_2\,,\\
 & \vdots \nonumber \\
 x_{\dim-1} &= f\sqrt{\left(\xi^2-1\right)\left(1-\eta^2\right)} \sin\phi_1\dots\cos\phi_{\dim-2} \\
 x_{\dim} &= f\sqrt{\left(\xi^2-1\right)\left(1-\eta^2\right)} \sin\phi_1\dots\sin\phi_{\dim-2}
\end{align}
\end{subequations}
where $-1<\eta<1;1<\xi<\infty;0\le\phi_{\dim-2}\le2\pi;0\le\phi_1,\dots,\phi_{\dim-3}\le\pi$ . 
$\xi$ and $\eta$ can be related to distance variables $r_1,r_2,\rel$ by using 
\begin{align}
 \xi & = \frac{r_1 + r_2}{\rel}\,,\quad\eta = \frac{r_1 - r_2}{\rel}\,,\quad f = \frac{\rel}{2}\,.
\end{align}
Now one can easily perform the transformation of the volume element which is given by
\begin{align}
 \text{d}^\dim x &= f^\dim (\xi^2 - \eta^2)\text{d}\xi\,\text{d}\eta\,\text{d}\Omega_{\dim-2}\,.
\end{align}
Thus we can perform an averaging over the $\dim-1$-dimensional subspace perpendicular to $\vnxa{12}$ by transforming the volume form 
\begin{align}
  & \int \text{d}^\dim x\,S_{i_1\dots i_k} (r_1, r_2, r_{12})\,n_1^{\perp\,i_1} \dots n_1^{\perp\,i_k} =
\int \text{d}\xi\,\text{d}\eta\,f^\dim (\xi^2 - \eta^2) S_{i_1\dots i_k} (\xi, \eta) \neanl & \times\int \text{d}\Omega_{\dim-2}\, n_1^{\perp\,i_1} \dots n_1^{\perp\,i_k}\,,\neanl &
 = \int \text{d}\xi\,\text{d}\eta\,\Omega_{\dim-2} f^\dim (\xi^2 - \eta^2) S_{i_1\dots i_k} (\xi, \eta)\, \langle n_1^{\perp\,i_1} \dots n_1^{\perp\,i_k} \rangle\,, \neanl &
 = \int \text{d}^\dim x\, S_{i_1\dots i_k} (r_1, r_2, r_{12})\, \langle n_1^{\perp\,i_1} \dots n_1^{\perp\,i_k} \rangle\,,
\end{align}
where $\langle n_1^{\perp\,i}\rangle = 0$ and $\langle n_1^{\perp\,i} n_1^{\perp\,j} \rangle = \tfrac{B^2}{\dim-1} \Nproj^{ij}$. 
Notice that $\Nproj^{ij}$ has the same properties like $\Aproj_{ij}$ and $\langle \dots \rangle$ the same like $\langle \dots \rangle_S$ 
in \ref{subsec:avgprocedures}. Further notice that the average in the integrand only depends on $r_1$, $r_2$, and $r_{12}$. 
Thus one is able to reduce all integrals of the
generalized Riesz-type containing $\vct{n}$-vectors to a linear combination of scalar integrals.
Notice that in the case $\gamma\ge0$
 the method from the present subsection is also
advantageous over rewriting $\vnxa{}$-vectors as derivatives. The latter can
lead to new singularities, logarithms, or even a need for a new kind of regulator, e.g.,
for an expression like
\begin{align}
 \int\text{d}^3 x\,\nxa{a}{i}\nxa{a}{j}\nxa{a}{k} r_a^{-3}\,.
\end{align}
By using the method of the present subsection these issues are avoided.

\subsection{Integration using Generalized Riesz-Formula}\label{subsec:genriesz}
After reduction to scalar integrals one is left with integrals of the form 
$\int \text{d}^3 x\, r_1^\alpha r_2^\beta s_{12}^\gamma$.
For integer $\gamma\ge0$ one can use \eqref{eq:sn} and end up with integrals whose solution were found by M.~Riesz for arbitrary $\dim$
\mycite{Riesz:1949,Riesz:1949:err}{Riesz:1949,*Riesz:1949:err}, namely
\begin{align}
\int \text{d}^\dim x \, r_1^{\alpha} r_2^{\beta} =
	\pi^{\dim/2} \frac{\Gamma(\frac{\alpha+\dim}{2})\Gamma(\frac{\beta+\dim}{2})\Gamma(-\frac{\alpha+\beta+\dim}{2})}
			{\Gamma(-\frac{\alpha}{2})\Gamma(-\frac{\beta}{2})\Gamma(\frac{\alpha+\beta+2\dim}{2})}
	\rel^{\alpha+\beta+\dim}\,.\label{eq:rieszformula}
\end{align}
For $\dim=3$ a generalization of the Riesz-formula was found by P. Jaranowski and G. Sch\"{a}fer in \cite{Jaranowski:Schafer:1998},
\begin{align}
 \int \text{d}^3 x\, r_1^\alpha r_2^\beta s_{12}^\gamma & = 
	2\pi \frac{\Gamma(\alpha+2)\Gamma(\beta+2)\Gamma(-\alpha-\beta-\gamma-4)}{\Gamma(-\gamma)}
	\biggl(
		I_{1/2}(\alpha+2,-\alpha-\gamma-2) \neanl &
		+I_{1/2}(\beta+2,-\beta-\gamma-2) \neanl &
		-I_{1/2}(\alpha+\beta+4,-\alpha-\beta-\gamma-4)
		-1
	\biggr) \rel^{\alpha+\beta+\gamma+3}\,,\label{eq:sintegral}
\end{align}
with $ I_{1/2}(x,y) := \tfrac{B_{1/2}(x,y)}{B(x,y)}$
the incomplete regularized Euler Beta function. One can express the incomplete Beta function (Euler integral of the first kind) 
$B_{1/2}(x,y)$ in terms of
the Gau\ss\ hypergeometric function ${}_2\! F_1$
\begin{align}
 B_{1/2}(x,y) &= \frac{1}{2^x x} {}_2\! F_1 \left(1-y,x;x+1;\frac{1}{2}\right) 
=\frac{1}{x}{}_2\! F_1\left(x+y,x;1+x;-1\right)\,.\label{eq:incompletebeta}
\end{align}
The function $B(x,y)$ represents the Euler Beta function or Euler integral of the second kind
with $B(x,y) = \tfrac{\Gamma(x+y)}{\Gamma(x)\Gamma(y)}$.
It turns out that the regularization procedure mentioned in \cite[Eq. (B21), (B23), and (B24)]{Jaranowski:Schafer:1998} only 
modifies $\alpha$ and $\beta$ to be non-integer via the introduced analytical regulators $\mu\epsilon$ and $\nu\epsilon$.
So one can in principle simplify 
the formula given above by the assumption that $\gamma\in\mathbb{Z}$, and $\alpha$ and $\beta$ being arbitrary. Positive 
integer powers of $s_{12}$ can be handled through \eqref{eq:sn} and \eqref{eq:rieszformula}. 
Thus the only relevant powers of $s_{12}$ are the negative integer ones.

Equation \eqref{eq:incompletebeta} leads 
to hypergeometric functions of the type ${}_2\! F_1(-\gamma,z;z+1;-1)$
in \eqref{eq:sintegral}, where $z$ depends on 
$\alpha$ and $\beta$, namely 
\begin{align}
\int \text{d}^3 x\,r_1^\alpha r_2^\beta s_{12}^{\gamma} &=
2 \pi \Gamma(2 + \alpha) \Gamma(2 + \beta ) \Gamma(-4 -  \alpha  -  \beta  -  \gamma ) 
\biggl\{
	-\frac{1}{\Gamma(- \gamma)} \neanl &
	+ \frac{
		{}_2\! F_1\bigl(- \gamma , 2 + \alpha; 3 + \alpha; -1\bigr)
	}{
		\Gamma\bigl(3 + \alpha \bigr)\Gamma(-2 -  \alpha  - \gamma ) 
	}   
	+ \frac{
		{}_2\! F_1\bigl(- \gamma , 2 + \beta; 3 + \beta; -1\bigr)
	}{
		\Gamma\bigl(3 + \beta \bigr)\Gamma(-2 -  \beta  - \gamma ) 
	} \neanl &
	-  \frac{
		{}_2\! F_1\bigl(- \gamma , 4 + \alpha  + \beta ; 5 + \alpha  + \beta ; -1\bigr)
	}{
		\Gamma\bigl(5 + \alpha  + \beta \bigr)\Gamma(-4 -  \alpha  -  \beta  -  \gamma ) 
	}  
\biggr\} r_{12}^{\alpha+\beta+\gamma+3}\,.
\end{align}
So for negative integer gamma $\gamma = -n$ one can express the solution of the integral in terms of 
${}_2\! F_1(n,z;z+1;-1)$. It is well-known that \cite{Prudnikov:Brychkov:Marichev:1986:vol3}
\begin{align}
 {}_2\! F_1 (0,z;z+1;-1) &= 1\,,\\
 {}_2\! F_1 (1,z;z+1;-1) &= \frac{z}{2}\left[\psi\left(\frac{z+1}{2}\right)-\psi\left(\frac{z}{2}\right)\right]\,,
\end{align}
where $\psi$ is the Digamma function.
From these two formulas and the contiguous relation of the Gau\ss\ hypergeometric function 
\cite{Klein:1933,Abramowitz:Stegun:1964,Olver:Lozier:Boisvert:Clark:2010}
\begin{align}
 0 &= (c-a){}_2\! F_1(a-1,b;c;z) + (2 a - c + (b-a)z) {}_2\! F_1(a,b;c;z)\neanl &
 + a (z-1) {}_2\! F_1(a+1,b;c;z)\,,
\end{align}
one obtains the recursive relation
\begin{align}
 {}_2\! F_1(n,z;z+1;-1) &=\frac{1}{2(n-1)}\biggl[(z-(n-2)){}_2\! F_1(n-2,z;z+1;-1) \neanl &
 + (3n - 2(2+z)) {}_2\! F_1(n-1,z;z+1;-1)\biggr]\label{eq:2F1recursion}\,.
\end{align}
From these relations one can show per induction the general structure of the hypergeometric function to be
\begin{align}
 {}_2\! F_1(n,z;z+1;-1) & = \sum_{k=0}^{n-1} a^{(n)}_k z^k
+ \frac{(-1)^{n-1}}{2(n-1)!} \left[\psi\left(\frac{z+1}{2}\right)-\psi\left(\frac{z}{2}\right)\right]\prod_{k=0}^{n-1} (z-k)\,.
\end{align}
Unfortunately the coefficients $a^{(n)}_{k}$ can only be obtained by complicated recursive relations between different $k$ and $n$
following from \eqref{eq:2F1recursion} and we can only give an explicit form for some of them
\begin{align}
 a_{0}^{(n)} &= 0\,,n\ge2\quad\text{and}\,a_0^{(0)} = 1\,,a_0^{(1)} = 0\,,\\
 a_{n-1}^{(n)} &= \frac{(-1)^n}{2(n-1)!}\,,n\ge2\,,\\
 a_{n-2}^{(n)} &= \frac{(-1)^n (1 + n - n^2)}{4(n-1)!}\,,n\ge3\,.
\end{align}
Nevertheless the mentioned recursion relations can be used to cache all Gau\ss\ hypergeometric functions. {\sc Mathematica} is able to handle limits,
series expansions and derivatives of the arising
Digamma functions very well and thus all occurring integrals in the binary Hamiltonian to linear order in the spin variables
can be solved.

%
%

\end{document}

%% file: macros.tex
\usepackage[normalem]{ulem}

\def\vct#1{{\mathbf{#1}}}
\def\defdby{:=} 
\def\bydefd{=:} 

\def\eanl{\\}
\def\neanl{\nonumber\\}
\def\nnl{\nonumber\\ & \quad}
\def\nlqn{\nnl}
\def\nln{\nonumber \\ &}

\def\dim{d}

\def\TT{{\text{TT}}}
\def\nonTT{{\text{non-TT}}}
\def\ADM{{\text{ADM}}}
\def\ADMTT{{\text{ADMTT}}}
\def\interaction{{\text{int}}}
\def\td{{(\text{td})}}
\def\ttd{{(\text{ttd})}}
\def\matter{{\text{matter}}}
\def\field{{\text{field}}}

\newcommand{\tproj}[2]{\mathcal{D}_{#1}^{#2}} 
\newcommand{\transv}[1]{\perp^{#1}} 
\newcommand{\longit}[1]{\parallel^{#1}} 

\newcommand{\TTproj}[2]{\delta^{\text{TT}\,#1}_{#2}}
\newcommand{\LTproj}[2]{\delta^{\text{LT}\,#1}_{#2}}
\newcommand{\Trproj}[2]{\delta^{\text{Tr}\,#1}_{#2}}

\def\canpi{\hat{\pi}}

\def\invlapl#1{\Delta^{-#1}}
\def\src#1{\mathcal{H}_{#1}^{\text{matter}}}

\def\srcfield#1{\mathcal{H}^{\field}_{#1}}
\def\htt{h^\TT}
\def\httdot{\dot{h}^\TT}
\def\httddot{\ddot{h}^\TT}
\def\pitt{\pi_{\TT}}

\def\picantt{\canpi_{\TT}}

\def\pimat{\pi_{\matter}}

\def\grav{G^{(\dim)}}
\def\gravthree{G}

\newcommand{\dunderline}[1]{{#1}}

\newcommand{\cInv}[1]{c^{-#1}}
\newcommand{\Order}[1]{\mathcal{O}({#1})}

\newcommand{\ricci}[2]{{{}^{#1} \text{R}_{#2}}}
\newcommand{\christoffel}[3]{{{}^{#1} \Gamma^{#2}_{#3}}}

\newcommand{\pitti}[1]{{\pitt^{#1}}} 
\newcommand{\picantti}[1]{{\picantt^{#1}}} 
\newcommand{\pimati}[1]{{\pimat^{#1}}} 
\newcommand{\pimatis}[2]{{\pi_{{(#1)}\matter}^{#2}}} 
\newcommand{\pia}[2]{\pi_{#1}^{#2}}
\newcommand{\picani}[1]{{\canpi}^{#1}}

\newcommand{\pitildei}[1]{{\tilde{\pi}^{#1}}} 
\newcommand{\pihati}[1]{{\breve{\pi}^{#1}}} 
\newcommand{\pihatmati}[1]{{\breve{\pi}^{#1}_{\matter}}} 

\newcommand{\momli}[1]{{\pitildei{#1}}} 

\newcommand{\phis}[1]{{\phi_{(#1)}{}}} 
\newcommand{\srcs}[1]{\src{(#1)}} 
\newcommand{\srcstt}[1]{\src{(#1) \TT}} 
\newcommand{\srcsnontt}[1]{\src{(#1) \nonTT}} 
\newcommand{\srcis}[2]{\src{#2 (#1)}} 
\newcommand{\srcisnontt}[2]{\src{#2 (#1) \nonTT}} 
\newcommand{\phibs}[1]{{{\bar{\phi}}_{(#1)}{}}} 
\newcommand{\momls}[2]{{{\tilde{\pi}^{#2}_{(#1)}{}}}}
\newcommand{\vpots}[2]{{V^{#2}_{(#1)}}} 
\newcommand{\pipots}[2]{{\tilde{\pi}^{#2}_{(#1)}}{}}

\newcommand{\Sfour}{{\bar{S}_{(4)}{}}}
\newcommand{\phibfo}{{\phibs{4\,2}{}}} 
\newcommand{\phibft}{{\phibs{4\,0}{}}} 

\newcommand{\momlones}[2]{{\tilde{\pi}^{#2}_{(#1)1}}{}}

\newcommand{\pdiffq}[2]{\frac{\partial #1}{\partial #2}}

\newcommand{\triad}[2]{e_{(#1)#2}}
\newcommand{\triads}[3]{e_{(#1)\,(#2)#3}}

\def\canmomp{p}
\def\canmom{\hat{\canmomp}}
\def\vcanmom{\hat{\vct{\canmomp}}}

\newcommand{\spin}[3]{\hat{S}_{#1\, (#2)(#3)}}
\newcommand{\kerrspin}[3]{\hat{a}_{#1\, (#2)(#3)}}
\newcommand{\mom}[2]{{\canmom}_{#1\,#2}}
\newcommand{\nmom}[1]{{{n\canmom}_{#1}}}

\newcommand{\nxa}[2]{{n}_{#1}^{#2}}
\newcommand{\nunit}[1]{\nxa{12}{#1}}
\newcommand{\nun}[1]{\nunit{#1}} 
\newcommand{\vnunit}{{\vnxa{12}}}
\newcommand{\vmom}[1]{{\vcanmom}_{#1}}
\newcommand{\vnxa}[1]{{\vct{n}}_{#1}}
\newcommand{\vnun}{{\vnunit}}
\newcommand{\dl}[1]{\delta_{#1}}
\newcommand{\scpm}[2]{(#1\,#2)}
\newcommand{\crpm}[2]{(#1\times #2)}
\newcommand{\trpm}[3]{\scpm{\crpm{#1}{#2}}{#3}}
\newcommand{\vspin}[1]{\hat{\vct{S}}_{#1}}
\newcommand{\vkerrspin}[1]{\hat{\vct{a}}_{#1}}
\newcommand{\vx}[1]{\hat{\vct{z}}_{#1}}
\newcommand{\xa}[2]{\hat{z}_{#1}^{#2}}

\newcommand{\relab}[1]{r_{#1}}
\newcommand{\rel}{\relab{12}}
\newcommand{\ang}{L}
\newcommand{\vang}{\vct{\ang}}

\def\Nproj{\mathfrak{P}}
\def\Aproj{\mathfrak{p}}

\newcommand{\shift}[1]{N_{#1}} 
\newcommand{\shiftup}[1]{N^{#1}} 
\newcommand{\lapse}{N} 


%% file: Technical3PNPaper_refs.tex
\providecommand{\href}[2]{#2}\begingroup\raggedright\endgroup

%% file: Technical3PNPaper.bbl
\begin{thebibliography}{100}

\bibitem{Einstein:1915:1}
A.~Einstein, ``{Z}ur allgemeinen {R}elativit{\"a}tstheorie,'' {\em Sitz.-Ber.
  Preu{\ss}. Akad. Wiss.} {\bf XLIV} (1915)  778--785.

\bibitem{Einstein:1915:2}
A.~Einstein, ``{Z}ur allgemeinen {R}elativit{\"a}tstheorie ({N}achtrag),'' {\em
  Sitz.-Ber. Preu{\ss}. Akad. Wiss.} {\bf XLVI} (1915)  799--801.

\bibitem{Einstein:1915:3}
A.~Einstein, ``{E}rkl{\"a}rung der {P}erihelbewegung des {M}erkur aus der
  allgemeinen {R}elativit{\"a}tstheorie,'' {\em Sitz.-Ber. Preu{\ss}. Akad.
  Wiss.} {\bf XLVII} (1915)  831--839.

\bibitem{Einstein:1915:4}
A.~Einstein, ``{D}ie {F}eldgleichungen der {G}ravitation,'' {\em Sitz.-Ber.
  Preu{\ss}. Akad. Wiss.} {\bf XLVIII} (1915)  844--847.

\bibitem{Einstein:1916}
A.~Einstein, ``{D}ie {G}rundlage der allgemeinen {R}elativit{\"a}tstheorie,''
  \href{http://dx.doi.org/10.1002/andp.200590044}{{\em Ann. Phys. (Berlin)}
  {\bf 354} (1916)  769--822}.

\bibitem{Stephani:Kramer:MacCallum:Hoenselaers:Herlt:2003}
H.~Stephani, D.~Kramer, M.~A.~H. MacCallum, C.~Hoenselaers, and E.~Herlt, {\em
  Exact Solutions of Einstein's Field Equations}.
\newblock Cambridge Monographs on Mathematical Physics. Cambridge University
  Press, The Edinburgh Building, Cambridge CB2 2RU, UK, 2nd~ed., 2003.

\bibitem{Griffiths:Podolsky:2009}
J.~B. Griffiths and J.~Podolsk{\'y}, {\em Exact Space-Times in Einstein's
  General Relativity}.
\newblock Cambridge Monographs on Mathematical Physics. Cambridge University
  Press, The Edinburgh Building, Cambridge CB2 2RU, UK, 2009.

\bibitem{Schwarzschild:1916}
K.~Schwarzschild, ``{\"U}ber das {G}ravitationsfeld eines {M}assenpunktes nach
  der {E}insteinschen {T}heorie,'' {\em Sitz.-Ber. Preu{\ss}. Akad. Wiss.}
  (1916)  189--196.

\bibitem{Kerr:1963}
R.~P. Kerr, ``Gravitational field of a spinning mass as an example of
  algebraically special metrics,''
\href{http://dx.doi.org/10.1103/PhysRevLett.11.237}{{\em Phys. Rev. Lett.} {\bf
  11} (1963)  237--238}.

\bibitem{Cutler:Apostolatos:Bildsten:others:1993}
C.~Cutler, T.~A. Apostolatos, L.~Bildsten, L.~S. Finn, {\'E}.~E. Flanagan,
  D.~Kennefick, D.~M. Markovic, A.~Ori, E.~Poisson, G.~J. Sussman, and K.~S.
  Thorne, ``{T}he last three minutes: {I}ssues in gravitational-wave
  measurements of coalescing compact binaries,''
  \href{http://dx.doi.org/10.1103/PhysRevLett.70.2984}{{\em Phys. Rev. Lett.}
  {\bf 70} (1993)  2984--2987},
  \href{http://arxiv.org/abs/astro-ph/9208005}{{\tt arXiv::astro-ph/9208005}}.

\bibitem{Reisswig:Husa:Rezzolla:Dorband:Pollney:Seiler:2009}
C.~Reisswig, S.~Husa, L.~Rezzolla, E.~N. Dorband, D.~Pollney, and J.~Seiler,
  ``{G}ravitational-wave detectability of equal-mass black-hole binaries with
  aligned spins,'' \href{http://dx.doi.org/10.1103/PhysRevD.80.124026}{{\em
  Phys. Rev. D} {\bf 80} (2009)  124026},
  \href{http://arxiv.org/abs/0907.0462}{{\tt arXiv:0907.0462 [gr-qc]}}.

\bibitem{GEO600home:2011}
``{GEO600 home page}.'' Electronic adress, May, 2011.
\newblock \url{http://www.geo600.org}.

\bibitem{VIRGOhome:2011}
``{VIRGO collaboration}.'' Electronic adress, Oct., 2011.
\newblock \url{https://wwwcascina.virgo.infn.it/}.

\bibitem{advancedLIGOhome:2011}
``{advanced LIGO home page}.'' Electronic adress, May, 2011.
\newblock \url{http://advancedligo.mit.edu}.

\bibitem{Hulse:Taylor:1975}
R.~A. Hulse and J.~H. Taylor, ``{D}iscovery of a pulsar in a binary system,''
  \href{http://dx.doi.org/10.1086/181708}{{\em ApJ} {\bf 195} (1975)
  L51--L53}.

\bibitem{Kramer:Stairs:Manchester:McLaughlin:Lyne:others:2006}
M.~Kramer, I.~H. Stairs, R.~N. Manchester, M.~A. McLaughlin, A.~G. Lyne, R.~D.
  Ferdman, M.~Burgay, D.~R. Lorimer, A.~Possenti, N.~D'Amico, J.~M. Sarkissian,
  G.~B. Hobbs, J.~E. Reynolds, P.~C.~C. Freire, and F.~Camilo, ``{T}ests of
  general relativity from timing the double pulsar,''
  \href{http://dx.doi.org/10.1126/science.1132305}{{\em Science} {\bf 314}
  (2006)  97--102},
\href{http://arxiv.org/abs/astro-ph/0609417}{{\tt arXiv:astro-ph/0609417}}.

\bibitem{Kramer:Wex:2009}
M.~Kramer and N.~Wex, ``The double pulsar system: a unique laboratory for
  gravity,''
\href{http://dx.doi.org/10.1088/0264-9381/26/7/073001}{{\em Class. Quant.
  Grav.} {\bf 26} (2009)  073001}.

\bibitem{Hermes:others:2012}
J.~J. Hermes, M.~Kilic, W.~R. Brown, D.~E. Winget, C.~A. Prieto, A.~Gianninas,
  A.~S. Mukadam, A.~{Cabrera-Lavers}, and S.~J. Kenyon, ``{R}apid orbital decay
  in the 12.75-minute {WD}+{WD} binary {J0651+2844},''
  \href{http://arxiv.org/abs/1208.5051}{{\tt arXiv:1208.5051 [astro-ph.SR]}}.

\bibitem{Schafer:1985}
G.~Sch{\"a}fer, ``The gravitational quadrupole radiation-reaction force and the
  canonical formalism of {ADM},''
\href{http://dx.doi.org/10.1016/0003-4916(85)90337-9}{{\em Ann. Phys. (N.Y.)}
  {\bf 161} (1985)  81--100}.

\bibitem{Lorentz:Droste:1917:1}
H.~A. Lorentz and J.~Droste, ``{De beweging van een stelsel lichamen onder den
  invloed van hunne onderlinge aantrekking, behandeld volgens de theorie van
  Einstein, I},'' {\em Versl. K. Akad. Wet. (Amsterdam)} {\bf 26} (1917)  392.

\bibitem{Lorentz:Droste:1917:2}
H.~A. Lorentz and J.~Droste, ``{De beweging van een stelsel lichamen onder den
  invloed van hunne onderlinge aantrekking, behandeld volgens de theorie van
  Einstein, II},'' {\em Versl. K. Akad. Wet. (Amsterdam)} {\bf 26} (1917)  649.

\bibitem{Einstein:Infeld:Hoffmann:1938}
A.~Einstein, L.~Infeld, and B.~Hoffmann, ``The gravitational equations and the
  problem of motion,'' \href{http://dx.doi.org/10.2307/1968714}{{\em Ann.
  Math.} {\bf 39} (1938)  65--100}.

\bibitem{Arnowitt:Deser:Misner:1962}
R.~L. Arnowitt, S.~Deser, and C.~W. Misner, ``The dynamics of general
  relativity,'' in {\em Gravitation: An Introduction to Current Research},
  L.~Witten, ed., pp.~227--265.
\newblock John Wiley, New York,
1962.
\newblock

\bibitem{Arnowitt:Deser:Misner:2008}
R.~L. Arnowitt, S.~Deser, and C.~W. Misner, ``Republication of: {T}he dynamics
  of general relativity,''
  \href{http://dx.doi.org/10.1007/s10714-008-0661-1}{{\em Gen. Relativ.
  Gravit.} {\bf 40} (2008)  1997--2027},
\href{http://arxiv.org/abs/gr-qc/0405109}{{\tt arXiv:gr-qc/0405109}}.

\bibitem{Ohta:Okamura:Kimura:Hiida:1974}
T.~Ohta, H.~Okamura, T.~Kimura, and K.~Hiida, ``Coordinate condition and higher
  order gravitational potential in canonical formalism,''
\href{http://dx.doi.org/10.1143/PTP.51.1598}{{\em Prog. Theor. Phys.} {\bf 51}
  (1974)  1598--1612}.

\bibitem{Damour:Schafer:1985}
T.~Damour and G.~Sch{\"a}fer, ``{L}agrangians for n point masses at the second
  post-{N}ewtonian approximation of general relativity,''
  \href{http://dx.doi.org/10.1007/BF00773685}{{\em Gen. Relativ. Gravit.} {\bf
  17} (1985)  879--905}.

\bibitem{Damour:Schafer:1988}
T.~Damour and G.~Sch{\"a}fer, ``{H}igher-order relativistic periastron advances
  and binary pulsars,'' \href{http://dx.doi.org/10.1007/BF02828697}{{\em Nuovo
  Cim. B} {\bf 101} (1988)  127--176}.

\bibitem{Schafer:1986}
G.~Sch{\"a}fer, ``The {ADM} {H}amiltonian at the postlinear approximation,''
  \href{http://dx.doi.org/10.1007/BF00765886}{{\em Gen. Relativ. Gravit.} {\bf
  18} (1986)  255--270}.

\bibitem{Jaranowski:Schafer:1998}
P.~Jaranowski and G.~Sch{\"a}fer, ``{T}hird post-{N}ewtonian higher order {ADM}
  {H}amilton dynamics for two-body point-mass systems,''
  \href{http://dx.doi.org/10.1103/PhysRevD.57.7274}{{\em Phys. Rev. D} {\bf 57}
  (1998)  7274--7291}, \href{http://arxiv.org/abs/gr-qc/9712075}{{\tt
  arXiv:gr-qc/9712075}}.

\bibitem{Kimura:Toiya:1972}
T.~Kimura and T.~Toiya, ``Potential in the canonical formalism of gravity,''
\href{http://dx.doi.org/10.1143/PTP.48.316}{{\em Prog. Theor. Phys.} {\bf 48}
  (1972)  316--328}.

\bibitem{Jaranowski:Schafer:1999}
P.~Jaranowski and G.~Sch{\"a}fer, ``Binary black-hole problem at the third
  post-{N}ewtonian approximation in the orbital motion: {S}tatic part,''
  \href{http://dx.doi.org/10.1103/PhysRevD.60.124003}{{\em Phys. Rev. D} {\bf
  60} (1999)  124003},
\href{http://arxiv.org/abs/gr-qc/9906092}{{\tt arXiv:gr-qc/9906092}}.

\bibitem{Damour:Jaranowski:Schafer:2000}
T.~Damour, P.~Jaranowski, and G.~Sch{\"a}fer, ``{P}oincar{\'e} invariance in
  the {ADM} {H}amiltonian approach to the general relativistic two-body
  problem,'' \href{http://dx.doi.org/10.1103/PhysRevD.62.021501}{{\em Phys.
  Rev. D} {\bf 62} (2000)  021501(R)},
\href{http://arxiv.org/abs/gr-qc/0003051}{{\tt arXiv:gr-qc/0003051}}.

\bibitem{Damour:Jaranowski:Schafer:2001}
T.~Damour, P.~Jaranowski, and G.~Sch{\"a}fer, ``{D}imensional regularization of
  the gravitational interaction of point masses,''
  \href{http://dx.doi.org/10.1016/S0370-2693(01)00642-6}{{\em Phys. Lett. B}
  {\bf 513} (2001)  147--155},
\href{http://arxiv.org/abs/gr-qc/0105038}{{\tt arXiv:gr-qc/0105038}}.

\bibitem{Jaranowski:Schafer:1997}
P.~Jaranowski and G.~Sch{\"a}fer, ``Radiative 3.5 post-{N}ewtonian {ADM}
  {H}amiltonian for many-body point-mass systems,''
\href{http://dx.doi.org/10.1103/PhysRevD.55.4712}{{\em Phys. Rev. D} {\bf 55}
  (1997)  4712--4722}.

\bibitem{Konigsdorffer:Faye:Schafer:2003}
C.~K{\"o}nigsd{\"o}rffer, G.~Faye, and G.~Sch{\"a}fer, ``{The binary black-hole
  dynamics at the third-and-a-half post-Newtonian order in the
  ADM-formalism},'' \href{http://dx.doi.org/10.1103/PhysRevD.68.044004}{{\em
  Phys. Rev. D} {\bf 68} (2003)  044004},
\href{http://arxiv.org/abs/gr-qc/0305048}{{\tt arXiv:gr-qc/0305048}}.

\bibitem{Blanchet:2006}
L.~Blanchet, ``Gravitational radiation from post-{N}ewtonian sources and
  inspiralling compact binaries,'' {\em Living Rev. Relativity} {\bf 9} (2006)
  4.
\url{http://www.livingreviews.org/lrr-2006-4}.

\bibitem{Futamase:Itoh:2007}
T.~Futamase and Y.~Itoh, ``The post-{N}ewtonian approximation for relativistic
  compact binaries,'' {\em Living Rev. Relativity} {\bf 10} (2007)  2.
\url{http://www.livingreviews.org/lrr-2007-2}.

\bibitem{Pati:Will:2000}
M.~E. Pati and C.~M. Will, ``Post-{N}ewtonian gravitational radiation and
  equations of motion via direct integration of the relaxed {E}instein
  equations: {F}oundations,''
  \href{http://dx.doi.org/10.1103/PhysRevD.62.124015}{{\em Phys. Rev. D} {\bf
  62} (2000)  124015},
\href{http://arxiv.org/abs/gr-qc/0007087}{{\tt arXiv:gr-qc/0007087}}.

\bibitem{Damour:EspositoFarese:1995}
T.~Damour and G.~Esposito-Farese, ``Testing gravity to second post-{N}ewtonian
  order: {A} field theory approach,''
  \href{http://dx.doi.org/10.1103/PhysRevD.53.5541}{{\em Phys. Rev. D} {\bf 53}
  (1996)  5541--5578},
\href{http://arxiv.org/abs/gr-qc/9506063}{{\tt arXiv:gr-qc/9506063 [gr-qc]}}.

\bibitem{Goldberger:Rothstein:2006}
W.~D. Goldberger and I.~Z. Rothstein, ``An effective field theory of gravity
  for extended objects,''
  \href{http://dx.doi.org/10.1103/PhysRevD.73.104029}{{\em Phys. Rev. D} {\bf
  73} (2006)  104029},
\href{http://arxiv.org/abs/hep-th/0409156}{{\tt arXiv:hep-th/0409156}}.

\bibitem{Gilmore:Ross:2008}
J.~B. Gilmore and A.~Ross, ``{E}ffective field theory calculation of second
  post-{N}ewtonian binary dynamics,''
  \href{http://dx.doi.org/10.1103/PhysRevD.78.124021}{{\em Phys. Rev. D} {\bf
  78} (2008)  124021}, \href{http://arxiv.org/abs/0810.1328}{{\tt
  arXiv:0810.1328 [gr-qc]}}.

\bibitem{Kol:Smolkin:2009}
B.~Kol and M.~Smolkin, ``Dressing the post-{N}ewtonian two-body problem and
  classical effective field theory,''
  \href{http://dx.doi.org/10.1103/PhysRevD.80.124044}{{\em Phys. Rev. D} {\bf
  80} (2009)  124044},
\href{http://arxiv.org/abs/0910.5222}{{\tt arXiv:0910.5222 [hep-th]}}.

\bibitem{Foffa:Sturani:2011}
S.~Foffa and R.~Sturani, ``{E}ffective field theory calculation of conservative
  binary dynamics at third post-{N}ewtonian order,''
  \href{http://dx.doi.org/10.1103/PhysRevD.84.044031}{{\em Phys. Rev. D} {\bf
  84} (2011)  044031},
\href{http://arxiv.org/abs/1104.1122}{{\tt arXiv:1104.1122 [gr-qc]}}.

\bibitem{Foffa:Sturani:2012}
S.~Foffa and R.~Sturani, ``{T}he dynamics of the gravitational two-body problem
  in the post-{N}ewtonian approximation at quadratic order in the {N}ewton's
  constant,'' \href{http://arxiv.org/abs/1206.7087}{{\tt arXiv:1206.7087
  [gr-qc]}}.

\bibitem{Barker:OConnell:1975}
B.~M. Barker and R.~F. O'Connell, ``Gravitational two-body problem with
  arbitrary masses, spins, and quadrupole moments,''
\href{http://dx.doi.org/10.1103/PhysRevD.12.329}{{\em Phys. Rev. D} {\bf 12}
  (1975)  329--335}.

\bibitem{DEath:1975}
P.~D. D'Eath, ``Interaction of two black holes in the slow-motion limit,''
  \href{http://dx.doi.org/10.1103/PhysRevD.12.2183}{{\em Phys. Rev. D} {\bf 12}
  (1975)  2183--2199}.

\bibitem{Barker:OConnell:1979}
B.~M. Barker and R.~F. O'Connell, ``The gravitational interaction: Spin,
  rotation, and quantum effects---a review,''
  \href{http://dx.doi.org/10.1007/BF00756587}{{\em Gen. Relativ. Gravit.} {\bf
  11} (1979)  149--175}.

\bibitem{Thorne:Hartle:1985}
K.~S. Thorne and J.~B. Hartle, ``Laws of motion and precession for black holes
  and other bodies,''
\href{http://dx.doi.org/10.1103/PhysRevD.31.1815}{{\em Phys. Rev. D} {\bf 31}
  (1985)  1815--1837}.

\bibitem{Goenner:Gralewski:Westpfahl:1967}
H.~Goenner, U.~Gralewski, and K.~Westpfahl, ``{G}ravitative {S}elbstkr{\"a}fte
  und {S}trahlungsverluste klassischer {S}pinteilchen (erste {N}{\"a}herung),''
  \href{http://dx.doi.org/10.1007/BF01326226}{{\em Z. Phys.} {\bf 207} (1967)
  186--208}.

\bibitem{Bennewitz:Westpfahl:1971}
F.~Bennewitz and K.~Westpfahl, ``{S}elbstwechselwirkung von
  {G}ravitationsfeldern schnell bewegter {P}ol-{D}ipolquellen,''
  \href{http://dx.doi.org/10.1007/BF01893619}{{\em Commun. math. Phys.} {\bf
  23} (1971)  296--318}.

\bibitem{Tagoshi:Ohashi:Owen:2001}
H.~Tagoshi, A.~Ohashi, and B.~J. Owen, ``Gravitational field and equations of
  motion of spinning compact binaries to 2.5 post-{N}ewtonian order,''
  \href{http://dx.doi.org/10.1103/PhysRevD.63.044006}{{\em Phys. Rev. D} {\bf
  63} (2001)  044006},
\href{http://arxiv.org/abs/gr-qc/0010014}{{\tt arXiv:gr-qc/0010014}}.

\bibitem{Faye:Blanchet:Buonanno:2006}
G.~Faye, L.~Blanchet, and A.~Buonanno, ``{H}igher-order spin effects in the
  dynamics of compact binaries. {I}. {E}quations of motion,''
  \href{http://dx.doi.org/10.1103/PhysRevD.74.104033}{{\em Phys. Rev. D} {\bf
  74} (2006)  104033},
\href{http://arxiv.org/abs/gr-qc/0605139}{{\tt arXiv:gr-qc/0605139}}.

\bibitem{Damour:Jaranowski:Schafer:2008:1}
T.~Damour, P.~Jaranowski, and G.~Sch{\"a}fer, ``{H}amiltonian of two spinning
  compact bodies with next-to-leading order gravitational spin-orbit
  coupling,'' \href{http://dx.doi.org/10.1103/PhysRevD.77.064032}{{\em Phys.
  Rev. D} {\bf 77} (2008)  064032},
\href{http://arxiv.org/abs/0711.1048}{{\tt arXiv:0711.1048 [gr-qc]}}.

\bibitem{Steinhoff:Hergt:Schafer:2008:2}
J.~Steinhoff, S.~Hergt, and G.~Sch{\"a}fer, ``Next-to-leading order
  gravitational spin(1)-spin(2) dynamics in {H}amiltonian form,''
  \href{http://dx.doi.org/10.1103/PhysRevD.77.081501}{{\em Phys. Rev. D} {\bf
  77} (2008)  081501(R)},
\href{http://arxiv.org/abs/0712.1716}{{\tt arXiv:0712.1716 [gr-qc]}}.

\bibitem{Steinhoff:Hergt:Schafer:2008:1}
J.~Steinhoff, S.~Hergt, and G.~Sch{\"a}fer, ``{S}pin-squared {H}amiltonian of
  next-to-leading order gravitational interaction,''
  \href{http://dx.doi.org/10.1103/PhysRevD.78.101503}{{\em Phys. Rev. D} {\bf
  78} (2008)  101503(R)},
\href{http://arxiv.org/abs/0809.2200}{{\tt arXiv:0809.2200 [gr-qc]}}.

\bibitem{Perrodin:2010}
D.~L. Perrodin, ``Subleading spin-orbit correction to the {N}ewtonian potential
  in effective field theory formalism,'' in {\em Proceedings of the 12th Marcel
  Grossmann Meeting on General Relativity}.
\newblock World Scientific, Singapore, 2011.
\newblock
\href{http://arxiv.org/abs/1005.0634}{{\tt arXiv:1005.0634 [gr-qc]}}.
\newblock

\bibitem{Porto:2010}
R.~A. Porto, ``Next to leading order spin-orbit effects in the motion of
  inspiralling compact binaries,''
  \href{http://dx.doi.org/10.1088/0264-9381/27/20/205001}{{\em Class. Quant.
  Grav.} {\bf 27} (2010)  205001},
\href{http://arxiv.org/abs/1005.5730}{{\tt arXiv:1005.5730 [gr-qc]}}.

\bibitem{Levi:2010}
M.~Levi, ``Next-to-leading order gravitational spin-orbit coupling in an
  effective field theory approach,''
  \href{http://dx.doi.org/10.1103/PhysRevD.82.104004}{{\em Phys. Rev. D} {\bf
  82} (2010)  104004},
\href{http://arxiv.org/abs/1006.4139}{{\tt arXiv:1006.4139 [gr-qc]}}.

\bibitem{Porto:Rothstein:2008:1}
R.~A. Porto and I.~Z. Rothstein, ``Spin(1)spin(2) effects in the motion of
  inspiralling compact binaries at third order in the post-{N}ewtonian
  expansion,'' \href{http://dx.doi.org/10.1103/PhysRevD.78.044012}{{\em Phys.
  Rev. D} {\bf 78} (2008)  044012},
\href{http://arxiv.org/abs/0802.0720}{{\tt arXiv:0802.0720 [gr-qc]}}.

\bibitem{Porto:Rothstein:2008:1:err}
R.~A. Porto and I.~Z. Rothstein, ``Erratum: {S}pin(1)spin(2) effects in the
  motion of inspiralling compact binaries at third order in the
  post-{N}ewtonian expansion,''
  \href{http://dx.doi.org/10.1103/PhysRevD.81.029904}{{\em Phys. Rev. D} {\bf
  81} (2010)  029904(E)}.

\bibitem{Levi:2008}
M.~Levi, ``Next-to-leading order gravitational spin1-spin2 coupling with
  {K}aluza-{K}lein reduction,''
  \href{http://dx.doi.org/10.1103/PhysRevD.82.064029}{{\em Phys. Rev. D} {\bf
  82} (2010)  064029},
\href{http://arxiv.org/abs/0802.1508}{{\tt arXiv:0802.1508 [gr-qc]}}.

\bibitem{Wang:Steinhoff:Zeng:Schafer:2011}
H.~Wang, J.~Steinhoff, J.~Zeng, and G.~Sch{\"a}fer, ``Leading-order spin-orbit
  and spin(1)-spin(2) radiation-reaction {H}amiltonians,''
  \href{http://dx.doi.org/10.1103/PhysRevD.84.124005}{{\em Phys. Rev. D} {\bf
  84} (2011)  124005}, \href{http://arxiv.org/abs/1109.1182}{{\tt
  arXiv:1109.1182 [gr-qc]}}.

\bibitem{Hergt:Schafer:2008:2}
S.~Hergt and G.~Sch{\"a}fer, ``Higher-order-in-spin interaction {H}amiltonians
  for binary black holes from source terms of {K}err geometry in approximate
  {ADM} coordinates,'' \href{http://dx.doi.org/10.1103/PhysRevD.77.104001}{{\em
  Phys. Rev. D} {\bf 77} (2008)  104001},
\href{http://arxiv.org/abs/0712.1515}{{\tt arXiv:0712.1515 [gr-qc]}}.

\bibitem{Hergt:Schafer:2008}
S.~Hergt and G.~Sch{\"a}fer, ``Higher-order-in-spin interaction {H}amiltonians
  for binary black holes from {P}oincar{\'e} invariance,''
  \href{http://dx.doi.org/10.1103/PhysRevD.78.124004}{{\em Phys. Rev. D} {\bf
  78} (2008)  124004},
\href{http://arxiv.org/abs/0809.2208}{{\tt arXiv:0809.2208 [gr-qc]}}.

\bibitem{Poisson:1998}
E.~Poisson, ``Gravitational waves from inspiraling compact binaries: {T}he
  quadrupole-moment term,''
  \href{http://dx.doi.org/10.1103/PhysRevD.57.5287}{{\em Phys. Rev. D} {\bf 57}
  (1998)  5287--5290},
\href{http://arxiv.org/abs/gr-qc/9709032}{{\tt arXiv:gr-qc/9709032}}.

\bibitem{Porto:Rothstein:2008:2}
R.~A. Porto and I.~Z. Rothstein, ``Next to leading order spin(1)spin(1) effects
  in the motion of inspiralling compact binaries,''
  \href{http://dx.doi.org/10.1103/PhysRevD.78.044013}{{\em Phys. Rev. D} {\bf
  78} (2008)  044013},
\href{http://arxiv.org/abs/0804.0260}{{\tt arXiv:0804.0260 [gr-qc]}}.

\bibitem{Porto:Rothstein:2008:2:err}
R.~A. Porto and I.~Z. Rothstein, ``Erratum: {N}ext to leading order
  spin(1)spin(1) effects in the motion of inspiralling compact binaries,''
  \href{http://dx.doi.org/10.1103/PhysRevD.81.029905}{{\em Phys. Rev. D} {\bf
  81} (2010)  029905(E)}.

\bibitem{Steinhoff:Schafer:2009:1}
J.~Steinhoff and G.~Sch{\"a}fer, ``Comment on two recent papers regarding
  next-to-leading order spin-spin effects in gravitational interaction,''
  \href{http://dx.doi.org/10.1103/PhysRevD.80.088501}{{\em Phys. Rev. D} {\bf
  80} (2009)  088501},
\href{http://arxiv.org/abs/0903.4772}{{\tt arXiv:0903.4772 [gr-qc]}}.

\bibitem{Hergt:Steinhoff:Schafer:2010:1}
S.~Hergt, J.~Steinhoff, and G.~Sch{\"a}fer, ``The reduced {H}amiltonian for
  next-to-leading-order spin-squared dynamics of general compact binaries,''
  \href{http://dx.doi.org/10.1088/0264-9381/27/13/135007}{{\em Class. Quant.
  Grav.} {\bf 27} (2010)  135007},
\href{http://arxiv.org/abs/1002.2093}{{\tt arXiv:1002.2093 [gr-qc]}}.

\bibitem{Schafer:1987}
G.~Sch{\"a}fer, ``Three-body {H}amiltonian in {G}eneral {R}elativity,''
  \href{http://dx.doi.org/10.1016/0375-9601(87)90389-6}{{\em Phys. Lett. A}
  {\bf 123} (1987)  336--339}.

\bibitem{Ohta:Kimura:Hiida:1975}
T.~Ohta, T.~Kimura, and K.~Hiida, ``Can the effect of distant matter on
  physical observables be observed?,''
  \href{http://dx.doi.org/10.1007/BF02726342}{{\em Nuovo Cim. B} {\bf 27}
  (1975)  103--120}.

\bibitem{Lousto:Nakano:2008}
C.~O. Lousto and H.~Nakano, ``Three-body equations of motion in successive
  post-{N}ewtonian approximations,''
  \href{http://dx.doi.org/10.1088/0264-9381/25/19/195019}{{\em Class. Quant.
  Grav.} {\bf 25} (2008)  195019},
\href{http://arxiv.org/abs/0710.5542}{{\tt arXiv:0710.5542 [gr-qc]}}.

\bibitem{Chu:2009}
Y.-Z. Chu, ``n-body problem in general relativity up to the second
  post-{N}ewtonian order from perturbative field theory,''
  \href{http://dx.doi.org/10.1103/PhysRevD.79.044031}{{\em Phys. Rev. D} {\bf
  79} (2009)  044031}, \href{http://arxiv.org/abs/0812.0012}{{\tt
  arXiv:0812.0012 [gr-qc]}}.

\bibitem{Hartung:Steinhoff:2010}
J.~Hartung and J.~Steinhoff, ``Next-to-leading order spin-orbit and
  spin(a)-spin(b) {H}amiltonians for $n$ gravitating spinning compact
  objects,'' \href{http://dx.doi.org/10.1103/PhysRevD.83.044008}{{\em Phys.
  Rev. D} {\bf 83} (2011)  044008},
\href{http://arxiv.org/abs/1011.1179}{{\tt arXiv:1011.1179 [gr-qc]}}.

\bibitem{Damour:Nagar:2009}
T.~Damour and A.~Nagar, ``Effective one body description of tidal effects in
  inspiralling compact binaries,''
  \href{http://dx.doi.org/10.1103/PhysRevD.81.084016}{{\em Phys. Rev. D} {\bf
  81} (2010)  084016},
\href{http://arxiv.org/abs/0911.5041}{{\tt arXiv:0911.5041 [gr-qc]}}.

\bibitem{Vines:Flanagan:2010}
J.~E. Vines and {\'E}.~E. Flanagan, ``Post-1-{N}ewtonian quadrupole tidal
  interactions in binary systems,''
\href{http://arxiv.org/abs/1009.4919}{{\tt arXiv:1009.4919 [gr-qc]}}.

\bibitem{Bini:Damour:Faye:2012}
D.~Bini, T.~Damour, and G.~Faye, ``{E}ffective action approach to higher-order
  relativistic tidal interactions in binary systems and their effective one
  body description,'' \href{http://dx.doi.org/10.1103/PhysRevD.85.124034}{{\em
  Phys. Rev. D} {\bf 85} (2012)  124034},
\href{http://arxiv.org/abs/1202.3565}{{\tt arXiv:1202.3565 [gr-qc]}}.

\bibitem{Steinhoff:Puetzfeld:2012}
J.~Steinhoff and D.~Puetzfeld, ``{I}nfluence of internal structure on the
  motion of test bodies in extreme mass ratio situations,''
  \href{http://dx.doi.org/10.1103/PhysRevD.86.044033}{{\em Phys. Rev. D} {\bf
  86} (2012)  044033},
\href{http://arxiv.org/abs/1205.3926}{{\tt arXiv:1205.3926 [gr-qc]}}.

\bibitem{Steinhoff:Schafer:2009:2}
J.~Steinhoff and G.~Sch{\"a}fer, ``Canonical formulation of self-gravitating
  spinning-object systems,''
  \href{http://dx.doi.org/10.1209/0295-5075/87/50004}{{\em Europhys. Lett.}
  {\bf 87} (2009)  50004},
\href{http://arxiv.org/abs/0907.1967}{{\tt arXiv:0907.1967 [gr-qc]}}.

\bibitem{Steinhoff:2011}
J.~Steinhoff, ``Canonical formulation of spin in general relativity,''
  \href{http://dx.doi.org/10.1002/andp.201000178}{{\em Ann. Phys. (Berlin)}
  {\bf 523} (2011)  296--353},
\href{http://arxiv.org/abs/1106.4203}{{\tt arXiv:1106.4203 [gr-qc]}}.

\bibitem{Steinhoff:Wang:2009}
J.~Steinhoff and H.~Wang, ``Canonical formulation of gravitating spinning
  objects at 3.5 post-{N}ewtonian order,''
  \href{http://dx.doi.org/10.1103/PhysRevD.81.024022}{{\em Phys. Rev. D} {\bf
  81} (2010)  024022},
\href{http://arxiv.org/abs/0910.1008}{{\tt arXiv:0910.1008 [gr-qc]}}.

\bibitem{Steinhoff:Schafer:Hergt:2008}
J.~Steinhoff, G.~Sch{\"a}fer, and S.~Hergt, ``{ADM} canonical formalism for
  gravitating spinning objects,''
  \href{http://dx.doi.org/10.1103/PhysRevD.77.104018}{{\em Phys. Rev. D} {\bf
  77} (2008)  104018},
\href{http://arxiv.org/abs/0805.3136}{{\tt arXiv:0805.3136 [gr-qc]}}.

\bibitem{Hartung:Steinhoff:2011:1}
J.~Hartung and J.~Steinhoff, ``Next-to-next-to-leading order post-{N}ewtonian
  spin-orbit {H}amiltonian for self-gravitating binaries,''
  \href{http://dx.doi.org/10.1002/andp.201100094}{{\em Ann. Phys. (Berlin)}
  {\bf 523} (2011)  783--790},
\href{http://arxiv.org/abs/1104.3079}{{\tt arXiv:1104.3079 [gr-qc]}}.

\bibitem{Hartung:Steinhoff:2011:2}
J.~Hartung and J.~Steinhoff, ``{N}ext-to-next-to-leading order post-{N}ewtonian
  spin(1)-spin(2) {H}amiltonian for self-gravitating binaries,''
  \href{http://dx.doi.org/10.1002/andp.201100163}{{\em Ann. Phys. (Berlin)}
  {\bf 523} (2011)  919--924},
\href{http://arxiv.org/abs/1107.4294}{{\tt arXiv:1107.4294 [gr-qc]}}.

\bibitem{Hartung:2012}
J.~Hartung, {\em {B}in{\"a}rsysteme kompakter {O}bjekte in hoher
  postnewtonscher {N}{\"a}herung in den {E}igendrehimpuls-{W}echselwirkungen}.
\newblock PhD thesis, Friedrich-Schiller-Universit{\"a}t Jena, 2012.

\bibitem{Levi:2011}
M.~Levi, ``{B}inary dynamics from spin1-spin2 coupling at fourth
  post-{N}ewtonian order,''
  \href{http://dx.doi.org/10.1103/PhysRevD.85.064043}{{\em Phys. Rev. D} {\bf
  85} (2012)  064043},
\href{http://arxiv.org/abs/1107.4322}{{\tt arXiv:1107.4322 [gr-qc]}}.

\bibitem{Hergt:2011}
S.~Hergt, {\em {H}amiltonsche {F}ormulierung und {B}ehandlung nichtlinearer
  {E}igendrehimpulsbeitr{\"a}ge in {B}in{\"a}rsystemen der {A}llgemeinen
  {R}elativit{\"a}tstheorie}.
\newblock PhD thesis, Friedrich-Schiller-Universit{\"a}t Jena, 2011.

\bibitem{Hergt:Steinhoff:Schafer:2011}
S.~Hergt, J.~Steinhoff, and G.~Sch{\"a}fer, ``Elimination of the spin
  supplementary condition in the effective field theory approach to the
  post-{N}ewtonian approximation,''
  \href{http://dx.doi.org/10.1016/j.aop.2012.02.006}{{\em Ann. Phys. (N.Y.)}
  {\bf 327} (2012)  1494--1537},
\href{http://arxiv.org/abs/1110.2094}{{\tt arXiv:1110.2094 [gr-qc]}}.

\bibitem{Marsat:Bohe:Faye:Blanchet:2012}
S.~Marsat, A.~Bohe, G.~Faye, and L.~Blanchet, ``Next-to-next-to-leading order
  spin-orbit effects in the equations of motion of compact binary systems,''
  \href{http://dx.doi.org/10.1088/0264-9381/30/5/055007}{{\em Class. Quant.
  Grav.} {\bf 30} (2013)  055007},
\href{http://arxiv.org/abs/1210.4143}{{\tt arXiv:1210.4143 [gr-qc]}}.

\bibitem{Bohe:Marsat:Faye:Blanchet:2012}
A.~Bohe, S.~Marsat, G.~Faye, and L.~Blanchet, ``Next-to-next-to-leading order
  spin-orbit effects in the near-zone metric and precession equations of
  compact binaries,''
\href{http://arxiv.org/abs/1212.5520}{{\tt arXiv:1212.5520 [gr-qc]}}.

\bibitem{Ledvinka:Schafer:Bicak:2008}
T.~Ledvinka, G.~Sch{\"a}fer, and J.~Bi\v{c}{\'a}k, ``Relativistic closed-form
  {H}amiltonian for many-body gravitating systems in the post-{M}inkowskian
  approximation,'' \href{http://dx.doi.org/10.1103/PhysRevLett.100.251101}{{\em
  Phys. Rev. Lett.} {\bf 100} (2008)  251101},
\href{http://arxiv.org/abs/0807.0214}{{\tt arXiv:0807.0214 [gr-qc]}}.

\bibitem{Tessmer:2011}
M.~Tessmer, {\em {M}otion and gravitational wave emission of spinning compact
  binaries}.
\newblock PhD thesis, Friedrich-Schiller-Universit{\"a}t Jena, 2011.

\bibitem{Damour:Deruelle:1985}
T.~Damour and N.~Deruelle, ``{G}eneral relativistic celestial mechanics of
  binary systems. {I}. the post-{N}ewtonian motion.,'' {\em Ann. Inst. H.
  Poincar{\'e} A} {\bf 43} (1985)  107--132.
  \url{http://www.numdam.org/item?id=AIHPA_1985__43_1_107_0}.

\bibitem{Schafer:Wex:1993}
G.~Sch{\"a}fer and N.~Wex, ``{S}econd post-{N}ewtonian motion of compact
  binaries,'' \href{http://dx.doi.org/doi:10.1016/0375-9601(93)90758-R}{{\em
  Phys. Lett. A} {\bf 174} (1993)  196--205}.

\bibitem{Schafer:Wex:1993:err}
G.~Sch{\"a}fer and N.~Wex, ``Erratum: {S}econd post-{N}ewtonian motion of
  compact binaries,''
  \href{http://dx.doi.org/10.1016/0375-9601(93)90980-E}{{\em Phys. Lett. A}
  {\bf 177} (1993)  461(E)}.

\bibitem{Memmesheimer:Gopakumar:Schafer:2004}
R.-M. Memmesheimer, A.~Gopakumar, and G.~Sch{\"a}fer, ``{T}hird
  post-{N}ewtonian accurate generalized quasi-{K}eplerian parametrization for
  compact binaries in eccentric orbits,''
  \href{http://dx.doi.org/10.1103/PhysRevD.70.104011}{{\em Phys. Rev. D} {\bf
  70} (2004)  104011},
\href{http://arxiv.org/abs/gr-qc/0407049}{{\tt arXiv:gr-qc/0407049}}.

\bibitem{Wex:1995}
N.~Wex, ``The second post-{N}ewtonian motion of compact binary-star systems
  with spin,'' \href{http://dx.doi.org/10.1088/0264-9381/12/4/009}{{\em Class.
  Quant. Grav.} {\bf 12} (1995)  983--1005}.

\bibitem{Konigsdorffer:Gopakumar:2005}
C.~K{\"o}nigsd{\"o}rffer and A.~Gopakumar, ``{P}ost-{N}ewtonian accurate
  parametric solution to the dynamics of spinning compact binaries in eccentric
  orbits: {T}he leading order spin-orbit interaction,''
  \href{http://dx.doi.org/10.1103/PhysRevD.71.024039}{{\em Phys. Rev. D} {\bf
  71} (2005)  024039}, \href{http://arxiv.org/abs/gr-qc/0501011}{{\tt
  arXiv:gr-qc/0501011}}.

\bibitem{Konigsdorffer:Gopakumar:2006}
C.~K{\"o}nigsd{\"o}rffer and A.~Gopakumar, ``{P}hasing of gravitational waves
  from inspiralling eccentric binaries at the third-and-a-half post-{N}ewtonian
  order,'' \href{http://dx.doi.org/10.1103/PhysRevD.73.124012}{{\em Phys. Rev.
  D} {\bf 73} (2006)  124012}, \href{http://arxiv.org/abs/gr-qc/0603056}{{\tt
  arXiv:gr-qc/0603056}}.

\bibitem{Tessmer:2009}
M.~Tessmer, ``{G}ravitational waveforms from unequal-mass binaries with
  arbitrary spins under leading order spin-orbit coupling,''
  \href{http://dx.doi.org/10.1103/PhysRevD.80.124034}{{\em Phys. Rev. D} {\bf
  80} (2009)  124034},
\href{http://arxiv.org/abs/0910.5931}{{\tt arXiv:0910.5931 [gr-qc]}}.

\bibitem{Tessmer:Hartung:Schafer:2010}
M.~Tessmer, J.~Hartung, and G.~Sch{\"a}fer, ``{M}otion and gravitational wave
  forms of eccentric compact binaries with orbital-angular-momentum-aligned
  spins under next-to-leading order in spin--orbit and leading order in
  spin(1)--spin(2) and spin-squared couplings,''
  \href{http://dx.doi.org/10.1088/0264-9381/27/16/165005}{{\em Class. Quant.
  Grav.} {\bf 27} (2010)  165005},
\href{http://arxiv.org/abs/1003.2735}{{\tt arXiv:1003.2735 [gr-qc]}}.

\bibitem{Tessmer:Hartung:Schafer:2012}
M.~Tessmer, J.~Hartung, and G.~Sch{\"a}fer, ``{A}ligned spins: Orbital
  elements, decaying orbits, and last stable circular orbit to high
  post-{N}ewtonian orders,''
  \href{http://dx.doi.org/10.1088/0264-9381/30/1/015007}{{\em Class. Quant.
  Grav.} {\bf 30} (2013)  015007}, \href{http://arxiv.org/abs/1207.6961}{{\tt
  arXiv:1207.6961 [gr-qc]}}.

\bibitem{Bonnor:1959}
W.~B. Bonnor, ``{S}pherical gravitational waves,''
  \href{http://dx.doi.org/10.1098/rsta.1959.0003}{{\em Phil. Trans. R. Soc. A}
  {\bf 251} (1959)  233--271}.

\bibitem{Thorne:1980}
K.~S. Thorne, ``Multipole expansions of gravitational radiation,''
\href{http://dx.doi.org/10.1103/RevModPhys.52.299}{{\em Rev. Mod. Phys.} {\bf
  52} (1980)  299--339}.

\bibitem{Blanchet:Damour:1986}
L.~Blanchet and T.~Damour, ``Radiative gravitational fields in general
  relativity {I}. {G}eneral structure of the field outside the source,'' {\em
  Phil. Trans. R. Soc. A} {\bf 320} (1986)  379--430.
  \url{http://www.jstor.org/stable/37878}.

\bibitem{Blanchet:Schafer:1989}
L.~Blanchet and G.~Sch{\"a}fer, ``{H}igher order gravitational radiation losses
  in binary systems,'' {\em Mon. Not. R. Astron. Soc.} {\bf 239} (1989)
  845--867. \url{http://adsabs.harvard.edu/abs/1989MNRAS.239..845B}.

\bibitem{Blanchet:Schafer:1989:err}
L.~Blanchet and G.~Sch{\"a}fer, ``{H}igher order gravitational radiation losses
  in binary systems: Erratum,'' {\em Mon. Not. R. Astron. Soc.} {\bf 242}
  (1990)  704.

\bibitem{Kidder:1995}
L.~E. Kidder, ``Coalescing binary systems of compact objects to
  (post)$^{5/2}$-{N}ewtonian order. {V}. {S}pin effects,''
  \href{http://dx.doi.org/10.1103/PhysRevD.52.821}{{\em Phys. Rev. D} {\bf 52}
  (1995)  821--847},
\href{http://arxiv.org/abs/gr-qc/9506022}{{\tt arXiv:gr-qc/9506022}}.

\bibitem{Blanchet:Buonanno:Faye:2006}
L.~Blanchet, A.~Buonanno, and G.~Faye, ``{H}igher-order spin effects in the
  dynamics of compact binaries. {II}. {R}adiation field,''
  \href{http://dx.doi.org/10.1103/PhysRevD.74.104034}{{\em Phys. Rev. D} {\bf
  74} (2006)  104034},
\href{http://arxiv.org/abs/gr-qc/0605140}{{\tt arXiv:gr-qc/0605140}}.

\bibitem{Blanchet:Buonanno:Faye:2006:err}
L.~Blanchet, A.~Buonanno, and G.~Faye, ``Erratum: {H}igher-order spin effects
  in the dynamics of compact binaries. {II}. {R}adiation field,''
  \href{http://dx.doi.org/10.1103/PhysRevD.75.049903}{{\em Phys. Rev. D} {\bf
  75} (2007)  049903(E)}.

\bibitem{Blanchet:Buonanno:Faye:2006:err:2}
L.~Blanchet, A.~Buonanno, and G.~Faye, ``{E}rratum: {H}igher-order spin effects
  in the dynamics of compact binaries. {II}. {R}adiation field,''
  \href{http://dx.doi.org/10.1103/PhysRevD.81.089901}{{\em Phys. Rev. D} {\bf
  81} (2010)  089901(E)}.

\bibitem{Buonanno:Faye:Hinderer:2012}
A.~Buonanno, G.~Faye, and T.~Hinderer, ``Spin effects on gravitational waves
  from inspiraling compact binaries at second post-{N}ewtonian order,''
\href{http://arxiv.org/abs/1209.6349}{{\tt arXiv:1209.6349 [gr-qc]}}.

\bibitem{Blanchet:Buonanno:Faye:2011}
L.~Blanchet, A.~Buonanno, and G.~Faye, ``{T}ail-induced spin-orbit effect in
  the gravitational radiation of compact binaries,''
  \href{http://dx.doi.org/10.1103/PhysRevD.84.064041}{{\em Phys. Rev. D} {\bf
  84} (2011)  064041},
\href{http://arxiv.org/abs/1104.5659}{{\tt arXiv:1104.5659 [gr-qc]}}.

\bibitem{Porto:Ross:Rothstein:2012}
R.~A. Porto, A.~Ross, and I.~Z. Rothstein, ``{S}pin induced multipole moments
  for the gravitational wave amplitude from binary inspirals to 2.5
  post-{N}ewtonian order,''
  \href{http://dx.doi.org/10.1088/1475-7516/2012/09/028}{{\em JCAP} {\bf 1209}
  (2012)  028},
\href{http://arxiv.org/abs/1203.2962}{{\tt arXiv:1203.2962 [gr-qc]}}.

\bibitem{Porto:Ross:Rothstein:2010}
R.~A. Porto, A.~Ross, and I.~Z. Rothstein, ``{S}pin induced multipole moments
  for the gravitational wave flux from binary inspirals to third
  post-{N}ewtonian order,''
  \href{http://dx.doi.org/10.1088/1475-7516/2011/03/009}{{\em JCAP} {\bf 1103}
  (2011)  009}, \href{http://arxiv.org/abs/1007.1312}{{\tt arXiv:1007.1312
  [gr-qc]}}.

\bibitem{Turner:Will:1978}
M.~Turner and C.~M. Will, ``{P}ost-{N}ewtonian gravitation bremsstrahlung,''
  \href{http://dx.doi.org/10.1086/155996}{{\em ApJ} {\bf 220} (1978)  1107}.

\bibitem{Galtsov:Matiukhin:Petukhov:1980}
D.~V. Gal'tsov, A.~A. Matiukhin, and V.~I. Petukhov, ``{R}elativistic
  corrections to the gravitational radiation of a binary system and the fine
  structure of the spectrum,''
  \href{http://dx.doi.org/10.1016/0375-9601(80)90728-8}{{\em Phys. Lett. A}
  {\bf 77} (1980)  387}.

\bibitem{Pierro:Pinto:Spallicci:Laserra:Recano:2001}
V.~Pierro, I.~M. Pinto, E.~Spallicci, E.~Laserra, and F.~Recano, ``{F}ast and
  accurate computational tools for gravitational waveforms from binary stars
  with any orbital eccentricity,''
  \href{http://dx.doi.org/10.1046/j.1365-8711.2001.04442.x}{{\em Mon. Not. R.
  Astron. Soc.} {\bf 325} (2001)  358--372}.

\bibitem{Junker:Schafer:1992}
W.~Junker and G.~Sch{\"a}fer, ``{B}inary systems - higher order gravitational
  radiation damping and wave emission,'' {\em Mon. Not. R. Astron. Soc.} {\bf
  254} (1992)  146--164.
  \url{http://adsabs.harvard.edu/abs/1992MNRAS.254..146J}.

\bibitem{MorenoGarrido:Buitrago:Mediavilla:1994}
C.~Moreno-Garrido, J.~Buitrago, and E.~Mediavilla, ``{S}pectral analysis of the
  gravitational radiation emitted by binary systems in moderately eccentric
  orbits - application to coalescing binaries,'' {\em Mon. Not. R. Astron.
  Soc.} {\bf 266} (1994)  16.
  \url{http://adsabs.harvard.edu/abs/1994MNRAS.266...16M}.

\bibitem{MorenoGarrido:Mediavilla:Buitrago:1995}
C.~Moreno-Garrido, E.~Mediavilla, and J.~Buitrago, ``{G}ravitational radiation
  from point masses in elliptical orbits: spectral analysis and orbital
  parameters,'' {\em Mon. Not. R. Astron. Soc.} {\bf 274} (1995)  115--126.
  \url{http://adsabs.harvard.edu/abs/1995MNRAS.274..115M}.

\bibitem{Gopakumar:Iyer:2002}
A.~Gopakumar and B.~R. Iyer, ``{S}econd post-{N}ewtonian gravitational wave
  polarizations for compact binaries in elliptical orbits,''
  \href{http://dx.doi.org/10.1103/PhysRevD.65.084011}{{\em Phys. Rev. D} {\bf
  65} (2002)  084011}, \href{http://arxiv.org/abs/0110100}{{\tt arXiv:0110100
  [gr-qc]}}.

\bibitem{Tessmer:Gopakumar:2006}
M.~Tessmer and A.~Gopakumar, ``{A}ccurate and efficient gravitational waveforms
  for certain galactic compact binaries,''
  \href{http://dx.doi.org/10.1111/j.1365-2966.2006.11179.x}{{\em Mon. Not. R.
  Astron. Soc.} {\bf 374} (2007)  721--728},
  \href{http://arxiv.org/abs/gr-qc/0610139}{{\tt arXiv:gr-qc/0610139}}.

\bibitem{Tessmer:Schafer:2010}
M.~Tessmer and G.~Sch{\"a}fer, ``{F}ull-analytic frequency-domain
  1p{N}-accurate gravitational wave forms from eccentric compact binaries,''
  \href{http://dx.doi.org/10.1103/PhysRevD.82.124064}{{\em Phys. Rev. D} {\bf
  82} (2010)  124064}, \href{http://arxiv.org/abs/1006.3714v2}{{\tt
  arXiv:1006.3714v2 [gr-qc]}}.

\bibitem{Tessmer:Schafer:2011}
M.~Tessmer and G.~Sch{\"a}fer, ``{F}ull-analytic frequency-domain gravitational
  wave forms from eccentric compact binaries to {2PN} accuracy,''
  \href{http://dx.doi.org/10.1002/andp.201100007}{{\em Ann. Phys. (Berlin)}
  {\bf 523} (2011)  813--864}, \href{http://arxiv.org/abs/1012.3894}{{\tt
  arXiv:1012.3894 [gr-qc]}}.

\bibitem{Gopakumar:Iyer:1997}
A.~Gopakumar and B.~R. Iyer, ``{Gravitational waves from inspiraling compact
  binaries: Angular momentum flux, evolution of the orbital elements, and the
  waveform to the second post-Newtonian order},''
  \href{http://dx.doi.org/10.1103/PhysRevD.56.7708}{{\em Phys. Rev. D} {\bf 56}
  (1997)  7708--7731}, \href{http://arxiv.org/abs/gr-qc/0110100}{{\tt
  arXiv:gr-qc/0110100}}.

\bibitem{Damour:Gopakumar:Iyer:2004}
T.~Damour, A.~Gopakumar, and B.~R. Iyer, ``{P}hasing of gravitational waves
  from inspiralling eccentric binaries,''
  \href{http://dx.doi.org/10.1103/PhysRevD.70.064028}{{\em Phys. Rev. D} {\bf
  70} (2004)  064028}, \href{http://arxiv.org/abs/gr-qc/0404128}{{\tt
  arXiv:gr-qc/0404128}}.

\bibitem{Chandrasekhar:Esposito:1970}
S.~Chandrasekhar and F.~P. Esposito, ``{The 2 1/2-Post-Newtonian Equations of
  Hydrodynamics and Radiation Reaction in General Relativity},''
  \href{http://dx.doi.org/10.1086/150414}{{\em ApJ} {\bf 160} (1970)  153}.

\bibitem{Finn:1992}
L.~S. Finn, ``{D}etection, measurement, and gravitational radiation,''
  \href{http://dx.doi.org/10.1103/PhysRevD.46.5236}{{\em Phys. Rev. D} {\bf 46}
  (1992)  5236--5249}, \href{http://arxiv.org/abs/gr-qc/9609027}{{\tt
  arXiv:gr-qc/9609027}}.

\bibitem{Apostolatos:1995}
T.~A. Apostolatos, ``{S}earch templates for gravitational waves from
  precessing, inspiraling binaries,''
  \href{http://dx.doi.org/10.1103/PhysRevD.52.605}{{\em Phys. Rev. D} {\bf 52}
  (1995)  605--620}.

\bibitem{Damour:Iyer:Sathyaprakash:2000}
T.~Damour, B.~R. Iyer, and B.~S. Sathyaprakash, ``{F}requency-domain
  p-approximant filters for time-truncated inspiral gravitational wave signals
  from compact binaries,''
  \href{http://dx.doi.org/10.1103/PhysRevD.62.084036}{{\em Phys. Rev. D} {\bf
  62} (2000)  084036}, \href{http://arxiv.org/abs/gr-qc/0001023}{{\tt
  arXiv:gr-qc/0001023}}.

\bibitem{Damour:Iyer:Sathyaprakash:2001}
T.~Damour, B.~R. Iyer, and B.~S. Sathyaprakash, ``{C}omparison of search
  templates for gravitational waves from binary inspiral,''
  \href{http://dx.doi.org/10.1103/PhysRevD.63.044023}{{\em Phys. Rev. D} {\bf
  63} (2001)  044023}, \href{http://arxiv.org/abs/gr-qc/0010009}{{\tt
  arXiv:gr-qc/0010009}}.

\bibitem{Damour:Iyer:Sathyaprakash:2001:err}
T.~Damour, B.~R. Iyer, and B.~S. Sathyaprakash, ``Erratum: {C}omparison of
  search templates for gravitational waves from binary inspiral,''
  \href{http://dx.doi.org/10.1103/PhysRevD.72.029902}{{\em Phys. Rev. D} {\bf
  72} (2005)  029902(E)}.

\bibitem{Ajith:Babak:Chen:others:2008}
P.~Ajith, S.~Babak, Y.~Chen, M.~Hewitson, B.~Krishnan, A.~M. Sintes, J.~T.
  Whelan, B.~Br{\"u}gmann, P.~Diener, N.~Dorband, J.~Gonzalez, M.~Hannam,
  S.~Husa, D.~Pollney, L.~Rezzolla, L.~Santamar{\'i}a, U.~Sperhake, and
  J.~Thornburg, ``{T}emplate bank for gravitational waveforms from coalescing
  binary black holes: {N}onspinning binaries,''
  \href{http://dx.doi.org/10.1103/PhysRevD.77.104017}{{\em Phys. Rev. D} {\bf
  77} (2008)  104017}, \href{http://arxiv.org/abs/0710.2335}{{\tt
  arXiv:0710.2335 [gr-qc]}}.

\bibitem{Ajith:Babak:Chen:others:2008:err}
P.~Ajith, S.~Babak, Y.~Chen, M.~Hewitson, B.~Krishnan, A.~M. Sintes, J.~T.
  Whelan, B.~Br{\"u}gmann, P.~Diener, N.~Dorband, J.~Gonzalez, M.~Hannam,
  S.~Husa, D.~Pollney, L.~Rezzolla, L.~Santamar{\'i}a, U.~Sperhake, and
  J.~Thornburg, ``Erratum: {T}emplate bank for gravitational waveforms from
  coalescing binary black holes: {N}onspinning binaries,''
  \href{http://dx.doi.org/10.1103/PhysRevD.79.129901}{{\em Phys. Rev. D} {\bf
  79} (2009)  129901(E)}.

\bibitem{Buonanno:Chen:Vallisneri:2003}
A.~Buonanno, Y.~Chen, and M.~Vallisneri, ``{D}etecting gravitational waves from
  precessing binaries of spinning compact objects: {A}diabatic limit,''
  \href{http://dx.doi.org/10.1103/PhysRevD.67.104025}{{\em Phys. Rev. D} {\bf
  67} (2003)  104025}, \href{http://arxiv.org/abs/gr-qc/0211087}{{\tt
  arXiv:gr-qc/0211087}}.

\bibitem{Buonanno:Chen:Vallisneri:2003:err}
A.~Buonanno, Y.~Chen, and M.~Vallisneri, ``Erratum: {D}etecting gravitational
  waves from precessing binaries of spinning compact objects: {A}diabatic
  limit,'' \href{http://dx.doi.org/10.1103/PhysRevD.74.029904}{{\em Phys. Rev.
  D} {\bf 74} (2006)  029904(E)}.

\bibitem{Damour:Nagar:2009:2}
T.~Damour and A.~Nagar, ``An improved analytical description of inspiralling
  and coalescing black-hole binaries,''
  \href{http://dx.doi.org/10.1103/PhysRevD.79.081503}{{\em Phys. Rev. D} {\bf
  79} (2009)  081503(R)},
\href{http://arxiv.org/abs/0902.0136}{{\tt arXiv:0902.0136 [gr-qc]}}.

\bibitem{Buonanno:etal:2009}
A.~Buonanno, Y.~Pan, H.~P. Pfeiffer, M.~A. Scheel, L.~T. Buchman, and L.~E.
  Kidder, ``Effective-one-body waveforms calibrated to numerical relativity
  simulations: {C}oalescence of non-spinning, equal-mass black holes,''
  \href{http://dx.doi.org/10.1103/PhysRevD.79.124028}{{\em Phys. Rev. D} {\bf
  79} (2009)  124028},
\href{http://arxiv.org/abs/0902.0790}{{\tt arXiv:0902.0790 [gr-qc]}}.

\bibitem{Pan:etal:2009}
Y.~Pan, A.~Buonanno, L.~T. Buchman, T.~Chu, L.~E. Kidder, H.~P. Pfeiffer, and
  M.~A. Scheel, ``Effective-one-body waveforms calibrated to numerical
  relativity simulations: {C}oalescence of nonprecessing, spinning, equal-mass
  black holes,'' \href{http://dx.doi.org/10.1103/PhysRevD.81.084041}{{\em Phys.
  Rev. D} {\bf 81} (2010)  084041},
\href{http://arxiv.org/abs/0912.3466}{{\tt arXiv:0912.3466 [gr-qc]}}.

\bibitem{Damour:2001}
T.~Damour, ``Coalescence of two spinning black holes: {A}n effective one-body
  approach,'' \href{http://dx.doi.org/10.1103/PhysRevD.64.124013}{{\em Phys.
  Rev. D} {\bf 64} (2001)  124013},
\href{http://arxiv.org/abs/gr-qc/0103018}{{\tt arXiv:gr-qc/0103018}}.

\bibitem{Barausse:Buonanno:2009}
E.~Barausse and A.~Buonanno, ``Improved effective-one-body {H}amiltonian for
  spinning black-hole binaries,''
  \href{http://dx.doi.org/10.1103/PhysRevD.81.084024}{{\em Phys. Rev. D} {\bf
  81} (2010)  084024},
\href{http://arxiv.org/abs/0912.3517}{{\tt arXiv:0912.3517 [gr-qc]}}.

\bibitem{Nagar:2011}
A.~Nagar, ``{E}ffective one-body {H}amiltonian of two spinning black holes with
  next-to-next-to-leading order spin-orbit coupling,''
  \href{http://dx.doi.org/10.1103/PhysRevD.84.084028}{{\em Phys. Rev. D} {\bf
  84} (2011)  084028},
\href{http://arxiv.org/abs/1106.4349}{{\tt arXiv:1106.4349 [gr-qc]}}.

\bibitem{Barausse:Buonanno:2011}
E.~Barausse and A.~Buonanno, ``{E}xtending the effective-one-body {H}amiltonian
  of black-hole binaries to include next-to-next-to-leading spin-orbit
  couplings,'' \href{http://dx.doi.org/10.1103/PhysRevD.84.104027}{{\em Phys.
  Rev. D} {\bf 84} (2011)  104027},
\href{http://arxiv.org/abs/1107.2904}{{\tt arXiv:1107.2904 [gr-qc]}}.

\bibitem{Taracchini:Pan:Buonanno:Barausse:Boyle:Chu:Lovelace:Pfeiffer:Scheel:2012}
A.~Taracchini, Y.~Pan, A.~Buonanno, E.~Barausse, M.~Boyle, T.~Chu, G.~Lovelace,
  H.~P. Pfeiffer, and M.~A. Scheel, ``{A} prototype effective-one-body model
  for non-precessing spinning inspiral-merger-ringdown waveforms,''
  \href{http://arxiv.org/abs/1202.0790}{{\tt arXiv:1202.0790 [gr-qc]}}.

\bibitem{Damour:Jaranowski:Schafer:2000:2}
T.~Damour, P.~Jaranowski, and G.~Sch{\"a}fer, ``{D}ynamical invariants for
  general relativistic two-body systems at the third post-{N}ewtonian
  approximation,'' \href{http://dx.doi.org/10.1103/PhysRevD.62.044024}{{\em
  Phys. Rev. D} {\bf 62} (2000)  044024},
  \href{http://arxiv.org/abs/gr-qc/9912092}{{\tt arXiv:gr-qc/9912092}}.

\bibitem{Damour:Jaranowski:Schafer:2000:3}
T.~Damour, P.~Jaranowski, and G.~Sch{\"a}fer, ``{D}etermination of the last
  stable orbit for circular general relativistic binaries at the third
  post-{N}ewtonian approximation,''
  \href{http://dx.doi.org/10.1103/PhysRevD.62.084011}{{\em Phys. Rev. D} {\bf
  62} (2000)  084011}, \href{http://arxiv.org/abs/gr-qc/0005034}{{\tt
  arXiv:gr-qc/0005034}}.

\bibitem{Faber:2009}
J.~Faber, ``{S}tatus of neutron star--black hole and binary neutron star
  simulations,'' \href{http://dx.doi.org/10.1088/0264-9381/26/11/114004}{{\em
  Class. Quant. Grav.} {\bf 26} (2009)  114004}.

\bibitem{Duez:2010}
M.~D. Duez, ``{N}umerical relativity confronts compact neutron star binaries: a
  review and status report,''
  \href{http://dx.doi.org/10.1088/0264-9381/27/11/114002}{{\em Class. Quant.
  Grav.} {\bf 27} (2010)  114002}, \href{http://arxiv.org/abs/0912.3529}{{\tt
  arXiv:0912.3529 [gr-qc]}}.

\bibitem{Rosswog:2010}
S.~Rosswog, ``{C}ompact binary mergers: an astrophysical perspective,''
  \href{http://arxiv.org/abs/1012.0912}{{\tt arXiv:1012.0912 [astro-ph.HE]}}.

\bibitem{Shibata:Uryu:2000}
M.~Shibata and K.~Ury\={u}, ``{S}imulation of merging binary neutron stars in
  full general relativity: ${\Gamma=2}$ case,''
  \href{http://dx.doi.org/10.1103/PhysRevD.61.064001}{{\em Phys. Rev. D} {\bf
  61} (2000)  064001}, \href{http://arxiv.org/abs/gr-qc/9911058}{{\tt
  arXiv:gr-qc/9911058}}.

\bibitem{Thierfelder:Bernuzzi:Brugmann:2011}
M.~Thierfelder, S.~Bernuzzi, and B.~Br{\"u}gmann, ``{N}umerical relativity
  simulations of binary neutron stars,''
  \href{http://dx.doi.org/10.1103/PhysRevD.84.044012}{{\em Phys. Rev. D} {\bf
  84} (2011)  044012}, \href{http://arxiv.org/abs/1104.4751}{{\tt
  arXiv:1104.4751 [gr-qc]}}.

\bibitem{Gold:Bernuzzi:Thierfelder:Brugmann:Pretorius:2011}
R.~Gold, S.~Bernuzzi, M.~Thierfelder, B.~Br{\"u}gmann, and F.~Pretorius,
  ``{E}ccentric binary neutron star mergers,''
  \href{http://arxiv.org/abs/1109.5128}{{\tt arXiv:1109.5128 [gr-qc]}}.

\bibitem{Bernuzzi:Nagar:Thierfelder:Brugmann:2012}
S.~Bernuzzi, A.~Nagar, M.~Thierfelder, and B.~Br{\"u}gmann, ``{T}idal effects
  in binary neutron star coalescence,''
  \href{http://arxiv.org/abs/1205.3403}{{\tt arXiv:1205.3403 [gr-qc]}}.

\bibitem{Pretorius:2005}
F.~Pretorius, ``{E}volution of binary black-hole spacetimes,''
  \href{http://dx.doi.org/10.1103/PhysRevLett.95.121101}{{\em Phys. Rev. Lett.}
  {\bf 95} (2005)  121101}, \href{http://arxiv.org/abs/gr-qc/0507014}{{\tt
  arXiv:gr-qc/0507014}}.

\bibitem{Baker:Centrella:Choi:Koppitz:vanMeter:2006}
J.~G. Baker, J.~Centrella, D.-I. Choi, M.~Koppitz, and J.~{van Meter},
  ``{G}ravitational-wave extraction from an inspiraling configuration of
  merging black holes,''
  \href{http://dx.doi.org/10.1103/PhysRevLett.96.111102}{{\em Phys. Rev. Lett.}
  {\bf 96} (2006)  111102}, \href{http://arxiv.org/abs/gr-qc/0511103}{{\tt
  arXiv:gr-qc/0511103}}.

\bibitem{Lousto:Zlochower:2008}
C.~O. Lousto and Y.~Zlochower, ``Foundations of multiple-black-hole
  evolutions,'' \href{http://dx.doi.org/10.1103/PhysRevD.77.024034}{{\em Phys.
  Rev. D} {\bf 77} (2008)  024034}, \href{http://arxiv.org/abs/0711.1165}{{\tt
  arXiv:0711.1165 [gr-qc]}}.

\bibitem{Campanelli:Lousto:Zlochower:2008}
M.~Campanelli, C.~O. Lousto, and Y.~Zlochower, ``Close encounters of three
  black holes,'' \href{http://dx.doi.org/10.1103/PhysRevD.77.101501}{{\em Phys.
  Rev. D} {\bf 77} (2008)  101501(R)},
  \href{http://arxiv.org/abs/0710.0879}{{\tt arXiv:0710.0879 [gr-qc]}}.

\bibitem{Galaviz:Brugmann:Cao:2010}
P.~Galaviz, B.~Br{\"u}gmann, and Z.~Cao, ``Numerical evolution of multiple
  black holes with accurate initial data,''
  \href{http://dx.doi.org/10.1103/PhysRevD.82.024005}{{\em Phys. Rev. D} {\bf
  82} (2010)  024005},
\href{http://arxiv.org/abs/1004.1353}{{\tt arXiv:1004.1353 [gr-qc]}}.

\bibitem{Galaviz:2011}
P.~Galaviz, ``{S}tability and chaos of hierarchical three black hole
  configurations,'' \href{http://dx.doi.org/10.1103/PhysRevD.84.104038}{{\em
  Phys. Rev. D} {\bf 84} (2011)  104038},
  \href{http://arxiv.org/abs/1108.4485}{{\tt arXiv:1108.4485 [gr-qc]}}.

\bibitem{LeTiec:Barausse:Buonanno:2012}
A.~{Le Tiec}, E.~Barausse, and A.~Buonanno, ``{G}ravitational self-force
  correction to the binding energy of compact binary systems,''
  \href{http://dx.doi.org/10.1103/PhysRevLett.108.131103}{{\em Phys. Rev.
  Lett.} {\bf 108} (2012)  131103}, \href{http://arxiv.org/abs/1111.5609}{{\tt
  arXiv:1111.5609 [gr-qc]}}.

\bibitem{Blanchet:Buonanno:LeTiec:2012}
L.~Blanchet, A.~Buonanno, and A.~{Le Tiec}, ``First law of mechanics for black
  hole binaries with spins,''
\href{http://arxiv.org/abs/1211.1060}{{\tt arXiv:1211.1060 [gr-qc]}}.

\bibitem{MartinGarcia:2002}
J.~M. Mart{\'i}n-Garc{\'i}a, {\em x{A}ct: Efficient Tensor Computer Algebra}.
\newblock \url{http://www.xact.es/}.

\bibitem{Wolfram:2003}
S.~Wolfram, {\em The Mathematica Book}.
\newblock Wolfram Media, Champaign, IL, 5th~ed., 2003.

\bibitem{MartinGarcia:2008}
J.~M. Mart{\'i}n-Garc{\'i}a, ``x{P}erm: fast index canonicalization for tensor
  computer algebra,'' \href{http://dx.doi.org/10.1016/j.cpc.2008.05.009}{{\em
  Comp. Phys. Commun.} {\bf 179} (2008)  597--603},
  \href{http://arxiv.org/abs/0803.0862}{{\tt arXiv:0803.0862 [cs.SC]}}.

\bibitem{Brizuela:MartinGarcia:MenaMarugan:2009}
D.~Brizuela, J.~M. Mart{\'i}n-Garc{\'i}a, and G.~A. {Mena Marug{\'a}n},
  ``x{P}ert: computer algebra for metric perturbation theory,''
  \href{http://dx.doi.org/10.1007/s10714-009-0773-2}{{\em Gen. Relativ.
  Gravit.} {\bf 41} (2009)  2415--2431},
\href{http://arxiv.org/abs/0807.0824}{{\tt arXiv:0807.0824 [gr-qc]}}.

\bibitem{DeWitt:1967}
B.~S. DeWitt, ``Quantum theory of gravity. {I}. {T}he canonical theory,''
\href{http://dx.doi.org/10.1103/PhysRev.160.1113}{{\em Phys. Rev.} {\bf 160}
  (1967)  1113--1148}.

\bibitem{Regge:Teitelboim:1974}
T.~Regge and C.~Teitelboim, ``Role of surface integrals in the {H}amiltonian
  formulation of general relativity,''
\href{http://dx.doi.org/10.1016/0003-4916(74)90404-7}{{\em Ann. Phys. (N.Y.)}
  {\bf 88} (1974)  286--318}.

\bibitem{Misner:Thorne:Wheeler:1973}
C.~W. Misner, K.~S. Thorne, and J.~A. Wheeler, {\em Gravitation}.
\newblock W. H. FREEMAN AND COMPANY, 41 Madison Avenue, New York, 21~ed., 1973.

\bibitem{Poisson:2002}
E.~Poisson, ``{An advanced course in general relativity}.'' Electronic adress,
  2002.
\newblock \url{http://www.physics.uoguelph.ca/poisson/research/agr.pdf}. Draft
  of lecture notes.

\bibitem{Gourgoulhon:2007}
{\'E}.~Gourgoulhon, ``3+1 formalism and bases of numerical relativity,''
\href{http://arxiv.org/abs/gr-qc/0703035}{{\tt arXiv:gr-qc/0703035}}.

\bibitem{Mathisson:1937}
M.~Mathisson, ``{N}eue {M}echanik materieller {S}ysteme,''
{\em Acta Phys. Pol.} {\bf 6} (1937)  163--200.

\bibitem{Mathisson:2010}
M.~Mathisson, ``Republication of: {N}ew mechanics of material systems,''
\href{http://dx.doi.org/10.1007/s10714-010-0939-y}{{\em Gen. Relativ. Gravit.}
  {\bf 42} (2010)  1011--1048}.

\bibitem{Papapetrou:1951}
A.~Papapetrou, ``{S}pinning test-particles in general relativity. {I},''
\href{http://dx.doi.org/10.1098/rspa.1951.0200}{{\em Proc. R. Soc. A} {\bf 209}
  (1951)  248--258}.

\bibitem{Tulczyjew:1959}
W.~M. Tulczyjew, ``{M}otion of multipole particles in general relativity
  theory,'' {\em Acta Phys. Pol.} {\bf 18} (1959)  393--409.

\bibitem{Dixon:1979}
W.~G. Dixon, ``Extended bodies in general relativity: {T}heir description and
  motion,'' in {\em Proceedings of the International School of Physics Enrico
  Fermi LXVII}, J.~Ehlers, ed., pp.~156--219.
\newblock North Holland, Amsterdam, 1979.

\bibitem{Westpfahl:1967}
K.~Westpfahl, ``{R}elativistische {B}ewegungsprobleme. {I}. {D}as freie
  {S}pinteilchen,'' \href{http://dx.doi.org/10.1002/andp.19674750302}{{\em Ann.
  Phys. (Berlin)} {\bf 475} (1967)  113--135}.

\bibitem{Westpfahl:1969:1}
K.~Westpfahl, ``Relativistische {B}ewegungsprobleme. {V}. {Z}ur
  allgemein-relativistischen {D}ynamik klassischer {S}pinteilchen,''
  \href{http://dx.doi.org/10.1002/andp.19694770705}{{\em Ann. Phys. (Berlin)}
  {\bf 477} (1969)  345--360}.

\bibitem{Goenner:Westpfahl:1967}
H.~Goenner and K.~Westpfahl, ``Relativistische {B}ewegungsprobleme. {II}. {D}er
  starre {R}otator,'' \href{http://dx.doi.org/10.1002/andp.19674750505}{{\em
  Ann. Phys. (Berlin)} {\bf 475} (1967)  230--240}.

\bibitem{Romer:Westpfahl:1969}
H.~R{\"o}mer and K.~Westpfahl, ``{R}elativistische {B}ewegungsprobleme. {IV}.
  {R}otator-{S}pinteilchen in schwachen {G}ravitationsfeldern,''
  \href{http://dx.doi.org/10.1002/andp.19694770506}{{\em Ann. Phys. (Berlin)}
  {\bf 477} (1969)  264--276}.

\bibitem{Westpfahl:1969:2}
K.~Westpfahl, ``Relativistische {B}ewegungsprobleme. {VI}.
  {R}otator-{S}pinteilchen und allgemeine {R}elativit{\"a}tstheorie,''
  \href{http://dx.doi.org/10.1002/andp.19694770706}{{\em Ann. Phys. (Berlin)}
  {\bf 477} (1969)  361--371}.

\bibitem{Hanson:Regge:1974}
A.~J. Hanson and T.~Regge, ``{T}he relativistic spherical top,''
\href{http://dx.doi.org/10.1016/0003-4916(74)90046-3}{{\em Ann. Phys. (N.Y.)}
  {\bf 87} (1974)  498--566}.

\bibitem{Bailey:Israel:1975}
I.~Bailey and W.~Israel, ``Lagrangian dynamics of spinning particles and
  polarized media in general relativity,''
  \href{http://dx.doi.org/10.1007/BF01609434}{{\em Commun. math. Phys.} {\bf
  42} (1975)  65--82}.

\bibitem{Faye:Jaranowski:Schafer:2004}
G.~Faye, P.~Jaranowski, and G.~Sch{\"a}fer, ``Skeleton approximate solution of
  the einstein field equations for multiple black-hole systems,''
  \href{http://dx.doi.org/10.1103/PhysRevD.69.124029}{{\em Phys. Rev. D} {\bf
  69} (2004)  124029}, \href{http://arxiv.org/abs/gr-qc/0311018}{{\tt
  arXiv:gr-qc/0311018}}.

\bibitem{Xanthopoulos:Zannias:1989}
B.~C. Xanthopoulos and T.~Zannias, ``{E}instein gravity coupled to a massless
  scalar field in arbitrary spacetime dimensions,''
  \href{http://dx.doi.org/10.1103/PhysRevD.40.2564}{{\em Phys. Rev. D} {\bf 40}
  (1989)  2564--2567}.

\bibitem{Landau:Lifshitz:Vol2:4}
L.~D. Landau and E.~M. Lifshitz, {\em The classical theory of fields}, vol.~2.
\newblock Pergamon Press, 1951.

\bibitem{York:1972}
J.~W. York, ``Role of conformal three-geometry in the dynamics of
  gravitation,''
\href{http://dx.doi.org/10.1103/PhysRevLett.28.1082}{{\em Phys. Rev. Lett.}
  {\bf 28} (1972)  1082--1085}.

\bibitem{Gibbons:Hawking:1977}
G.~W. Gibbons and S.~W. Hawking, ``Action integrals and partition functions in
  quantum gravity,''
\href{http://dx.doi.org/10.1103/PhysRevD.15.2752}{{\em Phys. Rev. D} {\bf 15}
  (1977)  2752--2756}.

\bibitem{Wald:1984}
R.~M. Wald, {\em General Relativity}.
\newblock The University of Chicago Press, Chicago 60637, 1984.

\bibitem{York:1986}
J.~W. York, ``Boundary terms in the action principles of general relativity,''
  \href{http://dx.doi.org/10.1007/BF01889475}{{\em Found. Phys.} {\bf 16}
  (1986)  249--257}.

\bibitem{Brown:York:1993}
J.~D. Brown and J.~W. York, ``Quasilocal energy and conserved charges derived
  from the gravitational action,''
  \href{http://dx.doi.org/10.1103/PhysRevD.47.1407}{{\em Phys. Rev. D} {\bf 47}
  (1993)  1407--1419},
\href{http://arxiv.org/abs/gr-qc/9209012}{{\tt arXiv:gr-qc/9209012}}.

\bibitem{Hawking:Horowitz:1996}
S.~W. Hawking and G.~T. Horowitz, ``The gravitational {H}amiltonian, action,
  entropy and surface terms,''
  \href{http://dx.doi.org/10.1088/0264-9381/13/6/017}{{\em Class. Quant. Grav.}
  {\bf 13} (1996)  1487--1498},
\href{http://arxiv.org/abs/gr-qc/9501014}{{\tt arXiv:gr-qc/9501014}}.

\bibitem{Galtsov:2002}
D.~V. Gal'tsov, ``{R}adiation reaction in various dimensions,''
  \href{http://dx.doi.org/10.1103/PhysRevD.66.025016}{{\em Phys. Rev. D} {\bf
  66} (2002)  025016}, \href{http://arxiv.org/abs/hep-th/0112110}{{\tt
  arXiv:hep-th/0112110}}.

\bibitem{Cardoso:Dias:Lemos:2003}
V.~Cardoso, {\'O}.~J.~C. Dias, and J.~P.~S. Lemos, ``{G}ravitational radiation
  in ${D}$-dimensional spacetimes,''
  \href{http://dx.doi.org/10.1103/PhysRevD.67.064026}{{\em Phys. Rev. D} {\bf
  67} (2003)  064026}, \href{http://arxiv.org/abs/hep-th/0212168}{{\tt
  arXiv:hep-th/0212168}}.

\bibitem{Blanchet:Damour:EspositoFarese:Iyer:2005}
L.~Blanchet, T.~Damour, G.~Esposito-Far{\`e}se, and B.~R. Iyer, ``{D}imensional
  regularization of the third post-{N}ewtonian gravitational wave generation
  from two point masses,''
  \href{http://dx.doi.org/10.1103/PhysRevD.71.124004}{{\em Phys. Rev. D} {\bf
  71} (2005)  124004}, \href{http://arxiv.org/abs/gr-qc/0503044}{{\tt
  arXiv:gr-qc/0503044}}.

\bibitem{Blanchet:Damour:EspositoFarese:2004}
L.~Blanchet, T.~Damour, and G.~Esposito-Far{\`e}se, ``Dimensional
  regularization of the third post-{N}ewtonian dynamics of point particles in
  harmonic coordinates,''
  \href{http://dx.doi.org/10.1103/PhysRevD.69.124007}{{\em Phys. Rev. D} {\bf
  69} (2004)  124007},
\href{http://arxiv.org/abs/gr-qc/0311052}{{\tt arXiv:gr-qc/0311052}}.

\bibitem{Iwanenko:Sokolow:1953}
D.~D. Iwanenko and A.~A. Sokolow, {\em Klassische Feldtheorie}.
\newblock Akademie-Verlag GmbH, Berlin, 1953.

\bibitem{Goldberger:2007}
W.~D. Goldberger, ``{L}es {H}ouches lectures on effective field theories and
  gravitational radiation,''
\href{http://arxiv.org/abs/hep-ph/0701129}{{\tt arXiv:hep-ph/0701129}}.

\bibitem{Galley:2012}
C.~R. Galley, ``The classical mechanics of non-conservative systems,''
\href{http://arxiv.org/abs/1210.2745}{{\tt arXiv:1210.2745 [gr-qc]}}.

\bibitem{Kibble:1963}
T.~W.~B. Kibble, ``Canonical variables for the interacting gravitational and
  {D}irac fields,'' \href{http://dx.doi.org/10.1063/1.1703923}{{\em J. Math.
  Phys.} {\bf 4} (1963)  1433--1437}.

\bibitem{Beig:OMurchadha:1987}
R.~Beig and N.~{{\'O} Murchadha}, ``{T}he {P}oincar{\'e} group as the symmetry
  group of canonical general relativity,''
  \href{http://dx.doi.org/10.1016/0003-4916(87)90037-6}{{\em Ann. Phys. (N.Y.)}
  {\bf 174} (1987)  463--498}.

\bibitem{Barausse:Racine:Buonanno:2009}
E.~Barausse, {\'E}.~Racine, and A.~Buonanno, ``{H}amiltonian of a spinning
  test-particle in curved spacetime,''
  \href{http://dx.doi.org/10.1103/PhysRevD.80.104025}{{\em Phys. Rev. D} {\bf
  80} (2009)  104025},
\href{http://arxiv.org/abs/0907.4745}{{\tt arXiv:0907.4745 [gr-qc]}}.

\bibitem{Jaranowski:Schafer:2012}
P.~Jaranowski and G.~Sch{\"a}fer, ``{T}owards the fourth post-{N}ewtonian
  {H}amiltonian for two-point-mass systems,''
  \href{http://dx.doi.org/10.1103/PhysRevD.86.061503}{{\em Phys. Rev. D} {\bf
  86} (2012)  061503},
\href{http://arxiv.org/abs/1207.5448}{{\tt arXiv:1207.5448 [gr-qc]}}.

\bibitem{Hinder:2010}
I.~Hinder, ``{T}he current status of binary black hole simulations in numerical
  relativity,'' \href{http://dx.doi.org/10.1088/0264-9381/27/11/114004}{{\em
  Class. Quant. Grav.} {\bf 27} (2010)  114004},
  \href{http://arxiv.org/abs/1001.5161}{{\tt arXiv:1001.5161 [gr-qc]}}.

\bibitem{Boos:Davydychev:1987}
E.~E. Boos and A.~I. Davydychev, ``{A} method for calculating vertex-type
  {F}eynman integrals.,'' {\em Vestn. Mosk. Univ. (Ser.3)} {\bf 28} (1987)
  8--12.

\bibitem{Fock:1960}
V.~A. Fock, {\em Theorie von Raum, Zeit und Gravitation}.
\newblock Akademie-Verlag GmbH, Leipziger Stra{\ss}e 3-4, Berlin W1, 1960.

\bibitem{Itoh:2004}
Y.~Itoh, ``{E}quation of motion for relativistic compact binaries with the
  strong field point particle limit: Third post-{N}ewtonian order,''
  \href{http://dx.doi.org/10.1103/PhysRevD.69.064018}{{\em Phys. Rev. D} {\bf
  69} (2004)  064018}, \href{http://arxiv.org/abs/gr-qc/0310029}{{\tt
  arXiv:gr-qc/0310029}}.

\bibitem{Jaranowski:1997}
P.~Jaranowski, ``Technicalities in the calculation of the 3rd post-{N}ewtonian
  dynamics,'' in {\em Mathematics of Gravitation, Part {II}: {G}ravitational
  Wave Detection}, A.~Kr{\'o}lak, ed., pp.~55--63.
\newblock Banach Center Publications, Vol.\ 41, Part II, Warszawa, 1997.

\bibitem{Infeld:Plebanski:1960}
L.~Infeld and J.~Pleba{\'n}ski, {\em {M}otion and relativity}.
\newblock Physical Monographs. Pergamon Press, 4\&5 Fitzroy Square, London.
  W.1., 1960.

\bibitem{Schafer:2009}
G.~Sch{\"a}fer, ``Post-{N}ewtonian methods: {A}nalytic results on the binary
  problem,'' in {\em Mass and Motion in General Relativity}, L.~Blanchet,
  A.~Spallicci, and B.~Whiting, eds.
\newblock Springer, Berlin, 2010.
\newblock
\href{http://arxiv.org/abs/0910.2857}{{\tt arXiv:0910.2857 [gr-qc]}}.
\newblock

\bibitem{Abramowitz:Stegun:1964}
M.~Abramowitz and I.~A. Stegun, {\em Handbook of Mathematical Functions with
  Formulas, Graphs, and Mathematical Tables}.
\newblock Dover Publications, 1964.

\bibitem{Riesz:1949}
M.~Riesz, ``L'int{\'e}grale de {R}iemann-{L}iouville et le probl{\`e}me de
  {C}auchy,'' \href{http://dx.doi.org/10.1007/BF02395016}{{\em Acta Math.} {\bf
  81} (1949)  1--222}.

\bibitem{Riesz:1949:err}
M.~Riesz, ``Erratum: {L}'int{\'e}grale de {R}iemann-{L}iouville et le
  probl{\`e}me de {C}auchy,'' \href{http://dx.doi.org/10.1007/BF02395017}{{\em
  Acta Math.} {\bf 81} (1949)  223(E)}.

\bibitem{Prudnikov:Brychkov:Marichev:1986:vol3}
A.~P. Prudnikov, Y.~A. Brychkov, and O.~I. Marichev, {\em More Special
  Functions}, vol.~3 of {\em Integrals and Series}.
\newblock Gordon and Breach, 1986.

\bibitem{Klein:1933}
F.~Klein, {\em Vorlesungen {\"u}ber die hypergeometrische Funktion}, vol.~39 of
  {\em Grundlehren der mathematischen Wissenschaften}.
\newblock Springer, 1933.

\bibitem{Olver:Lozier:Boisvert:Clark:2010}
F.~W.~J. Olver, D.~W. Lozier, R.~F. Boisvert, and C.~W. Clark, {\em {NIST}
  {H}andbook of {M}athematical {F}unctions}.
\newblock Cambridge University Press, Cambridge, 1st~ed., 2010.

\end{thebibliography}
